\documentclass[aps,twocolumn,prx,superscriptaddress,longbibliography]{revtex4-2}
\usepackage{amsmath}
\usepackage{amssymb}
\usepackage{enumerate}
\usepackage{mathrsfs}
\usepackage{cases}
\usepackage{color}
\usepackage{tikz}
\usepackage{appendix}
\usepackage{makecell}
\usepackage{multirow}
\usepackage{algpseudocode}
\usepackage{algorithmicx,algorithm}
\usepackage{graphicx}
\usepackage{subfig}
\definecolor{darkblue}{rgb}{0.0, 0.0, 0.55}
\definecolor{bordeaux}{rgb}{0.34, 0.01, 0.1}
\usepackage[colorlinks,bookmarks=false,citecolor=blue,linkcolor=blue,urlcolor=blue]{hyperref}
\usetikzlibrary{arrows,positioning}
\usepackage{siunitx}
\usepackage{xcolor,colortbl}
\usepackage{changes}
\usepackage{mathtools}

\usepackage[capitalise]{cleveref}

\def\bR{{\mathbb{R}}}
\def\C{{\mathbb{C}}}

\def\bc{{\mathbf{c}}}
\def\bd{{\mathbf{d}}}

\def\cB{{\mathcal{B}}}
\def\cP{{\mathcal{P}}}
\def\A{{\mathcal{A}}}

\def\M{{\mathcal{M}}}

\def\R{{\mathcal{R}}}
\def\MM{{\mathbf{M}}}

\def\mybar{{}}

\newcommand{\bra}[1]{\langle #1|}
\newcommand{\ket}[1]{|#1\rangle}

\newcommand{\expect}[1]{\langle #1\rangle}

\def\diag{\hbox{\rm{diag}}}

\def\NF{\hbox{\rm{NF}}}
\def\tr{\hbox{\rm{tr}}}

\def\int{\hbox{\rm{int}}}

\def\GS{\text{GS}}
\def\LB{\text{LB}}
\def\UB{\text{UB}}

\definecolor{dkgreen}{rgb}{0.0, 0.2, 0.13}

\definecolor{jacopino}{rgb}{0.65,0.42,1}

\newcommand{\revision}[1]{{{#1}}}

\newif\ifcomment
\commentfalse
\commenttrue
\begin{document}

\title{Certifying ground-state properties of \revision{quantum} many-body systems}

\author{Jie Wang}
\affiliation{Academy of Mathematics and Systems Science, Chinese Academy of Sciences, China}
\author{Jacopo Surace}
\affiliation{ICFO-Institut de Ciencies Fotoniques, The Barcelona Institute of Science and Technology, Av. Carl Friedrich Gauss 3, 08860 Castelldefels (Barcelona), Spain}
\affiliation{Perimeter Institute for Theoretical Physics, 31 Caroline Street North, Waterloo, ON N2L 2Y5, Canada}
\author{Ir\'en\'ee Fr\'erot}
\affiliation{Univ. Grenoble Alpes, CNRS, Grenoble INP, Institut N\'{e}el, 38000 Grenoble, France}
\affiliation{Laboratoire Kastler Brossel, Sorbonne Universit\'e, CNRS, ENS-PSL Research University, Coll\`ege de France, 4 Place Jussieu, 75005 Paris, France}
\author{Beno\^{\i}t Legat}
\affiliation{ESAT, KU Leuven, Leuven}
\author{Marc-Olivier Renou}
\affiliation{Inria Paris-Saclay, Bâtiment Alan Turing, 1, rue Honoré d’Estienne d’Orves – 91120 Palaiseau 
}
\affiliation{CPHT, Ecole polytechnique, Institut Polytechnique de Paris, Route de Saclay – 91128 Palaiseau
}
\author{Victor Magron}
\affiliation{LAAS-CNRS \& Institute of Mathematics from Toulouse, France}
\author{Antonio Ac\'{\i}n}
\affiliation{ICFO-Institut de Ciencies Fotoniques, The Barcelona Institute of Science and Technology, Av. Carl Friedrich Gauss 3, 08860 Castelldefels (Barcelona), Spain}
\affiliation{ICREA - Instituci\'{o} Catalana de Recerca i Estudis Avan\c{c}ats, 08010 Barcelona, Spain.}
\date{\today}

\begin{abstract}
A ubiquitous problem in quantum physics is to understand the ground-state properties of many-body systems. Confronted with the fact that exact diagonalisation quickly becomes impossible when increasing the system size, variational approaches are typically employed as a scalable alternative: energy is minimised over a subset of all possible states and then different physical quantities are computed over the solution state. Despite remarkable success, rigorously speaking, all what variational methods offer are upper bounds on the ground-state energy. On the other hand, so-called relaxations of the ground-state problem based on semidefinite programming represent a complementary approach, providing lower bounds to the ground-state energy. However, in their current implementation, neither variational nor relaxation methods offer provable bound on other observables in the ground state beyond the energy. In this work, we show that the combination of the two classes of approaches can be used to derive certifiable bounds on the value of any observable in the ground state, such as correlation functions of arbitrary order, structure factors, or order parameters. We illustrate the power of this approach in paradigmatic examples of 1D and 2D spin-one-half Heisenberg models. To improve the scalability of the method, we  exploit the symmetries and sparsity of the considered systems to reach sizes of hundreds of particles at much higher precision than previous works. Our analysis therefore shows how to obtain certifiable bounds on many-body ground-state properties beyond energy in a scalable way. 
\end{abstract}

\maketitle

\def\i{\hbox{\bf{i}}} 

\section{Introduction}
The quantitative description of many-body quantum systems is one of the most important challenges in physics. A standard formulation of the problem consists of $N$ particles each described by a Hilbert space of dimension $d$ with interactions encapsulated by a Hamiltonian $H$ acting on $\mathbb{C}^{d^N}$. Typical questions of interest are the study of the system Hamiltonian evolution, the computation of the energy spectrum, or the characterisation of thermal and ground-state properties. A brute force approach to these problems requires diagonalisation of the Hamiltonian, or more generally, dealing with matrices whose dimension, equal to $d^N$, grows exponentially with the number of particles. It is therefore intractable beyond small clusters of particles. 

Among quantum many-body problems, the study of ground states plays a central role due to its relevance for the understanding of the low-energy phases in the system, and in particular for the study of genuine quantum correlation properties without a classical analog~\cite{sachdev2011quantum}. Formally, a ground state $\ket{\psi_\GS}$ of a quantum Hamiltonian $H$ is a state of minimal energy, that is, a minimiser to the following problem:
\begin{equation}
\label{eq:gs}
E_\GS=\min_{\ket{\psi} \in \mathbb{C}^{d^N}} \bra\psi H \ket\psi .
\end{equation}
As mentioned, if $\ket{\psi}$ is decomposed in some given basis of the Hilbert space with an exponentially-growing number of parameters, the exact solution to this optimisation problem quickly becomes intractable when increasing the system size. In fact, the very enumeration of the exact ground-state coefficients in a given basis is out of reach. 

To solve this issue, the standard approach consists of finding approximations to the optimisation that provide a much better scaling in terms of computation. By far, the most popular approach is given by variational methods~\cite{wu2023variational}. There, the minimisation in \cref{eq:gs} is restricted to a subset of Ansatz states $\A$ for which the computation of the expectation value $\bra\psi H \ket\psi$ and its minimisation is scalable with the number of particles,
\begin{equation}
\label{eq:ansatz}
E_{\A}=\min_{\ket{\psi}\in\A} \bra\psi H \ket\psi .
\end{equation}
From an optimal solution state $\ket{\psi_\A}$, one then computes the value of some physically-relevant observables $o_\A=\bra{\psi_\A} O \ket{\psi_\A}$. 
In these methods, the hope is that the set of Ansatz states $\A$ is suitably chosen so that the obtained energy and state are close to the unknown exact values, $E_\A\sim E_\GS$ and $\ket{\psi_\A}\sim \ket{\psi_\GS}$, and so are other physically-relevant quantities, $o_\A \sim o_\GS=\bra{\psi_\GS} O \ket{\psi_\GS}$.

Despite remarkable success, variational methods suffer two major limitations: First, they provide no mathematical guarantee that the derived upper bound $E_\A$ is close to the exact ground-state energy $E_\GS$, although some physically-motivated criteria may be used, for instance involving the energy variance in the variational state \cite{wu2023variational}. Second, and more problematically, even with a promise that $E_\A$ is close to $E_\GS$, there is absolutely no guarantee that the state $\ket{\psi_\A}$ is close to the ground state $\ket{\psi_\GS}$ (unless one has more information about the system of interest, such as its energy gap).  Hence, for observables other than the energy, it is not known how the computed value $o_\A$ relates to the actual ground-state value $o_\GS$, and in particular whether $o_\A$ represents a lower or upper bound to $o_\GS$. As it turns out, $o_\A$ may significantly differ from $o_\GS$, as for instance strikingly observed in some fermionic Hubbard models~\cite{zhengetal2017}. These issues shall be discussed in more details in the core of the paper.

Complementary approaches to variational methods, and which are the focus of the present work, are so-called relaxations of the ground-state problem. The general idea is to minimise the energy over a set of parameters that contains all the physically-possible expectation values $\bra\psi H \ket\psi$, but also other values that are not allowed by quantum mechanics. As for variational methods, the optimisation in the relaxation has a much better scaling than exact diagonalisation and can be computed for larger system sizes. An example of this approach is given by the semidefinite programming (SDP) relaxation to non-commutative polynomial optimisation problems, which have been considered in many different ground-state problems, see for instance~\cite{Nakata01,Mazziotti01,BarthelH2012,Baumgratz_2012}, and formalised in~\cite{navascues2008convergent,pironio2010convergent}, see also~\cite{helton2004positivstellensatz,doherty08}. 
This relaxation, also known as the Navascu\'es-Pironio-Acin (NPA) hierarchy, plays a fundamental role in this article and is detailed below. By construction, as the minimisation is performed over a set of solutions that contains all the quantum physical values, but also other that do not have a quantum realization, any relaxation provides a lower bound of the exact ground-state energy $E_\R\leq E_\GS$. 
Relaxations can also provide estimates $o_\R$ to the expectation value of observables $O$ in the ground state, but in contrast to variational methods, one cannot guarantee that these estimates are compatible with some underlying quantum state. Furthermore, as for variational methods, there is no \textit{a priori} guarantee that these values are close to those of the ground state, $o_\R\sim o_\GS$, neither whether they represent upper or lower bounds to the ground-state values. In summary, combining the present techniques, all what can be certified about ground states is that its energy lies within the range $E_\GS\in[E_\R,E_\A]$.

In this work, we show that the combination of variational methods together with the NPA hierarchy is much richer than previously envisioned, and allows for deriving certified lower and upper bounds, $o_\LB$ and $o_\UB$, on the values of arbitrary observables in the ground state, $o_\GS\in[o_\LB,o_\UB]$. We achieve this by assisting the non-commutative polynomial relaxations with some available upper bound of the ground-state energy as given by variational methods, an approach also considered in~\cite{han2020quantum}. This allows for computing lower and upper bounds to any observable that can be expressed as a polynomial in a family of basic observables. Examples of these operators are correlation functions of arbitrary order, and structure factors which characterize long-range fluctuations in many-body systems. We apply this approach to several paradigmatic spin models. We focus on Heisenberg models with local interactions and translation symmetry in one and two spatial dimensions, and exploit these properties to construct SDP relaxations for systems of up to a hundred of particles, obtaining much better bounds to ground-state observables than achieved in previous works. Our approach therefore provides a scalable way to derive provable bounds on ground-state properties, beyond the energy.

The structure of the article is as follows. In Section \ref{sec_GS_problem} we introduce the quantum ground-state problem. We review both variational methods and relaxations, with an emphasis on SDP relaxations of non-commutative polynomial optimisation because of their central role in our analysis, and see how they provide, respectively, upper and lower bounds to the ground-state energy. In Section \ref{sec_GS_properties} we present our main idea and show how to derive certifiable bounds on the ground-state value of any polynomial observable when combining the two approaches. In Section \ref{sec_applications} we provide several applications and illustrations of our construction. We first introduce the general form of the considered models, and discuss how to exploit their symmetries and the sparsity of the Hamiltonian to reach systems of hundreds of particles. We then apply the method to different Heisenberg models in one- and two-dimensional lattices. In Section~\ref{sec_improvements} we present several directions for future work to improve the scalability and accuracy of the obtained bounds and we finally display our conclusions in Section \ref{sec_discussions}.

\section{Ground-state problem}
\label{sec_GS_problem}
We consider quantum systems composed of $N$ particles, whose interactions are described by a Hamiltonian operator. In what follows, and for simplicity, we are going to focus our discussion on spin systems of finite dimension $d$ on a lattice. The techniques we discuss also apply beyond this scenario, e.g. to fermions or boson models, once one takes into account the respective commutation-type properties of the creation and annihilation operators. We consider systems with local interactions so that it is possible to define a Hamiltonian for an arbitrary number of particles $H_N$ given by a linear combination of different tensor product terms $h_i$ acting on a subset of neighbouring particles, $H_N=\sum_i a_i h_i$. One is then interested in determining the value $o_\GS^{(N)}$ of relevant physical observables $O_N$ in the ground state, $\ket{\psi_\GS^{(N)}}$, see also \cref{eq:gs}, namely $o_\GS^{(N)}=\bra{\psi_\GS^{(N)}} O_N\ket{\psi_\GS^{(N)}}$. Often, one is also interested in the thermodynamic limit of an infinite number of particles, which are typically inferred using scaling considerations by studying the dependence of $o_\GS^{(N)}$ with $N$. In what follows, we often remove the dependence on $N$ to simplify the notation. 

Exactly solving the ground-state problem is computationally too costly already for systems of several tens of particles, as the Hilbert-space dimension of the systems ($d^N$) grows exponentially with the number of particles. Hence one should abandon looking for an exact solution to the problem and adopt approximations to it, such as variational methods or relaxations, which offer a much more favorable scaling with the number of particles $N$.

\subsection{Variational methods}

Variational methods restrict the ground-state optimisation to a subset of Ansatz states $\A$. It is then demanded that the number of parameters needed to specify an Ansatz state, $\ket{\psi_\A}$, scales polynomially with the number of particles, $N$, and that the computation of expectation values of the operators in the Hamiltonian, $\bra{\psi_\A} h_i\ket{\psi_\A}$, is efficient. This allows for solving the energy minimisation over Ansatz states, \cref{eq:ansatz}, for systems much larger than those for which an exact diagonalisation is possible.

Mean field is one of the simplest instances of a variational method, where the set of Ansatz states is defined by product states. Here, the number of parameters scales linearly with the system size, $Nd$, and the mean value of the local terms $h_i$ is easy to compute. The Density-Matrix-Renormalisation-Group (DMRG) approach has represented a breakthrough in the design of variational methods, for it often allows one to obtain good approximations to the ground-state energy of gapped 1D systems \cite{WhiteDMRG,schollwock_DMRG}. It is now well understood that the Ansatz states relevant in DMRG are the so-called Matrix-Product States (MPS), whose description requires $Nd\chi^2$ parameters \cite{OstlundRommer,SCHOLLWOCK201196,doi:10.1080/14789940801912366}.
Here $\chi$ is the so-called bond dimension that determines the entanglement properties of the MPS. 
In fact, product states, used in mean-field calculations, are MPS of bond dimension $\chi=1$.
It is known that DMRG works well for 1D systems because ground states of gapped 1D systems with local interactions can be approximated by MPS of fixed bond dimension, that is, DMRG is optimising over a set of states that contains a very good approximation to the unknown ground state~\cite{Ciracetal2021}. Using insights from entanglement theory, it was possible to generalise MPS to other subset of states, such as Projected-Entangled-Pair State (PEPS) \cite{ORUS2014117} or Multi-Entanglement-Renormalization-Ansatz~\cite{vidal1,vidal2,Evenbly_2014}, which may be viewed as special instances of the more general set of Tensor-Network states \cite{Ciracetal2021}. Other popular variational states not based on tensor networks include resonating valence bond states, introduced in the context of quantum magnetism \cite{RVB}, neural-network quantum states \cite{carleo_treuer}, or correlated-plaquette states~\cite{Mezzacapo_2009,PhysRevB.80.245116,PhysRevB.100.155148}.

It is not our purpose to review here all variational methods; instead, we emphasize that despite all their remarkable applications in the study of many-body quantum systems, variational methods do have intrinsic limitations. First of all, for many systems, it is not known whether the actual ground state can be well approximated by a state in the chosen set of Ansatz states. For instance, mean-field energy values can easily be computed for very large sizes, but it is expected that ground states of generic Hamiltonians are in fact entangled, so that all entanglement properties -- which sometimes form defining properties of the phase as in topological quantum matter \cite{LevinWen2006,wen2017} -- are inaccessible by construction. Second, even when Ansatz states properly approximate the ground state, the minimisation of the energy may remain hard. While an efficient algorithm exists for ground states of 1D gapped local Hamiltonians~\cite{Efficient1D}, it is expected that this is an exception. For instance, ground states of 2D gapped systems are known to be well approximated by PEPS, but the computation of expectation values of product operators with PEPS, including even the norm of the state, is $\#$P-hard in the number of tensors defining the state~\cite{ComplexPEPS}. Third, even if the ground state can be approximated by a given Ansatz state and the computation of the energy is scalable, its minimisation often presents many local minima and therefore one can never guarantee that a good approximation to the ground state has been achieved, $E_\A\sim E_\GS$. 
But most importantly, the situation is even worse for other relevant quantities computed from the Ansatz state resulting from the minimisation, as it is completely unknown how they compare to the actual values in the ground state. 
In fact, there exist simple paradigmatic models displaying a very complex low-energy landscape, with states very close in energy to the ground state, but with significantly different predictions for other quantities -- the fermionic Hubbard model being a prominent example \cite{zhengetal2017}.

In summary, variational methods have proven extremely useful to analyse ground-state problems; yet strictly speaking all what can be certified is an upper bound of the ground-state energy, $E_\GS\leq E_\A$.

\subsection{Relaxations}
So-called relaxations of the ground-state problem represent a complementary approach to variational methods. Quite generically, let us consider a given Hilbert space, and a Hamiltonian of the form $H=\sum_{i=1}^n a_i h_i$ with $h_i$ some (possibly non-commuting) quantum observables, and $a_i$ some coefficients. One may then consider the following problem: what are the possible combinations of mean values $(\langle h_1 \rangle, \langle h_2\rangle, \dots, \langle h_n \rangle$) that are allowed by quantum mechanics? The ground-state problem:
\begin{equation}
\label{eq:gs2}
E_\GS=\min_{\{\expect{h_i}\}_i \in \M_Q}\sum_i a_i \expect{h_i} ,
\end{equation}
where 
\begin{equation*}
\M_Q=\{\{\expect{h_i}\}_i: \exists\ket\psi \text{ such that } \forall i~ \expect{h_i}=\bra{\psi}h_i\ket{\psi}\}, 
\end{equation*}
can then be seen as a special case of this general problem, in which one searches for the minimal value $E_{\rm GS}$ of the linear combination $\sum_i a_i \langle h_i \rangle$: geometrically speaking, one goes as far as possible in the direction $-\vec a = -(a_1, \dots a_n)$ while remaining inside the allowed quantum region for $(\langle h_1\rangle, \dots, \langle h_n\rangle)$ \footnote{Notice that $\M_Q$ forms a convex set: this property follows directly from the convexity of the set of quantum density matrices.}. The expectation values $\expect{h_i}$ are often called \emph{moments} and therefore $\M_Q$ is the set of quantum physical moments. For many-body problems, where the Hilbert space dimension grows exponentially with the system size, the characterization of the set $\M_Q$ is generically very hard. Specifically, it contains as a special case the so-called quantum marginal problem, which is QMA-hard~\cite{QMA_Repr}. 

In such hard instances, one can however approximate the set $\M_Q$ of physical moments from the outside; this may be achieved for instance by deriving independent bounds on the spectra of the operators $h_i$. This is an example of a so-called \textit{relaxation} to the problem -- one relaxes certain constraints related, e.g. to the non-commutativity of the observables. When performing such a relaxation, one does not include all the constraints that stem from the Hilbert space structure of the problem, and from the algebraic relations between non-commuting observables. One only takes into account a subset of all such constraints. One defines in this way a superset $\M \supseteq \M_Q$ of moments compatible with such relaxed constraints. The set $\M$ contains all quantum physical expectation values $\expect{h_i}=\bra{\psi}h_i\ket{\psi}\in\M_Q$, but also other values that cannot be written in this form, and therefore do not have a quantum realization.

By construction, when the minimisation is performed over the larger set of moments $\M \supseteq \M_Q$, the obtained energy $E_\R$ cannot be larger than the actual quantum ground-state value, hence $E_\R\leq E_\GS$.

\subsubsection{A simple relaxation: Anderson's bound}
As the concept of relaxation is a central paradigm to this work, let us illustrate some of the above-mentioned basic aspects on a minimal example of three qubits, using a framework initially introduced by Anderson in the condensed-matter literature \cite{AndersonBound}. We consider the Hamiltonian $H = \vec \sigma_1 \cdot \vec \sigma_2 + \vec \sigma_2 \cdot \vec \sigma_3$, namely a so-called Heisenberg model for three qubits, $\vec \sigma_i = (\sigma_i^x, \sigma_i^y, \sigma_i^z)$ being the vector of Pauli matrices (Heisenberg models in different geometries will be the focus of our paper in Section \ref{sec_applications}). The Hamiltonian $H$ is of the form $H=h_1+h_2$, and the operators $h_1=\vec \sigma_1 \cdot \vec \sigma_2$ and $h_2=\vec \sigma_2 \cdot \vec \sigma_3$ do not commute. Anderson's bound states that $E_{\rm GS} \ge \min \expect{h_1} + \min\expect{h_2}$, where the $\min\expect{h_i}$ are independently obtained by minimizing over quantum states for two qubits, neglecting the constraints stemming from the appearance of the same operators $\vec\sigma_2$ in both $h_1$ and $h_2$ (or equivalently, neglecting the fact that $h_1$ and $h_2$ do not commute).

\begin{figure}[!ht]
	\centering
    \includegraphics[width=1\linewidth]{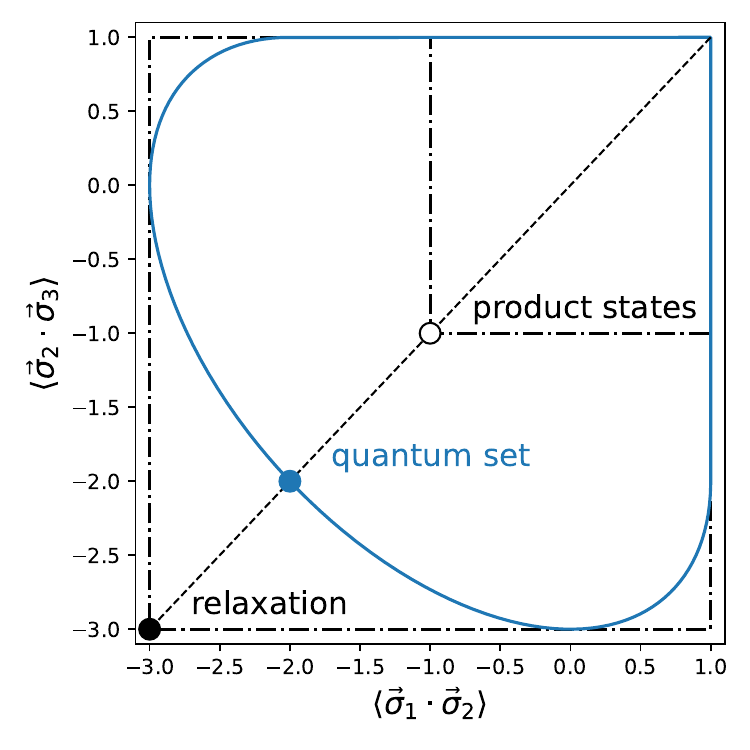}
	\caption{Relaxing the quantum ground-state problem with Anderson's bound: a three-qubit example (see text).}
 \label{fig_baby_heisenberg}
\end{figure}

 Figure \ref{fig_baby_heisenberg} shows the range of possible values for $(\langle h_1 \rangle, \langle h_2 \rangle)$ over three-qubit quantum states: the quantum set (in blue), as well as the outer-approximation given by Anderson's relaxation, which amounts to bound $\langle h_1 \rangle$ and $\langle h_2 \rangle$ independently [here: $\vec \sigma_1 \cdot \vec \sigma_2 = 2 \vec S_{\rm tot}^2 - 3$ with the total spin $\vec S_{\rm tot}=(\vec \sigma_1 + \vec \sigma_2)/2$, such that $\vec S_{\rm tot}^2$ takes its minimal value $0$ in the singlet state $(\ket{\uparrow\downarrow} - \ket{\downarrow\uparrow})/\sqrt2$, and its maximal value $2$ in the triplet states ($\ket{\uparrow\uparrow}, (\ket{\uparrow\downarrow} + \ket{\downarrow\uparrow})/\sqrt2, \ket{\downarrow\downarrow}$), hence the bounds $-3 \le \langle \vec \sigma_1 \cdot \vec \sigma_2  \rangle \le 1$, and similarly for $\langle \vec \sigma_2 \cdot \vec \sigma_3  \rangle$]. The Anderson's bound then gives $E_\R=-6 \le \langle H \rangle$ (black dot), a value which is not allowed in quantum mechanics, since there is no three-qubit state such that $\langle \vec \sigma_1 \cdot \vec \sigma_2  \rangle=\langle \vec \sigma_2 \cdot \vec \sigma_3  \rangle = -3$. In fact, the actual quantum bound is $E_{\rm GS} = -4$ (blue dot). Instead, optimizing over variational states, one would approximate the quantum set from the inside. For instance, in mean field, one mimimizes the energy over product states of the form $\ket\psi=\ket{\psi_1}\otimes\ket{\psi_2}\otimes\ket{\psi_3}$, which obey $-1 \le \langle \vec \sigma_i \cdot \vec \sigma_j  \rangle \le 1$, and therefore one obtains the upper bound $\langle H \rangle_{\rm min} \le E_\A=-2$ (open dot). As expected, one has $E_\R\leq E_\GS\leq E_\A$.

More generally, in order to obtain the Anderson's lower-bound to the ground-state energy of a translationally-invariant $N$-particle model, one performs exact diagonalisation of the same model but on a smaller size $K<N$, with open boundary conditions. As the set of moments compatible with an $N$-particle quantum state is strictly included in the one compatible with a quantum state of $K<N$ particles, it follows that the Anderson's bound is a lower-bound to the $N$-particle ground-state energy (see Appendix \ref{app_Anderson_bound} for further details).

\subsubsection{Relaxations based on semidefinite programming}

Relaxations of polynomial optimisation problems based on SDP provide another way to derive lower bounds using a similar argument. Because of the central role they play in our analysis, we first explain the main idea of the construction and then provide more details about its implementation, in particular in the context of ground-state problems. These relaxations apply to any optimisation problem of the form~\cite{pironio2010convergent}:
\begin{eqnarray}\label{eq:polopt}
p_{\min}=
&\min\limits_{\{\ket\psi,\mybar X\}}&\bra\psi p(\mybar X)\ket\psi\\
&\text{such that:} \quad& g_j(\mybar X) \succeq 0 \,\,\quad\quad\quad\hspace{0.2cm} j=1,\ldots ,m_1, \nonumber\\
&& \bra\psi h_k(\mybar X)\ket\psi\geq 0 \quad k=1,\ldots ,m_2.\nonumber
\end{eqnarray}
Here, $p$, $g_j$ and $h_k$ are polynomials defined over a set of $n$ operators $\mybar X=(X_1,\ldots,X_n)$, $\succeq$ refers to operator positivity, and $m_1$ and $m_2$ denote the number of polynomial constraints of each type in Eq.~\eqref{eq:polopt}. It is important to mention that the minimisation in Eq.~\eqref{eq:polopt} is over all possible tuples of operators $\mybar X$ satisfying the constraints, that is, it runs over all possible Hilbert spaces in which these operators can be defined.
For simplicity in the notation, we restrict the explanation to self-adjoint operators, although all what follows also applies to general operators. 
The solution to problem~\eqref{eq:polopt} may be hard. 
For instance, deciding whether the solution to this optimisation problem satisfies either $p_{\min}<0$ or $p_{\min}>1$ is Turing undecidable~\cite{slofstra_2019}. 

In Refs. \cite{navascues2008convergent,pironio2010convergent}, it was shown how to construct an infinite hierarchy, known as NPA, of  monotonically increasing lower bounds to the solution of the problem, $p^{(1)}\leq p^{(2)}\leq\cdots\leq p^{(\infty)}\leq p_{\min}$. 
Under some mild assumptions on the operators, that include the situation in which all the operators $X_i$ are bounded, the NPA hierarchy is convergent, that is $p^{(\infty)}= p_{\min}$. 
The main advantage of the hierarchy is that the computation of each lower bound defines a SDP instance and therefore can efficiently be performed. 
However, the size of the operators involved in each SDP step of the hierarchy is also monotonically growing and, in the limit, involves operators of infinite size.
Interestingly, 
for some problems, convergence is attained at a finite step in the hierarchy, while for others, first steps of the NPA hierarchy provide a good enough bound of the actual solution.  

Because of the generality of the formalism, many ground-state problems can be phrased in the language of polynomial optimisation. Fermionic Hamiltonians are often given by polynomials of creation and annihilation operators, which satisfy the anti-commutation relations in form of other polynomials, hence defining an instance of \eqref{eq:polopt}.  For spin-one-half systems, Hamiltonians are defined by polynomials of Pauli matrices on each site. 
Pauli matrices acting on the same site can be characterised by their algebra while operators on different sites commute, all these constraints having the form of polynomials [see Eq.~\eqref{ncpop_main}].  
Moreover, for the Pauli algebra, these constraints impose that the solution Hilbert space of Eq.~\eqref{eq:polopt} is of finite dimension, in particular of dimension two. 
All these ground-state problems define particular instances of the general optimisation in Eq.~\eqref{eq:polopt} where the polynomial to be minimised is defined by the Hamiltonian, $p(X)=H(X)$. The NPA relaxation therefore provides a rather versatile approach to derive an asymptotically convergent sequence of lower bounds to the ground-state energy of many models of interest, $E^{(1)}\leq E^{(2)}\leq\cdots\leq E^{(\infty)}=E_\GS$. In fact, as already mentioned, it has already been applied to different models, e.g., in the context of quantum chemistry~\cite{Nakata01,Mazziotti01}, many-body physics~\cite{Baumgratz_2012,BarthelH2012}, or conformal bootstrap~\cite{conformal}, often before or unaware of the general mathematical characterisation of non-commutative polynomial optimisation presented in~\cite{navascues2008convergent,pironio2010convergent}. As it happens for variational methods, relaxations also provide values for other observables beyond energy, $o_\R$, but again with no control about whether they are close to or bound in any way the value in the ground state. Therefore, when taken together, all what relaxations and variational methods define is an energy interval in which the searched ground-state energy lies, $E_\R\leq E_\GS\leq E_\A$. 

Before concluding this part, it is worth mentioning that SDP relaxations of polynomial optimisation problems have a long tradition in the context of classical spin systems. In this case, the mathematical framework is the one of commutative polynomial optimisation: 
\begin{eqnarray}
\label{eq:compolopt}
p_{\min}=&\min\limits_{\mybar x} &p(\mybar x) \nonumber\\
& \text{such that:} \quad &g_j(\mybar x) \geq 0 \quad j=1,\ldots ,m ,
\end{eqnarray}
where $p$ and $g_j$ are again polynomials but now over (classical, namely commuting) variables $\mybar x=(x_1,\ldots,x_n)$ and one deals with standard positivity constraints. Many classical spin systems have the form of polynomials over spin variables $\sigma_i$ such that $\sigma_i^2=1$, again a polynomial constraint. A hierarchy of SDP relaxations of commutative polynomial optimisation problems was introduced in~\cite{lasserre2001global} and, in fact, the formalism of~\cite{pironio2010convergent} used in this work can be understood as the extension of the construction in~\cite{lasserre2001global} to the non-commutative case, see~\cite{pironio2010convergent}.\\

\section{Certification of ground-state properties}
\label{sec_GS_properties}

Variational methods and relaxations of polynomial problems are often seen as two complementary approaches that allow one to bound the ground-state energy from above and below. The main point of our work is to show that their combination is much richer than expected, as together they can be used to derive certifiable bounds on any observable of interest in the ground state. 

The idea is quite simple and was also discussed in \cite{han2020quantum}. 
As mentioned above, the ground-state energy problem can be seen as an instance of polynomial optimisation because the Hamiltonian can be expressed as polynomials of some operators $X_i$. 
For instance, for finite dimensional systems, it is enough to take as $X_i$ a basis for the space of matrices at each site, say Pauli matrices for qubit systems. But
in fact, any observable of interest $O$ can be expressed as polynomial on these operators and bounds on it can be derived through the NPA formalism by taking $p(X)=O(X)$. A direct application of the method would provide rather trivial bounds, because the optimisation is not restricted to a region close to the ground state. To enforce this, one can use the best upper bound $E_{\A}$ to the ground-state energy derived through variational methods, as well as the best lower bound $E_\R$ derived through relaxations. This is because these two bounds also have a polynomial form, $\bra\psi E_{\A}-H(\mybar X)\ket\psi\geq 0$ and $\bra\psi H(\mybar X) - E_\R\ket\psi\geq 0$ for the upper and lower bounds, and can be added as additional constraints. The resulting optimisation reads, c.f.~\eqref{eq:polopt}:
\begin{eqnarray}
\label{eq:obsopt}
o_\LB=&\min\limits_{\{\ket\psi,\mybar X\}}&\bra\psi O(\mybar X)\ket\psi \nonumber\\
& \text{such that:} \quad &g_j(\mybar X) \succeq 0 \,\,\quad\quad\quad\hspace{0.2cm} j=1,\ldots ,m_1, \nonumber\\
& \quad\quad\quad &\bra\psi h_k(\mybar X)\ket\psi\geq 0 \quad k=1,\ldots ,m_2, \nonumber\\
& \quad\quad\quad &\bra\psi E_{\A}-H(\mybar X)\ket\psi\geq 0, \nonumber\\
& \quad\quad\quad &\bra\psi H(\mybar X) - E_\R\ket\psi\geq 0.
\end{eqnarray}
The different SDP relaxations of this minimisation provide a sequence of lower bounds to the actual value of the observable in the ground state, $o^{(1)}\leq\cdots\leq o^{(\infty)}\leq o_\GS$. Note that in this case, the asymptotic value $o^{(\infty)}$ is only guaranteed to be equal to the actual ground-state value $o_\GS$ if $E_{\A}=E_\GS$. 
Finally, upper bounds $o_\UB$ can be derived just by replacing the minimisation in \eqref{eq:obsopt} by a maximisation, obtaining the announced certifiable bounds for any observable in the ground state, $o_\GS\in[o_\LB,o_\UB]$.

To illustrate the power of this method, we apply it in what follows to several paradigmatic Heisenberg models for spin-$1/2$ systems.

\section{Applications and results}
\label{sec_applications}
We present several implementations of the method to obtain certified bounds on ground-state observables for various Heisenberg models in one and two spatial dimensions \footnote{Our codes for reproducing the results are available at \url{https://github.com/wangjie212/QMBCertify}. See \url{https://github.com/blegat/CondensedMatterSOS.jl} for other related codes.}. Generic Heisenberg models are defined by Hamiltonians of the form:
\begin{equation}
    H = (1/4)\sum_{i < j} J_{ij} \sum_{a\in\{x,y,z\}} \sigma_i^a \sigma_{j}^a ~,
    \label{eq_generic_Heisenberg}
\end{equation}
where $i\in\{1,2,\dots,N\}$ label the lattice sites, while the couplings $J_{ij}$ implicitly define the lattice geometry and $\sigma_i^a$ are the Pauli matrices acting on site $i$. The $1/4$ prefactor follows standard condensed-matter conventions, where Hamiltonians are typically defined in terms of spin operators $s_i^a = \sigma_i^a / 2$ instead of Pauli matrices. We shall consider four different geometries:
\begin{enumerate}
    \item The Heisenberg model with first-neighbour interactions on a 1D lattice, $J_{ij} = \delta_{j,i+1}$, with periodic boundary conditions (PBC), namely we use the convention that $N+1 \equiv 1$ for the $i,j$ labels.
    \item A 1D lattice with first- and second-neighbour couplings, $J_{ij} = \delta_{j,i+1} + J_2\delta_{j,i+2}$, where the $J_2$ term induces geometric frustration.
    \item A 2D square lattice with first-neighbour couplings. Here, lattice sites are labelled by $i=(x,y)$ with $x,y\in\{1,2,\dots, L\}$ (so that $N=L^2$), and couplings are of the form $J_{(x,y),(x',y')} = \delta_{y',y}\delta_{x',x+1} + \delta_{x',x} \delta_{y',y+1}$. We take PBC, namely $L+1 \equiv 1$ for both $x$ and $y$ labels.
    \item A 2D square lattice with first- and second-neighbour (frustration-inducing) couplings, where second neighbours are along the diagonal of elementary square plaquettes, namely, extra couplings of the form $J_2 [\delta_{(x',y'),(x+1,y+1)} + \delta_{(x',y'),(x+1,y-1)}]$.
\end{enumerate}
In all cases, we obtain certified lower bounds on the ground-state energy, as well as upper and lower bounds on relevant observables in the ground state, typically on spin-spin correlation functions. 

\subsection{Algorithmic considerations}
We begin with discussing the concrete SDP algorithm tailored to generic Heisenberg models [\cref{eq_generic_Heisenberg}]. In particular, we briefly discuss how to reduce the size of SDP relaxations by exploiting algebraic structures of the model, which is crucial in order to obtain the optimal results given some computational resource. More details are given in Appendix \ref{app_algorithmic_tricks}.

The ground-state energy of the Heisenberg model is the optimum of the following non-commutative polynomial optimisation problem:
\begin{equation}\label{ncpop_main}
    \begin{aligned}\min\limits_{\{\ket{\psi},\sigma_i^a\}}&\quad \bra\psi H(\{\sigma_i^a\}) \ket\psi\\
    \text{such that:}&\quad(\sigma_i^a)^2=1,\quad i=1,\ldots,N;~a\in\{x,y,z\},\\
    &\quad\sigma_i^x\sigma_i^y=\i\sigma_i^z,\quad\sigma_i^y\sigma_i^x=-\i\sigma_i^z,\quad i=1,\ldots,N,\\
    &\quad\sigma_i^y\sigma_i^z=\i\sigma_i^x,\quad\sigma_i^z\sigma_i^y=-\i\sigma_i^x,\quad i=1,\ldots,N,\\
    &\quad\sigma_i^z\sigma_i^x=\i\sigma_i^y,\quad\sigma_i^x\sigma_i^z=-\i\sigma_i^y,\quad i=1,\ldots,N,\\
    &\quad\sigma_i^a\sigma_j^b=\sigma_j^b\sigma_i^a,~ 1\le i < j\le N;~a,b\in\{x,y,z\}.
    \end{aligned}
\end{equation}
Again, this is because the Hamiltonian can be expressed as a polynomial over the Pauli matrices at each site, $\{\sigma_i^a\}_{i=1,\ldots,N;a\in\{x,y,z\}}$, that is, as a linear combination of some monomials over operators of the form $v_m = \sigma_{i_1}^{a_1}\dots\sigma_{i_{p_m}}^{a_{p_m}}$, where $p_m$ is the degree of the monomial, having $H(\{\sigma_i^a\})=\sum_m c_m v_m$. As explained above, the problem can equivalently be seen as a minimisation of the linear function $\langle H \rangle=\sum_m c_m \expect{v_m}$ over the set of quantum (Pauli) moments $\M_Q=\{\{\expect{v_i}\}_i: \exists\ket\psi \text{ such that } \forall i~ \expect{v_i}=\bra{\psi}v_i\ket{\psi}\}$.

Without entering into the details, the SDP relaxations of~\cite{navascues2008convergent,pironio2010convergent} replace the set of quantum moments $\M_Q$ in the optimisation by larger moment sets $\M$, hence providing lower bounds of the ground-state energy in~\cref{ncpop_main}. This is achieved by introducing the so-called moment matrices. Specifically, suppose that $\cB=\{v_m\}_m$ is a (Pauli) monomial list. For each quantum realisation (namely: for any given quantum state), the quantum moment matrix $\mathbf{M}$ indexed by $\cB$ is defined as $[\MM]_{vw}=\langle v^\dagger w\rangle$. The entries of these quantum moment matrices satisfy a series of linear relations resulting from the constraints in the optimisation, in our case coming from the Pauli algebra and the commutation of operators acting at different sites. It is rather easy to see that, for each choice of the monomial list $\cB$, the corresponding quantum moment matrix is positive, namely, $\mathbf{M}\succeq0$. 
However, the opposite is not necessarily true: there exist positive moment matrices satisfying the linear relations associated to the constraints that do not have a quantum realisation. The different lists of monomials, and corresponding moment matrices, define the different sets of moments $\M$ over which the relaxations are built. For that, it suffices to consider monomial lists such that the energy can be expressed as a linear combination of the entries of the corresponding moment matrix:
\begin{equation}
\langle H \rangle =  \sum_m c_m u_m = \tr(H_M \mathbf{M}) ,    
\end{equation}
for a given matrix $H_M$ depending on the coefficients $c_m$, and on the considered relaxation. 
Then a SDP relaxation to \cref{ncpop_main} is given by
\begin{equation}\label{cmom}
\begin{aligned}
    \min\limits_{\{\langle v^\dagger w\rangle\}_{v,w\in\cB}} &\quad
    \tr(H_M\mathbf{M})\\
    \text{such that:}&\quad\MM\succeq0, \\
    &\quad\MM \text{ obeys some moment replacement rules}.
\end{aligned}
\end{equation}
In particular, the equality constraints in \cref{ncpop_main} give rise to corresponding replacement rules on monomials, allowing one to reduce them to the {\em normal form} $\NF(u)\coloneqq c\sigma_{i_1}^{a_1}\sigma_{i_2}^{a_2}\cdots\sigma_{i_r}^{a_r}$ with $c\in\{1,-1,\i,-\i\}$, $1\le i_1<i_2<\cdots<i_r\le N$. It follows that the moment matrix $\MM$ satisfies the moment replacement rule: $\langle u\rangle=\langle\NF(u)\rangle$ for all entries $\langle u\rangle$ of $\MM$. Note that \eqref{cmom} is a complex SDP. To reformulate \eqref{cmom} as a SDP over real numbers, we refer the reader to \cite{wang2023efficient}. It can be seen that the more monomials we include in $\cB$, the larger size $\mathbf{M}$ has and the tighter lower bound \eqref{cmom} may provide.

Further symmetry considerations on the concrete considered models allow one to drastically reduce the size of the moment matrix $\mathbf{M}$ as well as the number of independent variables involved in \eqref{cmom}. For instance, given the symmetry of Heisenberg models under global rotations of the spins, correlations in the ground state are of the form $(1/4)\langle \sigma_i^a \sigma_j^b \rangle = \delta_{a,b} C_{ij}$. Another relevant symmetry is translation invariance which implies that correlation functions only depend on the relative position of the spins. Details on the technical implementation of those and other symmetries to reduce the computational complexity of the SDP algorithm are given in Appendix~\ref{app_algorithmic_tricks}.

\subsection{Heisenberg chain}
\label{sec:B1}

The Heisenberg chain is defined by the Hamiltonian:
\begin{equation}
    H = (1/4)\sum_{i=1}^N \sum_{a\in\{x,y,z\}} \sigma_i^a \sigma_{i+1}^a ~,
\end{equation}
where $N+1\equiv 1$ in order to implement PBC. The ground state is critical (namely: gapless in the thermodynamic limit) and displays antiferromagnetic correlations decaying as a power-law with distance, $\langle s_i^a s_{i+r}^a \rangle = (1/4)\langle \sigma_i^a \sigma_{i+r}^a \rangle = C_{r} \sim (-1)^r / r^\alpha$ with some exponent $\alpha$ \cite{giamarchi2003}.\\

\begin{figure}[!ht]
	\centering
\includegraphics[width=1\linewidth]{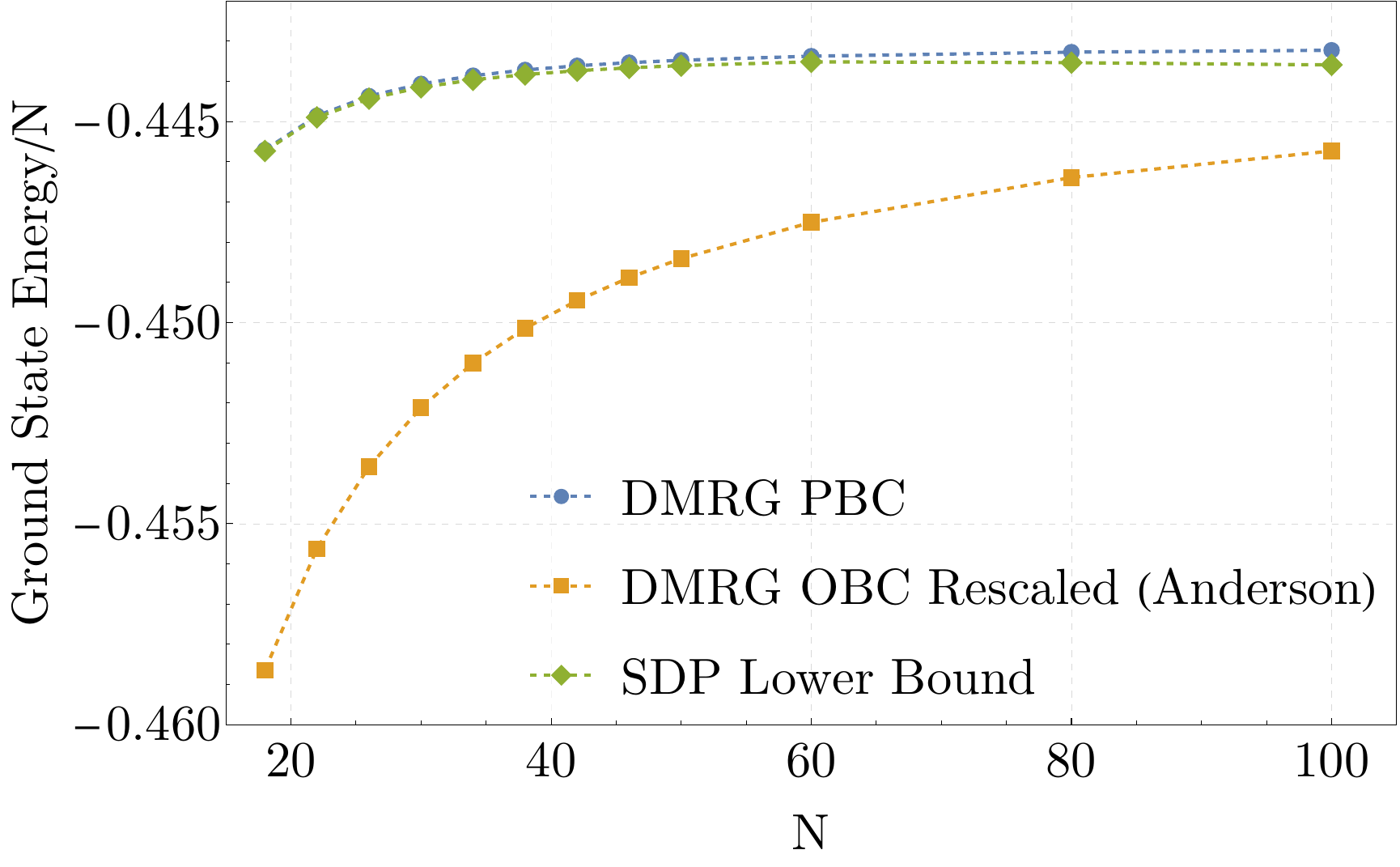}
	\caption{Ground-state energy per particle in the Heisenberg chain (data in Table \ref{Tab:B1-Energies}). Upper and lower bounds are derived, respectively, through DMRG and the implemented SDP relaxation. For comparison, we show the expected Anderson bound, that is, the lower bound one would obtain by exactly solving the same system with OBC. This estimation is also computed through DMRG, so that it is possible to plot system sizes that are out of reach for exact diagonalisation.}
	\label{fig:B1_fig1}
\end{figure}

\noindent\textit{Ground-state energy.--} 
The ground-state energy per spin is given by $e_{\rm PBC}(N) = \langle H \rangle / N = 3C_1$. In Fig.~\ref{fig:B1_fig1}, we plot the best lower bound of $e_{\rm PBC}$ as obtained by our SDP relaxation for up to $N=100$ spins. As a comparison, we plot the (quasi-)exact energy as obtained by DMRG simulations. It is also of interest to compare our SDP lower bound with the Anderson bound, which is obtained by exact diagonalisation on a system with open boundary conditions (OBC), $e_{\rm OBC}(K) \le e_{\rm PBC}(N)$ for all $N>K$ (see Appendix \ref{app_Anderson_bound} for details on the Anderson bound). In Fig.~\ref{fig:B1_fig1}, the SDP bound is seen to vastly outperform the Anderson bound (which is in fact estimated with DMRG for the sake of illustration for up to $N=100$, as beyond a few tens of qubits exact diagonalisation is out of reach). 

To build the moment matrix in the SDP construction, we use all monomials of the form: $1$, $\sigma_i^a$, $\sigma_{i}^a\sigma_{i+j}^b$, $\sigma_{i}^a \sigma_{i+1}^b \sigma_{i+2}^c$, $\sigma_{i}^a \sigma_{i+1}^b \sigma_{i+2}^c \sigma_{i+3}^d$ with $i\in\{1,\ldots,N\}, j \in \{1,2,\dots,r\}$ and $a,b,c,d\in\{x,y,z\}$ (all different monomials appearing only once). For each size $N$, we have chosen $r$ as large as possible compatible with memory limitations, namely $r=\frac{N}{2}$ for $N\le 60$, and $r=20$ for $N=80,100$. Furthermore for $N=100$ we discard all degree-four monomials $\sigma_{i}^a \sigma_{i+1}^b \sigma_{i+2}^c \sigma_{i+3}^d$ in order to allow for more degree-two monomials. For the sake of completeness, the data plotted in Fig.~\ref{fig:B1_fig1} are also reported in Table \ref{Tab:B1-Energies}. Combining both the DMRG upper bound $e_{\rm DMRG}$ and the SDP lower bound $e_{\rm SDP}$ allows us to sandwich the exact ground-state energy with a relative accuracy that remains below $10^{-3}$ up to $N=100$ spins. In contrast, previous works have achieved no better than a few percents accuracy for comparable system sizes \cite{BarthelH2012,Baumgratz_2012,haim2020variationalcorrelations}. The small energy gap between the DMRG (variational upper bound) and the SDP (certified lower bound) therefore certifies both the expected good performance of DMRG to approximate the actual 1D ground state and, in turn, also the good performance of the implemented SDP relaxation.


\subsection{Heisenberg chain with second-neighbour couplings}
\label{sec:B2}
\begin{figure}[!ht]
	\centering
\includegraphics[width=1\linewidth]{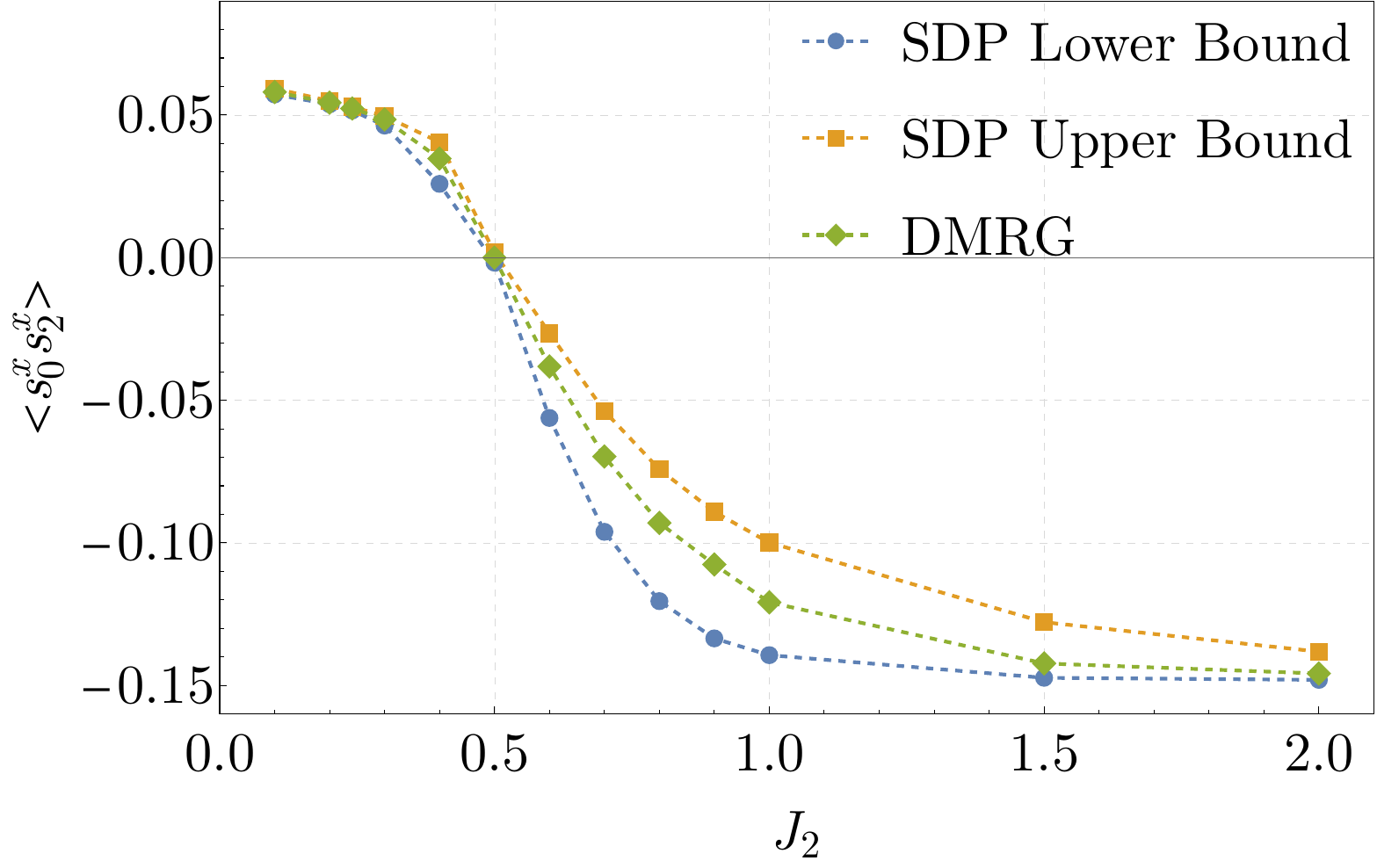}
	\caption{Second-neighbour spin correlations in the $J_1-J_2$ Heisenberg chain are certified to change from ferromagnetic ($C_2>0$) to antiferromagnetic ($C_2<0$) when crossing $J_2=0.5$ ($N=40$; data are given Table \ref{tab:B2-C2}). }
	\label{fig:B2_figC2}
\end{figure}

Our second application is the Heisenberg chain including both first- and second-neighbour couplings, namely the so-called $J_1-J_2$ Heisenberg model:
\begin{equation}
    H = (1/4)\sum_{i=1}^N \sum_{a\in\{x,y,z\}} [\sigma_i^a \sigma_{i+1}^a + J_2 \sigma_i^a \sigma_{i+2}^a] ~,
\end{equation}
with PBC (in our convention the first-neighbour coupling is $J_1=1$). The $J_2$ term induces geometric frustration, leading to the sign problem in quantum Monte Carlo methods and to a richer phase diagram. The model was investigated in early days of DMRG simulations \cite{white_affleck_1996}, and represents a cornerstone in the study of quantum magnetism, motivating the development of various variational wave-functions. In particular, it is predicted that for $J_2<J_{2,c}=0.241167\dots$, the spin correlation length is infinite, and correlations decay as a power-law as in the $J_2=0$ limit \cite{white_affleck_1996}. For $J_2>J_{2,c}$, a gap opens and the system spontaneously forms dimers among first neighbours. In particular, at $J_2=0.5$, the two exact ground states are products of Bell pairs among first neighbours \cite{majumdar1969next}. For larger values of $J_2$, more complex correlation patterns emerge, with both long-range dimer-dimer correlations and finite-range spiral spin correlations \cite{white_affleck_1996}. Those predictions are based on DMRG (hence, variational) numerical simulations. Here, in contrast, we investigate the ability of SDP techniques to offer relevant lower bounds to the ground-state energy, as well as certified bounds on spin correlations in the ground state -- something that, to our knowledge, no other approach can provide. In particular, we certify a change of sign for the second-neighbour spin correlations for $J_2>0.5$ (see Fig.~\ref{fig:B2_figC2}).\\

\noindent\textit{Ground state energy.--}
\begin{figure}[!ht]
	\centering
\includegraphics[width=1\linewidth]{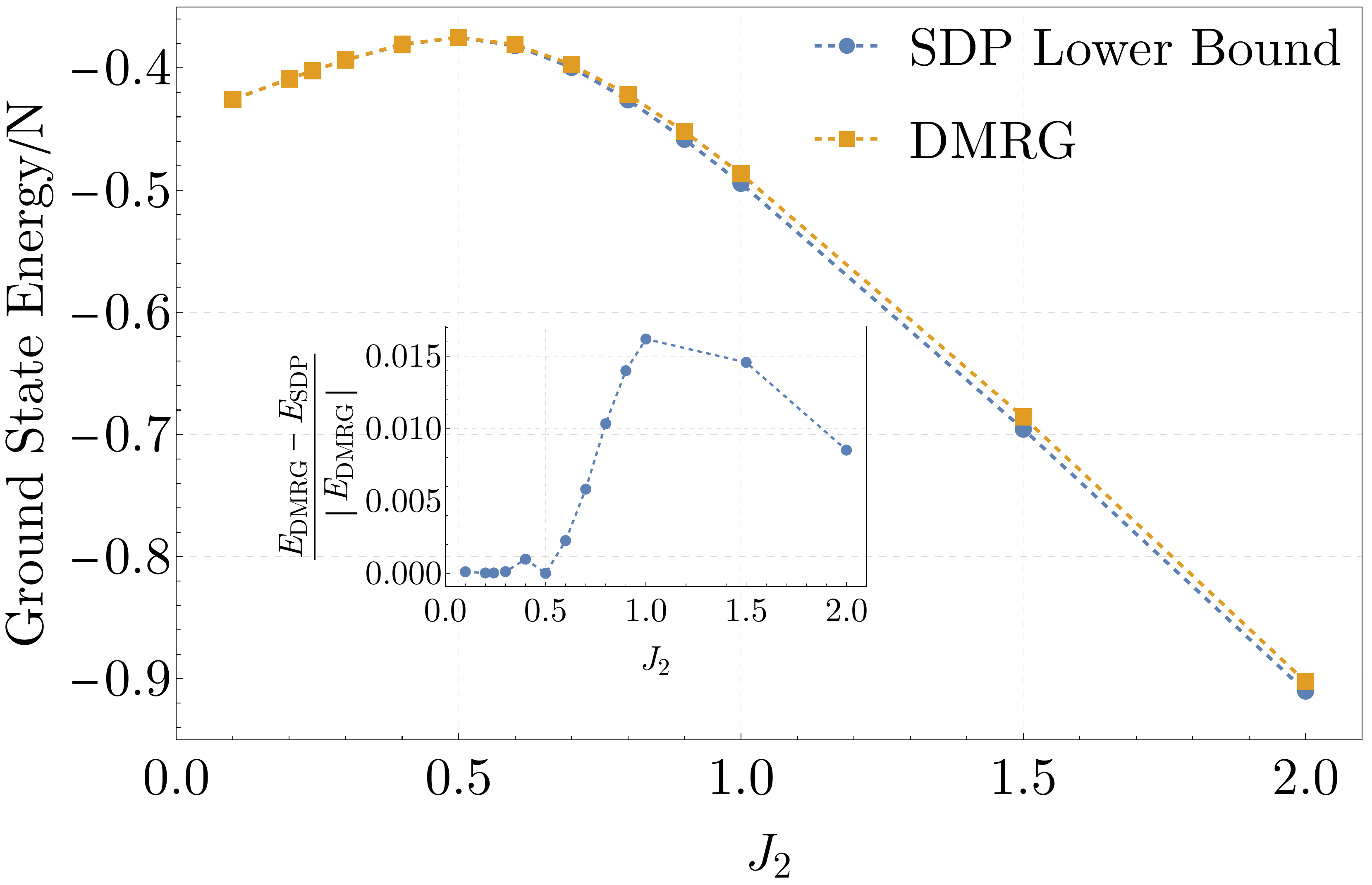}
	\caption{Ground-state energy per particle in the $J_1-J_2$ Heisenberg chain ($N=40$; data in Table \ref{Tab:Heisenberg-2nd-Energies-L40}). Upper and lower bounds are derived, respectively, through DMRG and the implemented SDP relaxation. Inset: the relative accuracy remains better than $0.016$ for all values of $J_2$.}
	\label{fig:B2_fig1}
\end{figure}	
In Fig.~\ref{fig:B2_fig1}, we plot the best lower bound of the ground-state energy for a system of size $N=40$, as compared to the DMRG value (the data are reported in Table \ref{Tab:Heisenberg-2nd-Energies-L40}; see also Table \ref{Tab:Heisenberg-2nd-Energies-L100} for the lower bounds computed for $N=100$). Our compromise for the choice of monomials is different for small and large values of $J_2$. For $J_2 \le 1$, the monomials are the same as for $J_2=0$, namely: 
\begin{align*}
1, \sigma_i^a, \sigma_{i}^a\sigma_{i+j}^b, \sigma_{i}^a \sigma_{i+1}^b \sigma_{i+2}^c, \sigma_{i}^a \sigma_{i+1}^b \sigma_{i+2}^c \sigma_{i+3}^d \,,
\end{align*}
with $i\in\{1,\ldots,N\}, j \in \{1,2,\dots,r\}$ and $a,b,c,d\in\{x,y,z\}$. For $J_2>1$, to better capture the effect of frustration, our (heuristic yet efficient) choice is: 
\begin{equation*}
1, \sigma_i^a, \sigma_{i}^a\sigma_{i+j}^b, \sigma_{i}^a \sigma_{i+2}^b \sigma_{i+4}^c, \sigma_{i}^a \sigma_{i+1}^b \sigma_{i+2}^c \sigma_{i+3}^d
\end{equation*}
with $i\in\{1,\ldots,N\}, j \in \{1,2,\dots,r\}$ and $a,b,c,d\in\{x,y,z\}$. 

As can be seen in Table \ref{Tab:Heisenberg-2nd-Energies-L40}, for $J_2 \lesssim 0.5$ we obtain a relative accuracy of $10^{-3}$, and for all values of $J_2$ the relative accuracy remains better than $0.016$. As expected, the largest gap appears at $J_2=1.0$, where the two couplings become comparable 
and there is competition between them.\\

\begin{figure}[!ht]
	\centering
\includegraphics[width=1\linewidth]{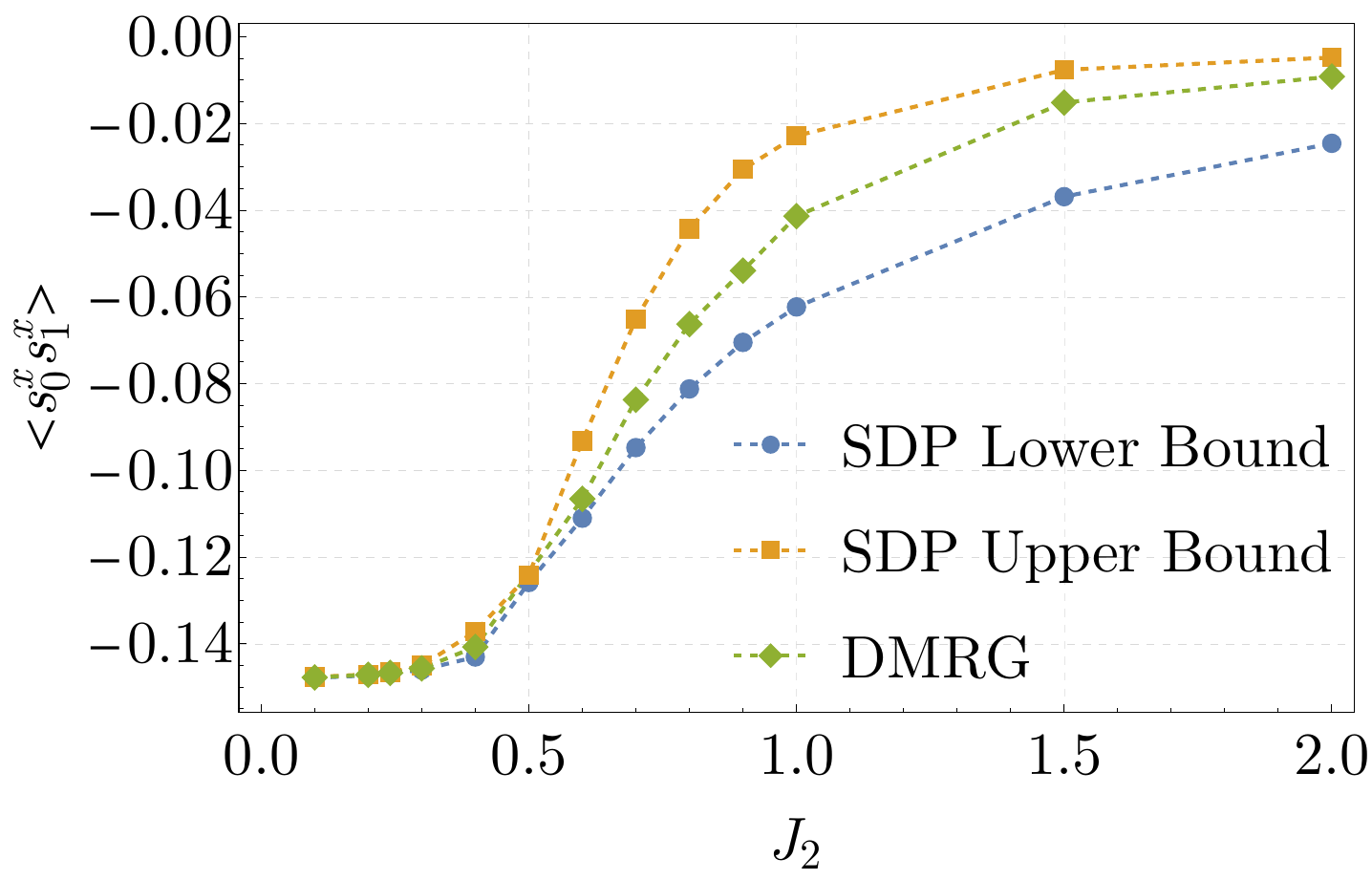}
	\caption{First-neighbour spin correlations in the $J_1-J_2$ Heisenberg chain are certified to remain antiferromagnetic ($\langle s_0^x s_1^x\rangle =C_1<0$) for all values of $J_2$ ($N=40$; data are given in Table \ref{tab:B2-C1}).}
	\label{fig:B2_figC1}
\end{figure}

\noindent\textit{Individual terms in the Hamiltonian.--} 
As mentioned, the SDP approach allows one to obtain certified bounds on relevant observables in the ground-state beyond the energy. In order to do so, we constrain the energy to lie in-between the DMRG upper bound and the SDP lower bound. As a first application, we compute bounds on the first-neighbour spin correlations $C_1$ (see Fig.~\ref{fig:B2_figC1}), as well as the second-neighbour spin correlation $C_2$ (see Fig.~\ref{fig:B2_figC2}), namely both individual terms composing the Hamiltonian. For the sake of comparison, we also plot the results obtained through DMRG calculations, which are expected to be very close to the exact value. The derived lower and upper bounds certify that:
\begin{itemize}
    \item First-neighbour correlations remain antiferromagnetic, $C_1<0$, for all values of $J_2$, as its upper bound is always negative (see Fig.~\ref{fig:B2_figC1} and Table \ref{tab:B2-C1}); 
    \item At $J_2 = 0.5$, the second-neighbour correlations change from ferromagnetic ($C_2>0$) to antiferromagnetic ($C_2<0$) (see Fig.~\ref{fig:B2_figC2} and Table \ref{tab:B2-C2}). This is a non-trivial qualitative information, illustrating the competition between the $J_1$ term which favors staggered correlations among first neighbours (namely $C_1<0$ and $C_2>0$) and the $J_2$ term which favors staggered correlations among second neighbours (namely $C_2<0$). 
\end{itemize}
These findings are fully compatible with the DMRG results, but recall that the latter cannot provide any certification about these properties. This is our first illustration of how physically relevant correlation properties in the ground state can be certified using SDP relaxations. \\


\noindent\textit{Spin correlations.--} 
\begin{figure}[!ht]
	\centering
\includegraphics[width=1\linewidth]{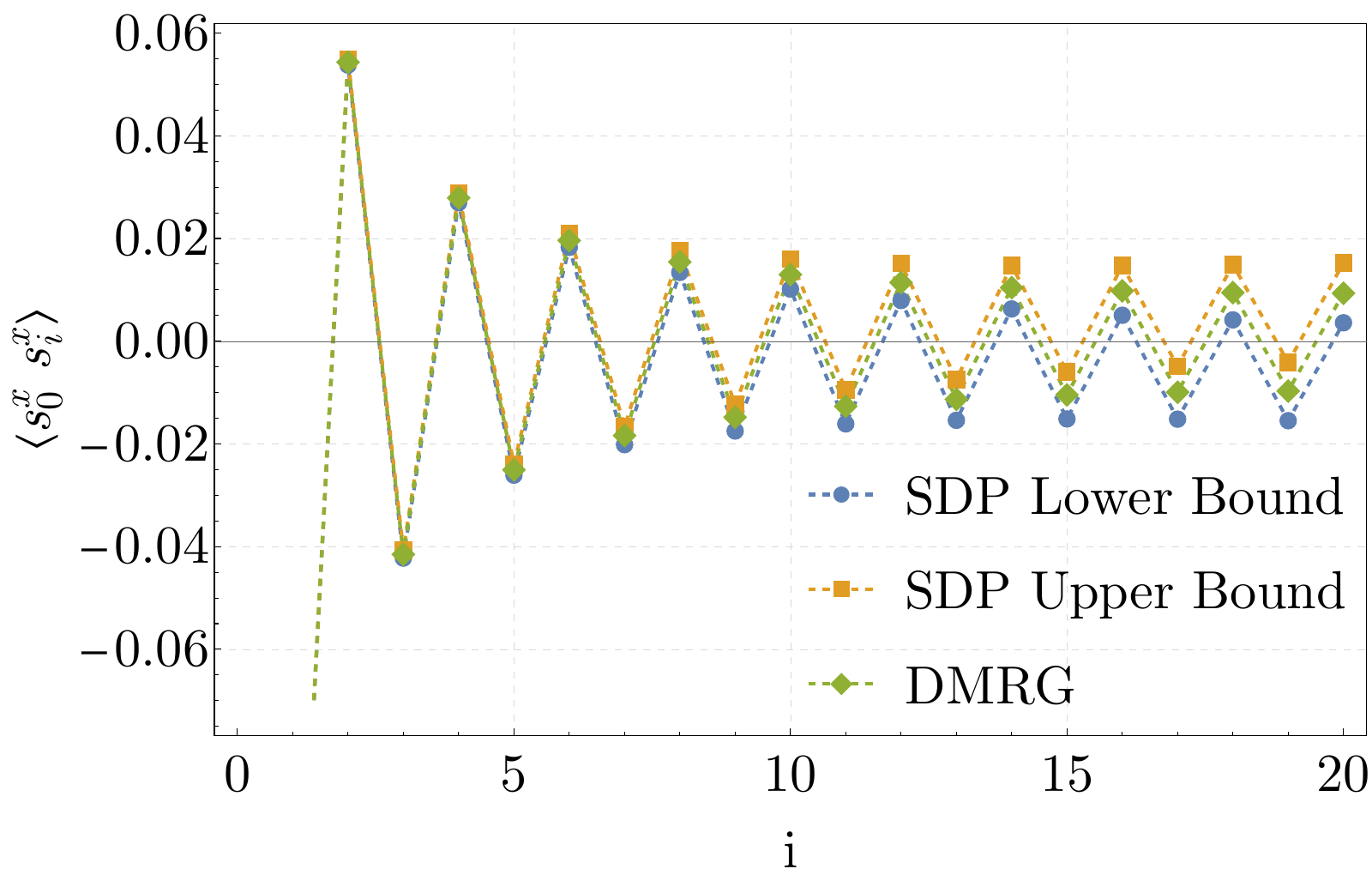}
	\caption{Spin-spin correlation in the $J_1-J_2$ Heisenberg chain for $J_2=0.2$ and system size $N=40$ (data in table \ref{tab:B2-spinspin-J202}). The staggered sign structure is certified at all distances.}
	\label{fig:B2_spinspin-J202}
\end{figure}	
\begin{figure}[!ht]
	\centering
\includegraphics[width=1\linewidth]{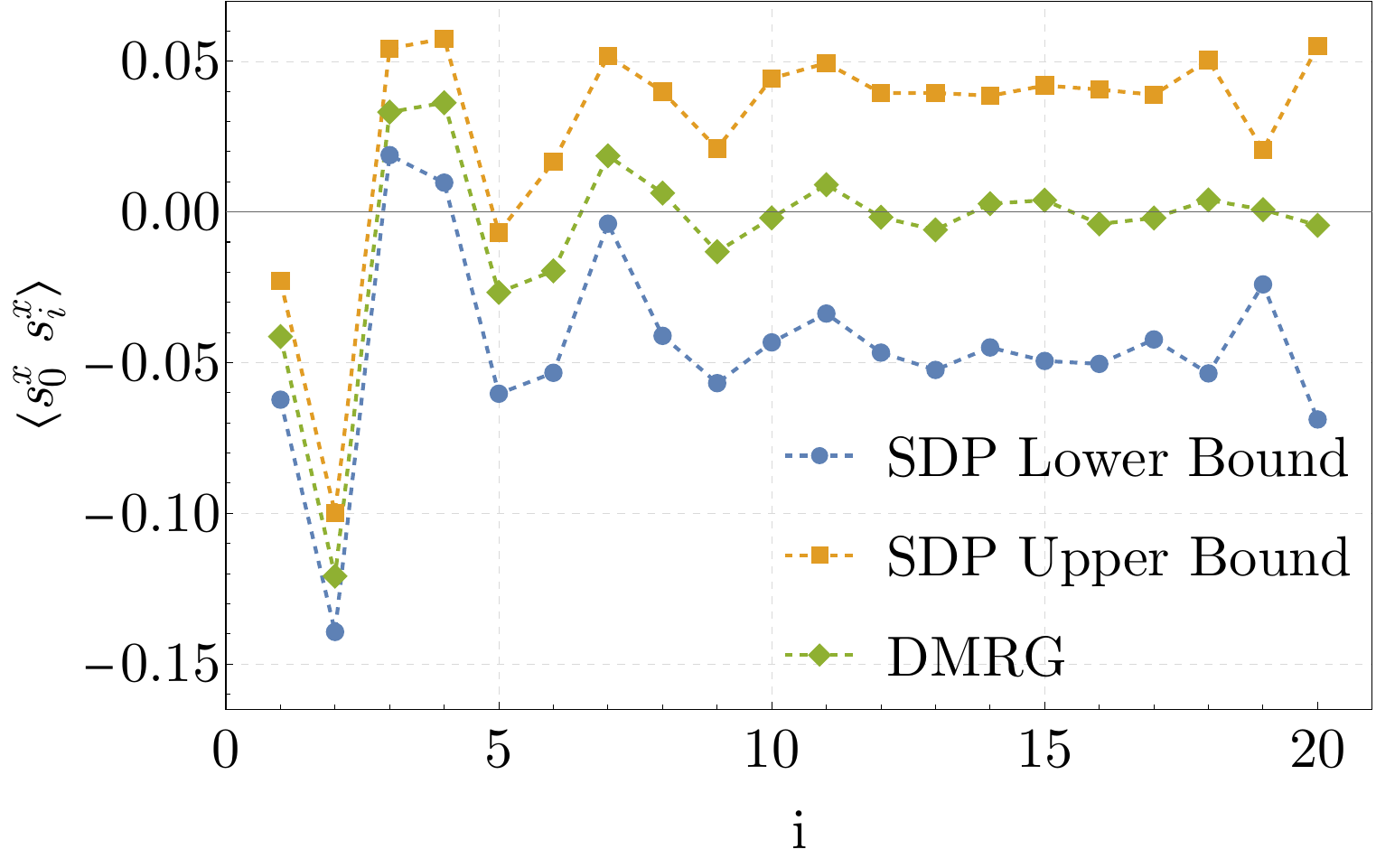}
	\caption{Spin-spin correlation in the $J_1-J_2$ Heisenberg chain for $J_2=1.0$ and system size $N=40$ (data in table \ref{tab:B2-spinspin-J21}). Note that the bounds at $i=19$ show a remarkable and surprising improvement. We have however verified that the obtained SDP solutions define feasible points and hence provide valid lower and upper bounds.}
	\label{fig:B2_spinspin-J21}
\end{figure}	
We then study the ability of the SDP approach to bound the spin correlation function at larger distance. 
In particular, as mentioned, for values of $J_2 < J_{2,c}$, one expects that the system develops antiferromagnetic (that is, staggered) spin correlations which decay as a power-law with distance, as in the $J_2=0$ limit \cite{white_affleck_1996}. In order to explore the potentiality of SDP relaxations to capture such quasi-long-range order in the ground state, we compute bounds on the spin-spin correlations as a function of distance for a fixed system size of $N=40$. We consider both $J_2=0.2<J_{2,c}$ (Fig.~\ref{fig:B2_spinspin-J202} and Table \ref{tab:B2-spinspin-J202}) and $J_2=1.0>J_{2,c}$ (Fig.~\ref{fig:B2_spinspin-J21} and Table \ref{tab:B2-spinspin-J21}). As can be seen, the SDP upper and lower bounds tightly sandwich the DMRG value at small distances, while they become looser at larger distances. Yet, one sees that
\begin{itemize}
    \item For $J_2=0.2$, the SDP bounds are tight enough to certify the staggered sign structure of the correlation function up to the maximal distance $i=N/2$ (see Fig.~\ref{fig:B2_spinspin-J202} and Table \ref{tab:B2-spinspin-J202}). Indeed, both the lower and upper bounds change sign with the distance.
    \item For $J_1=1.0$, the SDP bounds certify instead a qualitatively different spatial structure of spin correlations at short distance, while they become much looser at larger distance. 
\end{itemize} 

It is important to remark that we have not attempted here to optimise the choice of monomials to best capture the correlation function at large distance; instead we have kept the same monomials as for tightly bounding the energy, which especially constrain correlations at short distances. One can expect to get tighter bounds by tailoring the monomial list to the observable to be certified. We come back to this point below.

\subsection{Square lattice Heisenberg model}
\label{sec:B3}

\begin{figure}[!ht]
	\centering
\includegraphics[width=1\linewidth]{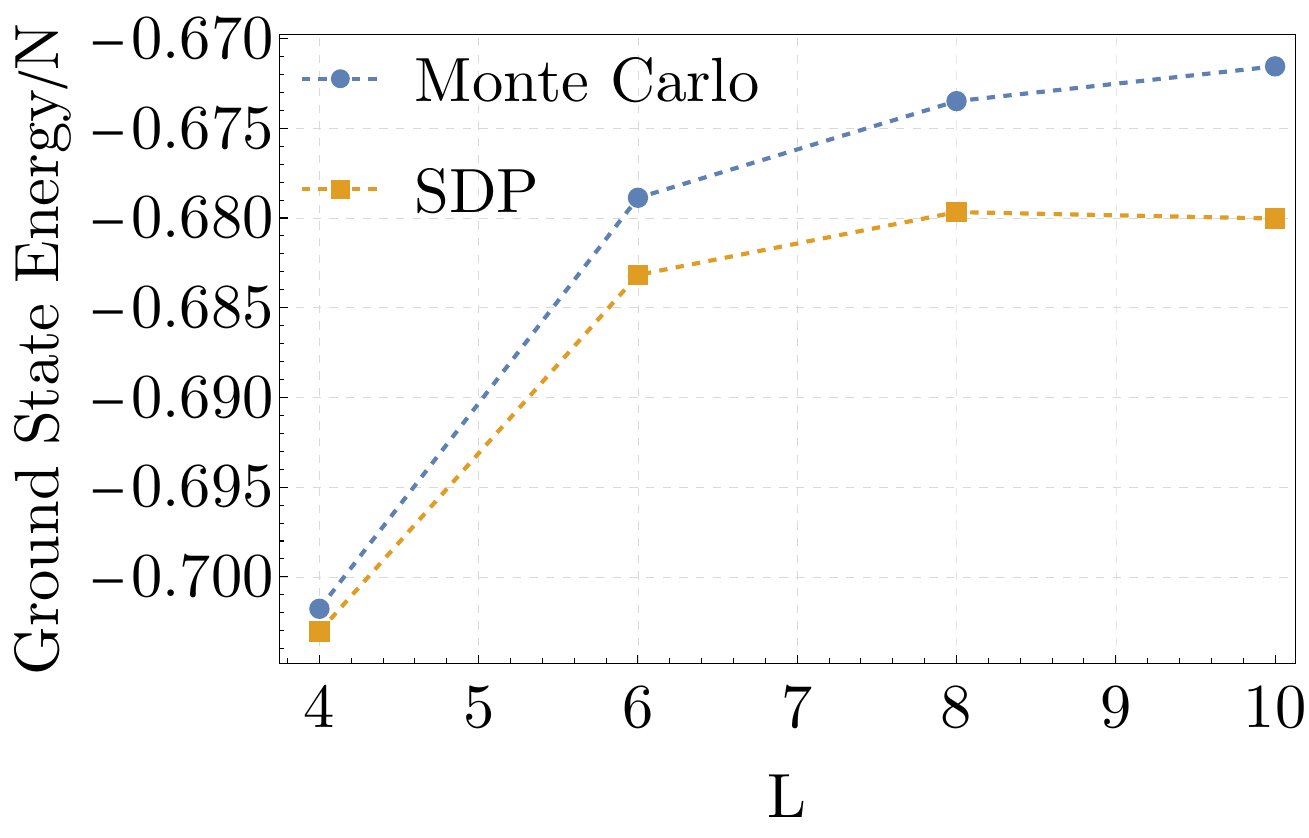}
	\caption{Ground-state energy in the square lattice Heisenberg model, and comparison to quantum Monte Carlo (data in Table \ref{tab:B3-Energies}). }
	\label{fig:B3_Energies}
\end{figure}	
We now move to the most challenging case of two-dimensional systems. As above, we start with the Heisenberg model, but now on a square lattice:
\begin{equation}
    H = (1/4)\sum_{i=1}^L \sum_{j=1}^L \sum_{a\in\{x,y,z\}} \sigma_{(i,j)}^a [\sigma_{(i+1,j)}^a + \sigma_{(i,j+1)}^a] ~,
\end{equation}
where $(i,j)$ label the position of the spins on a square lattice with PBC ($L+1 \equiv 1$). In contrast to the 1D model, it is expected that the square-lattice Heisenberg model spontaneously breaks the $SU(2)$ symmetry in the thermodynamic limit and displays true long-range antiferromagnetic order in the ground state. In particular, this implies that $C(L/2,L/2) \to {\rm cst.} >0$ for $L \to \infty$, where we define $C(i,j)=(1/4)\langle \sigma_{(0,0)}^x \sigma_{(i,j)}^x \rangle$. As further discussed below, while we do certify this property for $L\le 8$, we cannot reliably extrapolate the obtained SDP bounds to the thermodynamic limit. \\

\textit{Ground-state energy.--} 
We first use our SDP algorithm to compute lower bounds on the ground-state energy, as done for the previous models. Our choice of monomials is as follows: 
\begin{align*}
&1, \sigma_{(i,j)}^a, \sigma_{(i,j)}^a\sigma_{(i+r_1,j+r_2)}^b,\\
&\sigma_{(i,j)}^a \sigma_{(i,j+1)}^b \sigma_{(i+1,j+1)}^c, \sigma_{(i,j)}^a \sigma_{(i,j+1)}^b \sigma_{(i-1,j+1)}^c,\\
&\sigma_{(i,j)}^a \sigma_{(i+1,j)}^b \sigma_{(i+1,j+1)}^c, \sigma_{(i,j)}^a \sigma_{(i-1,j)}^b \sigma_{(i-1,j+1)}^c,\\
&\sigma_{(i,j)}^a \sigma_{(i+1,j)}^b \sigma_{(i+2,j)}^c, \sigma_{(i,j)}^a \sigma_{(i,j+1)}^b \sigma_{(i,j+2)}^c,\\
&\sigma_{(i,j)}^a \sigma_{(i+1,j)}^b \sigma_{(i,j+1)}^c \sigma_{(i+1,j+1)}^d
\end{align*}
with $i,j \in \{1,2,\dots,L\}$, $r_1,r_2 \in \{-3,-2,\ldots,3\}$ and $a,b,c,d\in\{x,y,z\}$. For $L=10$, we discard all degree-four monomials $\sigma_{(i,j)}^a \sigma_{(i+1,j)}^b \sigma_{(i,j+1)}^c \sigma_{(i+1,j+1)}^d$.
We consider systems of linear size $L=4,6,8,10$. 

We compare the derived bounds to the quantum Monte Carlo data from~\cite{sandvik1997}, which are expected to be equal to the exact values (up to statistical error bars which are negligible on the scale of our comparison) (see Fig.~\ref{fig:B3_Energies} and Table \ref{tab:B3-Energies}). The energy gaps are now larger than in the one-dimensionl case, but the relative accuracy of the SDP lower bound of the energy is still about $0.01$ as compared to the quantum Monte Carlo result. 

\begin{figure}[!ht]
	\centering
\includegraphics[width=1\linewidth]{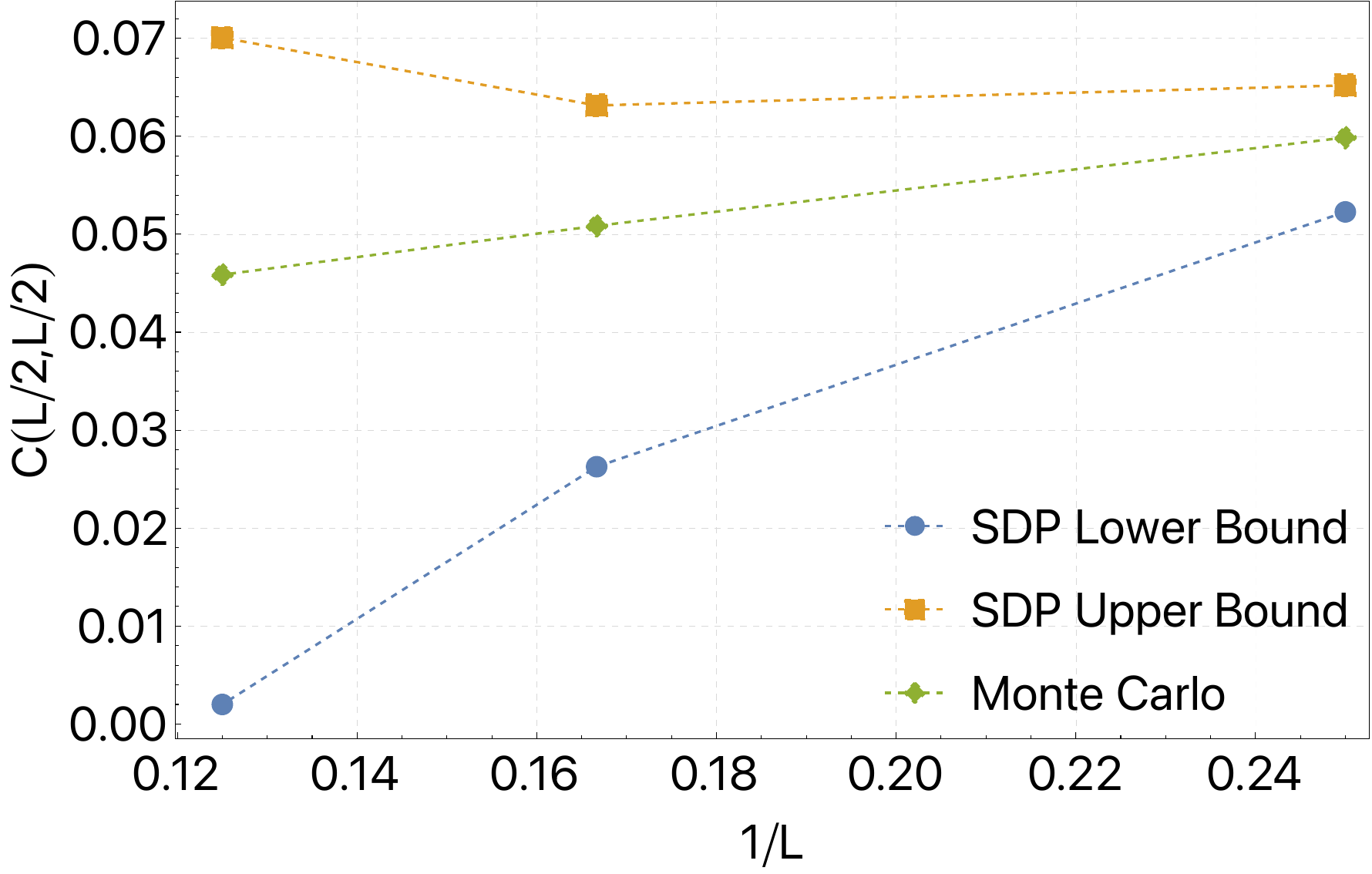}
	\caption{Spin correlation at maximal distance in the square lattice Heisenberg model as compared to Monte Carlo computations (data in Table \ref{tab:B3-CL2}).}
	\label{fig:B3_CL2}
\end{figure}	
 
\textit{Long-range order.--} We now focus on other ground-state properties beyond energy and, in particular, on long-range correlations. In order to investigate the possibility to certify spontaneous symmetry breaking and the associated long-range antiferromagnetic order in the ground state, we compute bounds on the correlation at maximal distance $C(L/2,L/2)$ (see Fig.~\ref{fig:B3_CL2} and Table \ref{tab:B3-CL2}). For all the computed sizes, the lower bound on $C(L/2,L/2)$ remains positive, hence certifying the presence of long-range order on those sizes. However, similarly to the case of the $J_1-J_2$ model at $J_2=0.2$ (Section \ref{sec:B2}), for increasing system size the SDP bounds on $C(L/2,L/2)$ become increasingly looser. It is therefore not possible to argue that $C(L/2,L/2)$ will remain positive for $L>8$ (corresponding to $1/L<0.125$ on the figure) from scaling arguments on the derived lower bounds. 

Note again that, to derive these bounds on long-range correlations, we have used the same monomial list as for optimising the ground state energy. In future works one may instead tailor the choice of the monomials to better bound the correlation function at large distance, which can be expected to offer some improvement.

\subsection{$J_1-J_2$ square lattice Heisenberg model}
\label{sec:B4}
As a last example, we consider the $J_1-J_2$ Heisenberg model on a square lattice:
\begin{eqnarray}
    H = (1/4)\sum_{i=1}^L \sum_{j=1}^L \sum_{a\in\{x,y,z\}} \sigma_{(i,j)}^a \left[\sigma_{(i+1,j)}^a + \sigma_{(i,j+1)}^a \right.\nonumber\\
    \left. + J_2(\sigma_{(i+1,j+1)}^a + \sigma_{(i+1,j-1)}^a)\right] ~,
\end{eqnarray}
with PBC. The $J_2$ terms favors antiferromagnetic correlations along the diagonals of the square lattice, which are incompatible with the correlations favored by the first-neighbour $J_1=1$ term and leads to frustration. As for the $J_1-J_2$ Heisenberg chain, this model is not amenable to quantum Monte Carlo due to the sign problem. Several variational methods based on Ansatz wavefunctions have however been applied to this paradigmatic model of frustrated quantum magnetism, sometimes obtaining conflicting results due to a complex energy landscape with various ground-state candidates which are close in energy yet with incompatible forms of order \cite{zhitomirskyU1996,capriottiS2000,jiangetal2012,wangetal2013,gongetal2014,chooetal2019}. The emerging consensus regarding the nature of the ground-state is that increasing the ratio $J_2/J_1$, the ground states displays a long-range Néel order for $J_2/J_1 \lesssim 0.45$, followed by a spin-liquid without any form of long-range order ($0.45 \lesssim J_2/J_1 \lesssim 0.55$), then long-range dimer correlations ($0.55 \lesssim J_2/J_1 \lesssim 0.61$), and finally long-range columnar antiferromagnetic order for $0.61 \lesssim J_2/J_1$ \cite{PhysRevLett.121.107202,PhysRevX.11.031034,PhysRevB.102.014417,LIU20221034} (the precise boundaries of these phases remain debated). All those results are established by variational methods on different geometries.\\

\begin{figure}[!ht]
	\centering
\includegraphics[width=1\linewidth]{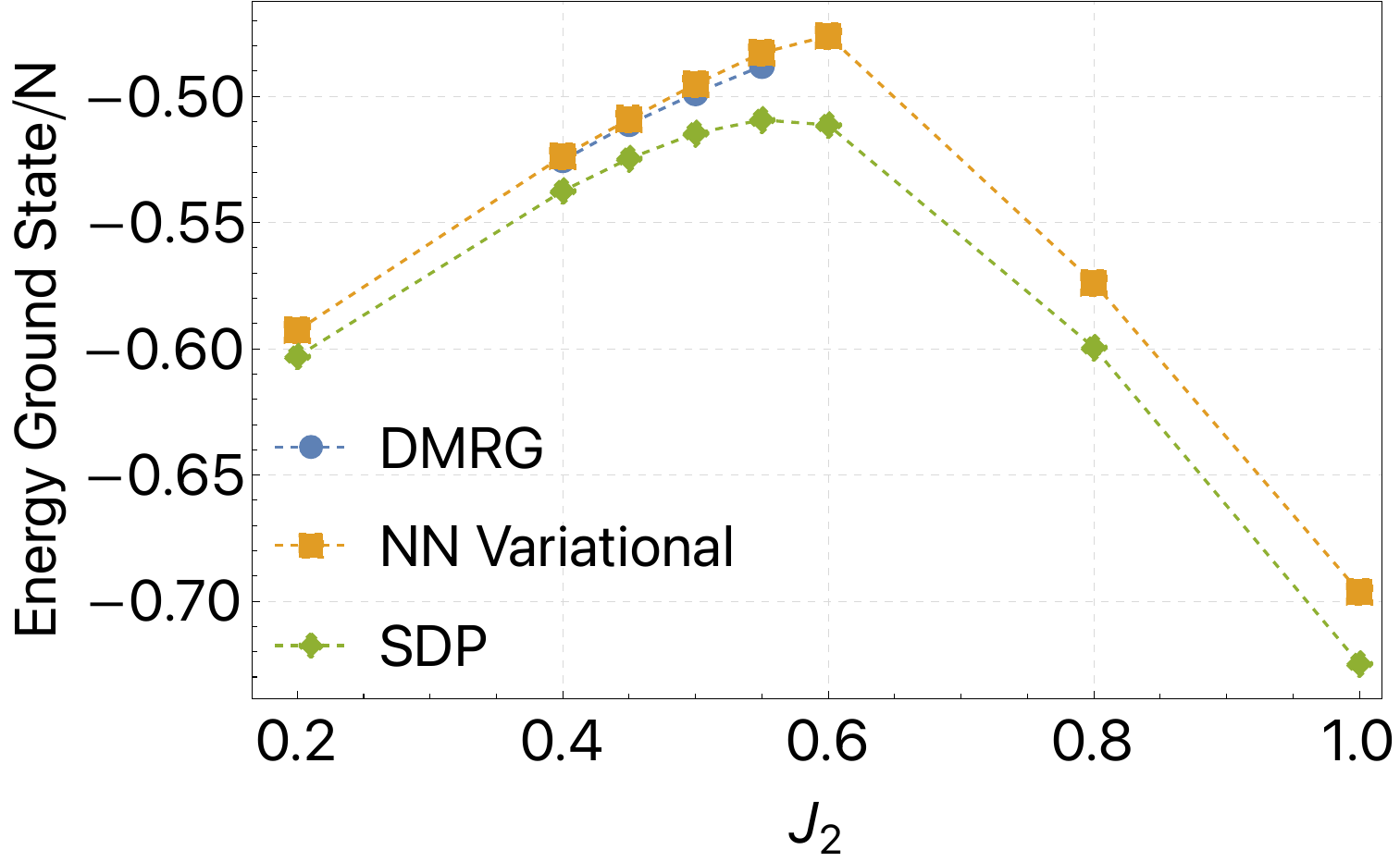}
	\caption{Energy lower bounds for the $2D$ $J_1-J_2$ Heisenberg model on a square with $L=10$ (data in table \ref{tab:B4-L10}). 
 }
	\label{fig:B4_3}
\end{figure}

\textit{Ground-state energy.--} Again, we first compute SDP lower bounds on the energy, which complement variational methods. We present results for $L=6,8$ in the Appendix \ref{app_B4} (see Figures \ref{fig:B4_1} and \ref{fig:B4_2}), and for $L=10$ in Figure \ref{fig:B4_3} (data are respectively given in Tables \ref{tab:B4-L6}, \ref{tab:B4-L8} and \ref{tab:B4-L10}). We compare them with state-of-the-art upper bounds obtained by DMRG \cite{gongetal2014} and neural-network wave functions \cite{chooetal2019} (very similar upper bounds where recently obtained also by machine-learning inspired variational states \cite{PhysRevX.11.031034}). Notice in particular that size $L=10$ (namely, $N=100$ qubits) is not achievable with exact methods, so that combining upper- and lower bounds become very relevant to constrain ground-state properties. Combining variational upper bounds and SDP lower bounds allows us to sandwich the true ground-state energy with a few percent of relative accuracy (Table \ref{tab:B4-L10}).\\

\textit{Spin correlations.--} Finally, the SDP approach can also be applied to deliver certified bounds on relevant observables in regimes inaccessible to exact numerical methods, such as the $J_1-J_2$ model on a square lattice for $L=10$. Constraining the energy to lie in-between the maximal lower bound (as obtained by the SDP) and the minimal upper bound (as obtained using variational methods), we obtain certified bounds on first- and second-neighbour (diagonal) correlations in the exact ground state. The monomial list that we use to bound both the energy and spin correlations is the same as the one in Section~\ref{sec:B3} for the square lattice Heisenberg model. 

\begin{figure}[!ht]
	\centering
\includegraphics[width=1\linewidth]{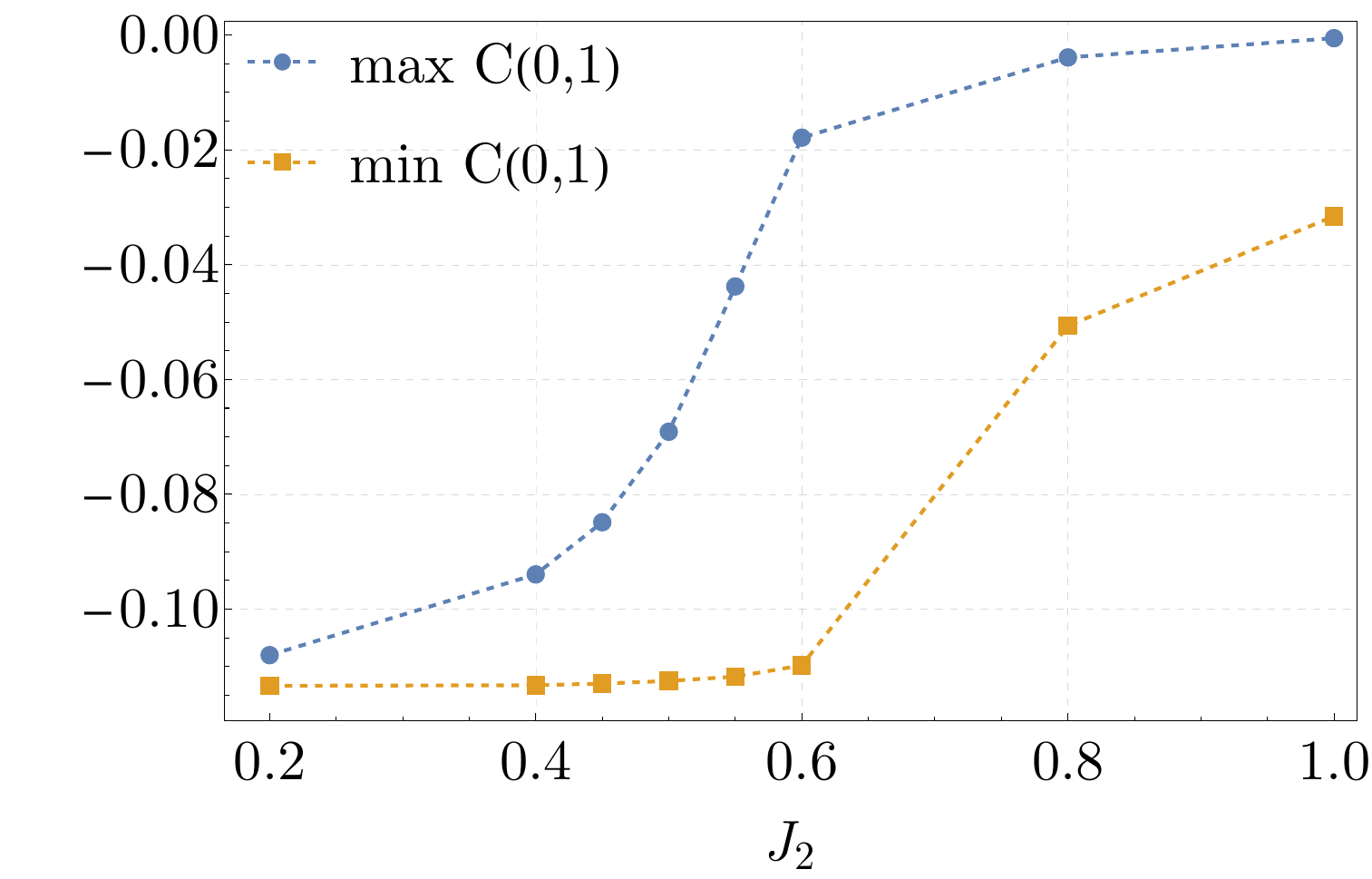}
	\caption{Bounds for the correlations $C(0,1)$ of the $2D$ $J_1-J_2$ Heisenberg model on a square lattice of dimension $L=10$ (data in table \ref{tab:B4-C01-L10}).}
	\label{fig:B4_bounds-C01-L10}
\end{figure}

\begin{figure}[!ht]
	\centering
\includegraphics[width=1\linewidth]{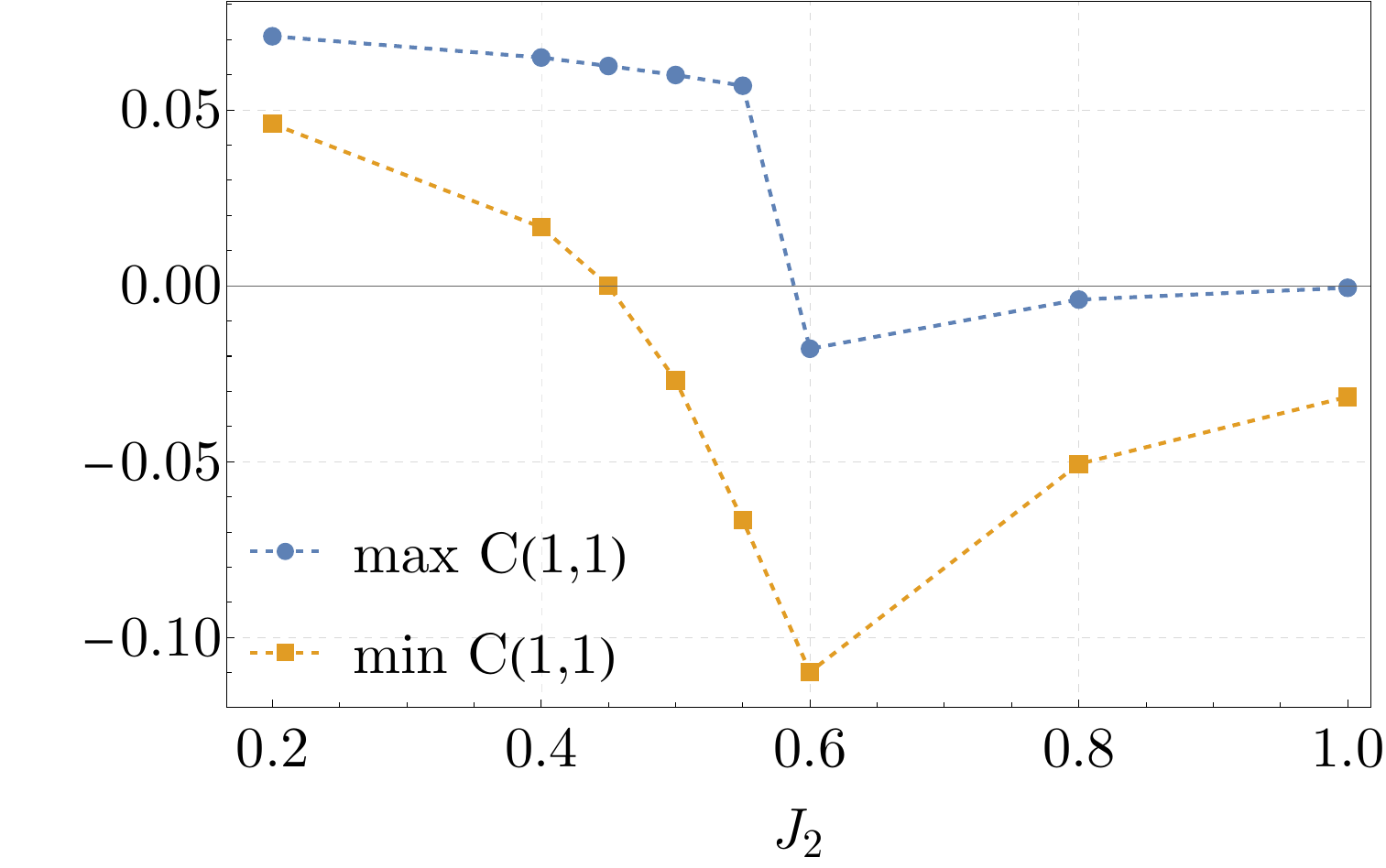}
	\caption{Bounds for the correlations $C(1,1)$ of the $2D$ $J_1-J_2$ Heisenberg model on a square lattice of dimension $L=10$ (data in table \ref{tab:B4-C11-L10}).}
	\label{fig:B4_bounds-C11-L10}
\end{figure}

The results are respectively displayed in Fig.~\ref{fig:B4_bounds-C01-L10} and Fig.~\ref{fig:B4_bounds-C11-L10}  (data in Tables \ref{tab:B4-C01-L10} and \ref{tab:B4-C11-L10}). We emphasize that this frustrated model for $N=100$ spins is well beyond the capabilities of known exact methods such as exact diagonalisation, so that the certified bounds offered by the SDP approach are especially insightful. We notice in particular that SDP bounds are sufficiently accurate to certify:
\begin{itemize}
    \item Correlations $C(0,1)$ remain antiferromagnetic for all studied values of $J_2$, as the computed upper bound is always negative.
    \item Second-neighbour correlations experience a change of sign while varying the $J_2/J_1$ ratio, a behavior reminiscent of the 1D model studied in Section \ref{sec:B2}. The transition occurs for a value of $J_2$ in the range $(0.45,0.6)$. 
\end{itemize}
Again, this type of certification is impossible with previous approaches.

\section{Venues for improvements}
\label{sec_improvements}

In the landscape of numerical methods to address the quantum many-body problem, SDP-based approaches are conceptually unique in that they are the only ones to offer certified results, both on the energy of the ground state and on its properties. 
While the bounds we obtained for 1D systems are very accurate, they are looser for the more challenging situation of 2D systems, yet promising. Note also that in comparison with variational approaches, SDP relaxations are still in their infancy. Considerable improvement in the tightness of the delivered bounds can be expected in future implementations, as it happened in the classical context, where SDP techniques recently yielded spectacularly high accurate results in bounding correlation functions in two- and three-dimensional Ising models using the bootstrap framework~\cite{cho2022bootstrapping}.
In fact, our work already improves previous results using SDP relaxations in a many-body context by at least an order of magnitude~\cite{BarthelH2012,Baumgratz_2012,haim2020variationalcorrelations}. 

Our considered SDP hierarchies apply to any generic polynomial optimisation problem. However, tighter bounds can be obtained by adapting the relaxation to the specific model under study. While in our work we took advantage of several aspect of the structure of the tackled problem, we discuss in what follows several avenues that deserve further investigation in future works.

\emph{Sparsity and symmetry:} A systematic study of the symmetries of the models would result in tighter bounds. Our work already exploits the sparsity and some symmetries of the considered models to significantly improve the scalability. However, we do not exhaust the structures of the models yet. For instance, we do not perform block-diagonalisation by exploiting permutation/mirror/rotation symmetries of the models. Also, we do not exploit the fact that any entry of the moment matrix is either a real or a pure imaginary number. 

\emph{Choice of monomials:} 
An optimization of the best choice of monomials to the observable of interest would also result in tighter bounds.
In our work, we choose the monomials based on the local operators appearing in the Hamiltonian. We heuristically optimize the monomial choice to tighten the bound on the ground-state energy. However, once the bounds on the energy are obtained, one could have modified the monomials in the SDP when studying other quantities, such as long-range correlations. In fact, it is reasonable to expect that there is a different optimal choice of monomials for each observable. Preliminary numerical studies suggest that the bounds on observables can be significantly improved by tailoring the monomial choice.
We also note that machine learning can be employed to optimize the choice of monomials, as done in~\cite{MLbounds}.

\emph{Optimality constraints:} The solution to optimisation problems have additional constraints that can be incorporated to the hierarchy. For instance, in many situations, such as for energy minimisation, the searched state is an eigenstate of the polynomial to be optimised, the Hamiltonian in our case. This can be used to enforce additional polynomial constraints that are satisifed by the ground state, e.g. $\langle [H,O]\rangle_\GS=0$ for any operator $O$. In this sense, recent works, \cite{fawzi2023certified} and \cite{araujo2023karushkuhntucker}, appearing after the completion of a first version of the present manuscript, provide other ground-state relaxations that exploit the optimality constraints. Interestingly, these constraints allow deriving certifiable bounds on the value of any observable without the need for upper bounds on the energy. Ref.~\cite{araujo2023karushkuhntucker} is in fact more general and shows how to extend the Karush-Kuhn-Tucker optimality conditions, well known in classical optimisation, to the non-commuting case. It is also reported in \cite{fawzi2023entropy} that adding entropy constraints could strengthen the SDP bounds on ground-state problems. Note that as the SDP hieararchy, when increasing the relaxation order, converges to the exact ground-state solution where these constraints are automatically satisfied, such additional constraints are in principle not needed. However, in practice, imposing them may significantly improve the results at finite steps of the relaxation.

\emph{Variational information:} A better use of all the information available in the variational state, beyond energy upper bound as considered here, may improve the bounds.
For instance, from the variational result and the fact that the ground state is an eigenstate, one not only knows that $\expect H\leq E_\A$, but also all its moments, that is, $\expect{H^n}\leq E_\A^n$ for all $n$ (for that one requires that the Hamiltonian is positive, which can be enforced by using any ground-state energy lower bound). These are polynomial constraints, hence can be implemented in SDP hierarchies.
Other information about the state resulting from the variational method could also be used. A remarkable example is provided in Ref.~\cite{kull2022lower}, where the tensor resulting from DMRG optimisations is used to improve the lower bounds on ground-state energies.

\emph{Dimension constraints:} 
Adapting the method to take advantage of the finite dimensionality of each systems could also result in tighter bounds.
Here we have used the NPA hierarchy for polynomial optimisations, where operators are defined for arbitrary dimension, and combined it with upper bounds from variational approaches. 
However, most Hamiltonians are defined by polynomials of operators acting over a given finite dimensional space.
As explained, while for the spin-one-half models studied here the dimension constraint is eventually recovered from convergence and the Pauli relations, for practical purposes it may be convenient to exploit it at each finite step. There indeed exist SDP hierarchies that are tailored to a given dimension, a remarkable example being the one introduced in Ref.~\cite{kull2022lower}.


\emph{Numerical methods:} On the numerical side, some of the lower bounds based on a solution returned by a SDP solver may come together with unsatisfying numerical feasibility status. Therefore, another interesting research direction is to obtain truly certified lower bounds based on exact rational arithmetic. 
For this, one could design a post-processing method, relying either on rigorous interval arithmetic or rounding-projection techniques, in the same spirit as in \cite{magron2015,cafuta2015rational}. On the other hand, the SDP in our work involves complex numbers and we reformulate it as a real SDP to feed a SDP solver designed only for real numbers. However, SDP solvers natively supporting complex numbers may handle it more efficiently. Also, the SDPs arising from the NPA hierarchy possess some special structures (e.g., low-rank optimal solutions, unit diagonal) which could be exploited to design more efficient SDP algorithms as in~\cite{wang2023solving}. We can thus rely on a structure-exploiting SDP solver to obtain tighter bounds and to approach models of larger size in the future.

Investigating all these ideas would most probably lead to significant improvement on the tightness of the SDP bounds, especially in the most challenging case of 2D models. It is worth mentioning that many of the previous ingredients and constructions can often be arranged into a single SDP relaxation that combines the benefits of each of them.

All the previous points have focused on methods and ideas to improve the bounds. We would like to conclude this section presenting a couple of applications of relaxations that go beyond what studied here.

\emph{Thermodynamic limit:}
While our work has focused on systems consisting of a finite number of particles, the thermodynamic limit can also be tackled by SDP relaxations. The constructions of Refs.~\cite{kull2022lower,fawzi2023certified,araujo2023karushkuhntucker} mentioned above are in fact presented in this limit. But our approach can also be easily adapted to this situation using similar arguments as in~\cite{kull2022lower}.
Remarkably, Ref.~\cite{fawzi2023entropy} goes beyond ground-state problems and provides SDP relaxations converging to any finite temperature equilibrium state in the thermodynamic limit, hence obtaining bounds valid for any equilibrium state.

\emph{Spontaneous symmetry breaking from local observables bounds:} 
One could identify spontaneous symmetry breaking by looking at the bounds on local observables, instead of long-range correlations as attempted in the present work. For instance, in the considered Heisenberg models, one could study lower and upper bounds on the local magnetization $m=\langle \sigma_i^z \rangle$. In the absence of spontaneous symmetry breaking, $m=0$; while $m=\pm m_0$ signals the onset of long-range order in the thermodynamic limit. We expect that the latter case would manifest itself in the upper (lower) bound saturating to $m_0$ ($-m_0$) while increasing the level of the SDP hieararchy. Such behavior is indeed reported in a recent and related study in the case of the transverse-field Ising model \cite{fawzi2023certified}. Investigating the nature of the ground state of, e.g., the $J_1-J_2$ Heisenberg model in the highly frustrated parameter regime (Section \ref{sec:B4}) represents a very exciting and relevant application of the SDP approach.

\section{Conclusion}
\label{sec_discussions}

In this work we have shown how SDP relaxations of polynomial optimisation problems when combined with upper bounds obtained through variational methods can provide certifiable bounds on ground-state properties beyond energy. We have illustrated the potentialities of the method in 1D and 2D Heisenberg models. The choice is motivated by their rich phenomenology, the existence of previous results to benchmark our results, and their symmetries, which allow us reaching large system sizes. However, the method is general and can be applied to essentially any many-body Hamiltonian. In fact, the method can be combined with any variational approach, including upper bounds to ground-state energies obtained using variational quantum hardware or simulators, to provide certifiable bounds on any other observable of interest.

There are many natural continuations of our results that are worth considering: 1) apply the introduced approach to other relevant models in physics, for instance fermions; 2) validate the output of existing quantum simulators on specific finite-size instances of quantum many-body problems; 3) establish phase diagrams of quantum many-body Hamiltonians directly in the thermodynamic limit; 4) improve the scalability of the method, especially by making use of the physical properties of the considered Hamiltonian model. While being at the moment more limited in terms of scalability, it is our strong belief that the considered techniques will play an important role in complementing the results of variational methods and, therefore, become a central tool to understand the physics of interacting many-body quantum systems.





\section{Acknowledgements}
We thank Miguel Navascu\'es, Luca Tagliacozzo and Filippo Vicentini for discussions.
This work is supported by the ERC AdG CERQUTE, the Government of Spain (NextGenerationEU PRTR-C17.I1 and Quantum in Spain, Severo Ochoa CEX2019-000910-S), Fundaci\'{o} Cellex, Fundaci\'{o} Mir-Puig, Generalitat de Catalunya (CERCA programme), the AXA Chair in Quantum Information Science, EU Quantera project Veriqtas and PASQUANS2, the National Natural Science Foundation of China under Grant No. 12201618, the European Union Horizon's 2020 research and innovation programme under the Marie Sklodowska Curie grant agreement No 101031549 (QuoMoDys), the NSF grant OAC-1835443 and the ERC Adv. Grant grant agreement No 885682. 
Research at the Perimeter Institute for Theoretical Physics is supported by the Government of Canada through the Department of Innovation, Science and Economic Development Canada and by the Province of Ontario through the Ministry of Research, Innovation and Science.
VM was supported by the FastQI grant funded by the Institut Quantique Occitan, the PHC Proteus grant
46195TA, the European Union’s Horizon 2020 research and innovation programme under the Marie Sk{\l}odowska-Curie Actions, grant agreement 813211 (POEMA), by the AI Interdisciplinary Institute ANITI funding, through the French ``Investing for the Future PIA3'' program under the Grant agreement n${}^\circ$ ANR-19-PI3A-0004 as well as by the National Research Foundation, Prime Minister’s Office, Singapore under its Campus for Research Excellence and Technological Enterprise (CREATE) programme. This work was partially performed using HPC resources from CALMIP (Grant 
2023-P23035).

\clearpage
\begin{appendix}

\section{Anderson bound}
\label{app_Anderson_bound}
In this appendix, we illustrate the concept of Anderson bound to obtain lower bounds on ground-state energies. For the sake of concreteness and simplicity, we consider a one-dimensional model with translation invariance and PBC, although the idea is straightforward to extend to other cases. 
The Hamiltonian is of the form $H = \sum_{i=1}^N h_i$ with $h_i$ acting in the neighbourhood of site $i$. 
We then define the restricted Hamiltonian $H_i(K) = \sum_{j=i}^K h_j$ for $K<N$. We may rewrite the full Hamiltonian as $H=K^{-1}\sum_{i=1}^N H_i(K)$, with the $1/K$ prefactor compensating for the fact that all individual terms $h_i$ are repeated $K$ times in the sum. The Anderson bound is then obtained by noting that for any state $|\psi\rangle$, $\langle \psi | H_i(K) |\psi \rangle$ cannot be smaller than the smallest eigenvalue of $H_i(K)$, namely, to the ground-state energy of $H_i(K)$. This holds in particular when $|\psi\rangle$ is the ground state of $H$. As $H_i(K)$ describes the initial model on a cluster of $K$ sites with OBC, we conclude that:
\begin{equation}
    E_{\rm PBC}(N) \ge \frac{N}{K} E_{\rm OBC}(K) ~,
\end{equation}
or equivalently:
\begin{equation}
    e_{\rm PBC}(N) \ge e_{\rm OBC}(K) ~.
\end{equation}

\section{SDP reductions by exploiting structure}
\label{app_algorithmic_tricks}
In this appendix, we further discuss how to implement symmetries of the problem in the SDP algorithm to reduce the number of free variables in the implementation. We start by restating the main problem for completeness.

The ground state energy of the Heisenberg model is the optimum of the following non-commutative polynomial optimisation problem:
\begin{equation}\label{ncpop}
    \begin{aligned}\min\limits_{\{\ket{\psi},\sigma_i^a\}}&\quad \bra\psi H \ket\psi\\
    \text{such that:}&\quad(\sigma_i^a)^2=1,\quad i=1,\ldots,N;a\in\{x,y,z\},\\
    &\quad\sigma_i^x\sigma_i^y=\i\sigma_i^z,\quad\sigma_i^y\sigma_i^x=-\i\sigma_i^z,\quad i=1,\ldots,N,\\
    &\quad\sigma_i^y\sigma_i^z=\i\sigma_i^x,\quad\sigma_i^z\sigma_i^y=-\i\sigma_i^x,\quad i=1,\ldots,N,\\
    &\quad\sigma_i^z\sigma_i^x=\i\sigma_i^y,\quad\sigma_i^x\sigma_i^z=-\i\sigma_i^y,\quad i=1,\ldots,N,\\
    &\quad\sigma_i^a\sigma_j^b=\sigma_j^b\sigma_i^a,\quad 1\le i\ne j\le N;a,b\in\{x,y,z\}.
    \end{aligned}
\end{equation}
The NPA hierarchy \cite{pironio2010convergent} can be then applied to \eqref{ncpop} yielding a non-decreasing sequence of lower bounds on the ground state energy. Specifically, suppose that $\cB_d$ is a monomial basis (i.e., a subset of monomials w.r.t. the non-commutating variables $\{\sigma_i^a\}_{i=1,\ldots,N,a\in\{x,y,z\}}$) up to degree $d$. Then the $d$-th order moment relaxation of the NPA hierarchy for \eqref{ncpop} is given by
\begin{equation}\label{mom}
\begin{aligned}
    \min\limits_{\{\langle v^\dagger w\rangle\}} &\quad\langle H\rangle\\
    \text{such that:}&\quad\MM_{d}\succeq0, \\
    &\quad\MM_{d} \text{ obeys some moment replacement rules},
\end{aligned}
\end{equation}
where $\mathbf{M}_{d}$ is the moment matrix indexed by $\cB_d$ with $[\MM_{d}]_{vw}=\langle v^\dagger w\rangle$. \revision{The decision variables of \eqref{mom} are the moments $\langle v^\dagger w\rangle, v,w\in\cB_d$.} Note that the equality constraints in \eqref{ncpop} give rise to the following replacement rules on monomials:
\begin{subequations}
\begin{align}
    (\sigma_i^a)^2\quad&\longrightarrow\quad1,\quad i=1,\ldots,N,a\in\{x,y,z\},\\
    \sigma_i^x\sigma_i^y\quad&\longrightarrow\quad\i\sigma_i^z,\quad i=1,\ldots,N,\\
    \sigma_i^y\sigma_i^x\quad&\longrightarrow\quad-\i\sigma_i^z,\quad i=1,\ldots,N,\\
    \sigma_i^y\sigma_i^z\quad&\longrightarrow\quad\i\sigma_i^x,\quad i=1,\ldots,N,\\
    \sigma_i^z\sigma_i^y\quad&\longrightarrow\quad-\i\sigma_i^x,\quad i=1,\ldots,N,\\
    \sigma_i^z\sigma_i^x\quad&\longrightarrow\quad\i\sigma_i^y,\quad i=1,\ldots,N,\\
    \sigma_i^x\sigma_i^z\quad&\longrightarrow\quad-\i\sigma_i^y,\quad i=1,\ldots,N,\\
    \sigma_i^a\sigma_j^b\quad&\longrightarrow\quad\sigma_j^b\sigma_i^a,\quad 1\le i\ne j\le N,a,b\in\{x,y,z\}.
\end{align}
\end{subequations}
For any monomial $u$, by applying the above replacement rules, we can reduce it to the {\em normal form} $\NF(u)\coloneqq c\sigma_{i_1}^{a_1}\sigma_{i_2}^{a_2}\cdots\sigma_{i_r}^{a_r}$ with $c\in\{1,-1,\i,-\i\}$, $1\le i_1<i_2<\cdots<i_r\le N$. It follows that the moment matrix $\MM_{d}$ satisfies the moment replacement rule: $\langle u\rangle=\langle\NF(u)\rangle$ for all entries $\langle u\rangle$ of $\MM_{d}$.

\subsection{Sparsity}
In order to exploit the sparsity of the Heisenberg model, for each degree $d$, we pick monomials that are supported on contiguous sites. Specifically, for the 1D Heisenberg model, we let
\begin{align*}
    \cP_d\coloneqq&\,\{\sigma_i^{a_1}\sigma_{i+1}^{a_2}\cdots\sigma_{i+d-1}^{a_d}\mid i\in\{1,\ldots,N\},\\
    &\,\,\quad a_j\in\{x,y,z\},j=1,\ldots,d\}.
\end{align*}
Then at relaxation order $d$, we use the sparse monomial basis $\cB_d=\cup_{i=0}^d\cP_i$ instead of the full monomial basis. 
Moreover, to capture long-range correlations, we also include the monomials of form $\sigma_{i}^a\sigma_{i+j}^b$ with $j=2,\ldots,r$ and $a,b\in\{x,y,z\}$ in the monomial basis $\cB_d$.
\revision{For the 2D Heisenberg model, we use the following sparse monomial basis ($d\coloneqq4$)
\begin{align*}
    \cB_d=\{&1, \sigma_{(i,j)}^a, \sigma_{(i,j)}^a\sigma_{(i+r_1,j+r_2)}^b, \sigma_{(i,j)}^a \sigma_{(i,j+1)}^b \sigma_{(i+1,j+1)}^c, \\
    &\sigma_{(i,j)}^a \sigma_{(i,j+1)}^b \sigma_{(i-1,j+1)}^c,\sigma_{(i,j)}^a \sigma_{(i+1,j)}^b \sigma_{(i+1,j+1)}^c,\\
    &\sigma_{(i,j)}^a \sigma_{(i-1,j)}^b \sigma_{(i-1,j+1)}^c,\sigma_{(i,j)}^a \sigma_{(i+1,j)}^b \sigma_{(i+2,j)}^c,\\
    &\sigma_{(i,j)}^a \sigma_{(i,j+1)}^b \sigma_{(i,j+2)}^c,\sigma_{(i,j)}^a \sigma_{(i+1,j)}^b \sigma_{(i,j+1)}^c \sigma_{(i+1,j+1)}^d\\
    &\mid i,j \in \{1,2,\dots,L\},r_1,r_2\in\{-3,-2,\ldots,3\},\\
    &\,\,\,\,\,a,b,c,d\in\{x,y,z\}\}.
\end{align*}}

The resulting SDP relaxations are more efficient to solve but possibly lead to more conservative lower bounds.
We emphasize that similar reduction of the monomial basis already appeared in the related literature. 
In the context of quantum information theory, we refer to \cite{pal2009quantum} where the authors obtain upper bounds on maximal violations of Bell inequalities after random selection of a subset of monomials with given degrees.  
For general (non-)commutative polynomial optimisation problems, one can exploit either correlative sparsity \cite{lasserre2006convergent,klep2021sparse}, occurring when there are few correlations between the variables of the input problem, or term sparsity \cite{wang2021tssos,wang2021exploiting}, occurring when there are a small number of terms involved in the input problem by comparison with the fully dense case. 
The interested reader is referred to \cite{magron2023sparse} for a recent monograph on this topic.



\subsection{Symmetry}

\subsubsection{Sign symmetry of the model}\label{ss1}
We can observe that the feasible set of
\eqref{ncpop} is invariant under the substitution of two of the three variables, e.g., $\sigma_i^x$ and $\sigma_i^y$ of a given site into their opposite, e.g., $-\sigma_i^x$ and $-\sigma_i^y$.
In order for any objective functions of the form \eqref{eq_generic_Heisenberg}
to also be invariant, we need to consider the same substitutions for all the sites.
There are therefore three substitutions:
\begin{subequations}\label{eq:ss1}
\begin{align}
    s_{xy} : & \, (\sigma_i^x,\sigma_i^y,\sigma_i^z)_{i=1}^N\longrightarrow (-\sigma_i^x,-\sigma_i^y,\sigma_i^z)_{i=1}^N,\\
    s_{yz} : & \, (\sigma_i^x,\sigma_i^y,\sigma_i^z)_{i=1}^N\longrightarrow (\sigma_i^x,-\sigma_i^y,-\sigma_i^z)_{i=1}^N,\\
    s_{zx} : & \, (\sigma_i^x,\sigma_i^y,\sigma_i^z)_{i=1}^N\longrightarrow (-\sigma_i^x,\sigma_i^y,-\sigma_i^z)_{i=1}^N.
\end{align}
\end{subequations}
Note that $s_{zx}$ is the composition of $s_{xy}$ and $s_{yz}$ so we only need to consider the invariance under two of the three substitutions.

For each monomial $m$, $s_{xy}(m)$ (resp. $s_{yz}(m)$) is either $m$ or $-m$.
Similarly to \cite[Section~III.C]{Lofberg2009},
for each monomial $m$, we consider its \emph{signature} as the vector
$(s_{xy}(m)/m, s_{yz}(m)/m) \in \{-1, 1\}^2$.
Suppose that the monomial basis $\cB_d$ is arranged such that each of the four groups of monomials with the same signature appears contiguously.
Notice that a product of two monomials is invariant under the sign symmetries if and only if it is the product of two monomials of the same signature.
The moments of symmetric monomials therefore form a block diagonal structure of 4 blocks in the moment matrix.
Thus, we can reduce the large positive semidefinite matrix $\mathbf{M}_{d}$ into 4 smaller positive semidefinite submatrices as shown in \cite[Theorem~4]{Lofberg2009} in the commutative case, \revision{each indexed by monomials with the same signature. For the 1D model and $d=4$, the partition of monomials is given in \cref{tab:sign2}. The 2D case is similar.}


\begin{table}[!ht]
\centering\revision{
\begin{tabular}{|c|c|}
\hline
Signature & Monomials\\
\hline
\multirow{3}{*}{$(1, 1)$}& $1,\sigma_i^a\sigma_{i+j}^a,\sigma_i^a\sigma_{i+1}^b\sigma_{i+2}^c,\sigma_i^a\sigma_{i+1}^a\sigma_{i+2}^a\sigma_{i+3}^a$,\\
&$\sigma_i^a\sigma_{i+1}^a\sigma_{i+2}^b\sigma_{i+3}^b,\sigma_i^a\sigma_{i+1}^b\sigma_{i+2}^a\sigma_{i+3}^b$,$\sigma_i^a\sigma_{i+1}^b\sigma_{i+2}^b\sigma_{i+3}^a$,\\
&$a\ne b\ne c\in\{x,y,z\},i=1,\ldots,N,j=1,\ldots,r$\\
\hline
\multirow{6}{*}{$(1, -1)$}& $\sigma_i^z,\sigma_i^a\sigma_{i+j}^b,\sigma_i^z\sigma_{i+1}^z\sigma_{i+2}^z,\sigma_i^z\sigma_{i+1}^a\sigma_{i+2}^a$,\\
&$\sigma_i^a\sigma_{i+1}^z\sigma_{i+2}^a,\sigma_i^a\sigma_{i+1}^a\sigma_{i+2}^z,\sigma_i^z\sigma_{i+1}^z\sigma_{i+2}^a\sigma_{i+3}^b$,\\
&$\sigma_i^z\sigma_{i+1}^a\sigma_{i+2}^z\sigma_{i+3}^b,\sigma_i^z\sigma_{i+1}^a\sigma_{i+2}^b\sigma_{i+3}^z$,$\sigma_i^a\sigma_{i+1}^z\sigma_{i+2}^z\sigma_{i+3}^b$,\\
&$\sigma_i^a\sigma_{i+1}^z\sigma_{i+2}^b\sigma_{i+3}^z,\sigma_i^a\sigma_{i+1}^b\sigma_{i+2}^z\sigma_{i+3}^z$,$\sigma_i^a\sigma_{i+1}^b\sigma_{i+2}^b\sigma_{i+3}^b$,\\
&$\sigma_i^b\sigma_{i+1}^a\sigma_{i+2}^b\sigma_{i+3}^b,\sigma_i^b\sigma_{i+1}^b\sigma_{i+2}^a\sigma_{i+3}^b$,$\sigma_i^b\sigma_{i+1}^b\sigma_{i+2}^b\sigma_{i+3}^a$,\\
&$a\ne b\in\{x,y\},i=1,\ldots,N,j=1,\ldots,r$\\
\hline
\multirow{6}{*}{$(-1, 1)$}& $\sigma_i^x,\sigma_i^a\sigma_{i+j}^b,\sigma_i^x\sigma_{i+1}^x\sigma_{i+2}^x,\sigma_i^x\sigma_{i+1}^a\sigma_{i+2}^a$,\\
&$\sigma_i^a\sigma_{i+1}^x\sigma_{i+2}^a,\sigma_i^a\sigma_{i+1}^a\sigma_{i+2}^x,\sigma_i^x\sigma_{i+1}^x\sigma_{i+2}^a\sigma_{i+3}^b$,\\
&$\sigma_i^x\sigma_{i+1}^a\sigma_{i+2}^x\sigma_{i+3}^b,\sigma_i^x\sigma_{i+1}^a\sigma_{i+2}^b\sigma_{i+3}^x$,$\sigma_i^a\sigma_{i+1}^x\sigma_{i+2}^x\sigma_{i+3}^b$,\\
&$\sigma_i^a\sigma_{i+1}^x\sigma_{i+2}^b\sigma_{i+3}^x,\sigma_i^a\sigma_{i+1}^b\sigma_{i+2}^x\sigma_{i+3}^x$,$\sigma_i^a\sigma_{i+1}^b\sigma_{i+2}^b\sigma_{i+3}^b$,\\
&$\sigma_i^b\sigma_{i+1}^a\sigma_{i+2}^b\sigma_{i+3}^b,\sigma_i^b\sigma_{i+1}^b\sigma_{i+2}^a\sigma_{i+3}^b$,$\sigma_i^b\sigma_{i+1}^b\sigma_{i+2}^b\sigma_{i+3}^a$,\\
&$a\ne b\in\{y,z\},i=1,\ldots,N,j=1,\ldots,r$\\
\hline
\multirow{6}{*}{$(-1, -1)$}& $\sigma_i^y,\sigma_i^a\sigma_{i+j}^b,\sigma_i^y\sigma_{i+1}^y\sigma_{i+2}^y,\sigma_i^y\sigma_{i+1}^a\sigma_{i+2}^a$,\\
&$\sigma_i^a\sigma_{i+1}^y\sigma_{i+2}^a,\sigma_i^a\sigma_{i+1}^a\sigma_{i+2}^y,\sigma_i^y\sigma_{i+1}^y\sigma_{i+2}^a\sigma_{i+3}^b$,\\
&$\sigma_i^y\sigma_{i+1}^a\sigma_{i+2}^y\sigma_{i+3}^b,\sigma_i^y\sigma_{i+1}^a\sigma_{i+2}^b\sigma_{i+3}^y$,$\sigma_i^a\sigma_{i+1}^y\sigma_{i+2}^y\sigma_{i+3}^b$,\\
&$\sigma_i^a\sigma_{i+1}^y\sigma_{i+2}^b\sigma_{i+3}^y,\sigma_i^a\sigma_{i+1}^b\sigma_{i+2}^y\sigma_{i+3}^y$,$\sigma_i^a\sigma_{i+1}^b\sigma_{i+2}^b\sigma_{i+3}^b$,\\
&$\sigma_i^b\sigma_{i+1}^a\sigma_{i+2}^b\sigma_{i+3}^b,\sigma_i^b\sigma_{i+1}^b\sigma_{i+2}^a\sigma_{i+3}^b$,$\sigma_i^b\sigma_{i+1}^b\sigma_{i+2}^b\sigma_{i+3}^a$,\\
&$a\ne b\in\{x,z\},i=1,\ldots,N,j=1,\ldots,r$\\
\hline
\end{tabular}}
\caption{Table of monomials indexing the sign symmetry blocks when $d = 4$.}
\label{tab:sign2}
\end{table}

\subsubsection{Sign symmetry of the Hamiltonian}\label{ss2}

The Hamiltonian $H$ in \eqref{ncpop} may have more sign symmetries. For example, consider the Heisenberg model, the Hamiltonian $H$ is invariant under the substitutions:
\begin{subequations}\label{eq:ss2}
\begin{align}
(\sigma_i^x,\sigma_i^y,\sigma_i^z)_{i=1}^N\longrightarrow (-\sigma_i^x,\sigma_i^y,\sigma_i^z)_{i=1}^N,\\
(\sigma_i^x,\sigma_i^y,\sigma_i^z)_{i=1}^N\longrightarrow (\sigma_i^x,-\sigma_i^y,\sigma_i^z)_{i=1}^N,\\
(\sigma_i^x,\sigma_i^y,\sigma_i^z)_{i=1}^N\longrightarrow (\sigma_i^x,\sigma_i^y,-\sigma_i^z)_{i=1}^N.
\end{align}
\end{subequations}

Besides the zero entries given in Section~\ref{ss1}, these additional sign symmetries of the Hamiltonian yield extra zero entries of the moment matrix: $\langle u\rangle=0$ if $\NF(u)$ is variant under the transformations \eqref{eq:ss2}.


Note that symmetry reduction usually applies to any convex problem with symmetric objective and symmetric feasible set.
In this case, the symmetry of the feasible set is not apparent because the replacement rules are not symmetric, e.g., replacing $\sigma_i^x\sigma_i^y$ by $\i \sigma_i^z$ is not symmetric under any of the substitutions~\eqref{eq:ss2}.
We can show the symmetry of the set of sums of Hermitian squares using (1) their connection with the set of strictly positive Hermitian elements over the quotient ring (detailed below) and (2) the fact that the Hamiltonian is invariant under the substitutions~\eqref{eq:ss2}.

Indeed, let us denote the non-commutative polynomial ring by $\C\langle\{\sigma_i^x,\sigma_i^y,\sigma_i^z\}_{i=1}^N\rangle$ and the ideal generated by the equality constraints of \eqref{ncpop} by $I$. Let $\Omega\coloneqq\{\mathrm{NF}(u)\mid u \text{ is a monomial in } \{\sigma_i^x,\sigma_i^y,\sigma_i^z\}_{i=1}^N\}$. Then the optimization problem \eqref{ncpop} is equivalent to the unconstrained optimization problem: $\min\limits_{\{\ket{\psi},\sigma_i^a\}}\bra\psi H \ket\psi$, considered in the quotient ring $\C\langle\{\sigma_i^x,\sigma_i^y,\sigma_i^z\}_{i=1}^N\rangle/I\cong\C\langle\Omega\rangle$. 
Let us denote by $S$ the group of additional sign symmetries given by \eqref{eq:ss2} and by $\Sigma_S$ the set of sums of Hermitian squares of $\C\langle\Omega\rangle$, that are invariant under $S$ after conversion to normal form, i.e., elements of the form $h = \sum_j p_j^\dagger p_j$, $p_j \in \C\langle\Omega\rangle$, such that $s(\NF(h)) = \NF(h)$ for any $s\in S$. 
Then one can show that any strictly positive Hermitian element of $\C\langle\Omega\rangle$ that is invariant under $S$ lies in $\Sigma_S$ by \cite{helton2004positivstellensatz}, 
the proof being very similar to the one of \cite[Proposition 3.1]{klep2023}. 
Here ``strict positivity'' should be understood as strict positivity over all possible evaluations in Hilbert spaces, as detailed, e.g., in~\cite[Section 2.2]{klep2023}. 
By duality one considers optimization over linear functionals nonnegative on $\Sigma_S$, which leads to an equivalent formulation of the above unconstrained optimization problem:
\begin{equation}\label{momsym}
\begin{aligned}
    \min\limits_{\text{linear } \ell : \C\langle\Omega\rangle \to \C} &\quad \ell(H)\\
    \text{such that:} &\quad \ell(q) \geq 0 \quad \forall q \in \Sigma_S \,,\\
    & \quad \ell(p^\dagger) = \ell(p)^* \quad \forall p \in \C\langle\Omega\rangle \,, \\
    & \quad \ell(1)=1 \,.
\end{aligned}
\end{equation}

From any $\ell$ feasible for the above problem~\eqref{momsym}, let us define the linear functional $\ell_S : \C\langle\Omega\rangle \to \C$ by $\ell_S(p) = 1/|S| \sum_{s \in S} \ell(s(p))$ for all $p \in \C\langle\Omega\rangle$. 
Then it is clear that $\ell_S(H) = \ell(H)$ since $H$ is invariant under the action of $S$. 
In addition, $\ell_S(1)=1$, $\ell_S (p^\dagger) = \ell(p)^*$ for all $p \in \C\langle\Omega\rangle$ and $\ell_S (q) \geq 0$ for all $q \in \Sigma_S$. 
To prove the latter fact, we used the fact that one has $s(q) = q$ for any $s\in S$ and any $q \in \Sigma_S$ written in normal form. 
Overall this shows that $\ell_S$ is feasible for \eqref{momsym} and yields the same objective value than the one with $\ell$. 

To conclude, at each relaxation~\eqref{mom} we can restrict ourselves to optimizing over linear functionals vanishing on variant elements of $\C\langle\Omega\rangle$ under the transformations \eqref{eq:ss2}. 
This boils down to setting every entry of the Hermitian moment matrix $\MM = [\ell(v^\dagger w)]_{v,w \in \Omega}$ from~\eqref{mom} to $\langle u\rangle=0$ if $\NF(u)$ is variant under the transformations~\eqref{eq:ss2}. 

\subsubsection{Translation symmetry}\label{ts}
The translation symmetry of \eqref{ncpop} comes from that the Hamiltonian $H$ is invariant under any translation of sites, which implies
\begin{equation}
    \langle\upsilon(u)\rangle=\langle u\rangle,
\end{equation}
where $\upsilon\colon i\longrightarrow i+k$ denotes a translation of sites with $k\in\{1,\ldots,L\}$ ($L\coloneqq N$ for the 1D Heisenberg model with $N$ sites).
This together with the PBC imposes a block structure on the moment matrix $\MM_{d}$ where each block is a circulant matrix of size $L$ as long as the monomial basis $\cB_d$ is appropriately sorted \cite{Baumgratz_2012}. 
For example, consider in the 1D Heisenberg model the submatrix $T$ of the moment matrix $\MM_{d}$ indexed by $\{\sigma_i^x\}_{i=1}^N$. The translation symmetry implies $T_{i,j}=T_{j,i}=\langle\sigma_i^x\sigma_j^x\rangle=\langle\sigma_1^x\sigma_{j-i+1}^x\rangle$. Therefore, $T$ is a symmetric circulant matrix.
Any circulant matrix of size $L$ can be diagonalised by a discrete Fourier transform $P\in\C^{L\times L}$ with
\begin{equation}
    P_{i,j}=\frac{1}{\sqrt{L}}\mathrm{e}^{-2\pi\i(i-1)(j-1)/L},\quad i,j=1,\ldots,L.
\end{equation} 
By virtue of this fact, we are able to further block-diagonalise each block of the moment matrix $\MM_{d}$ provided in Section~\ref{ss1}. 

\revision{
Specifically, suppose that the monomial basis $\cB_d$ is arranged such that monomials of the same type with varying site label $i$ appear contiguously for $i=1,\ldots,L$ (e.g., $\sigma_1^x\sigma_{2}^x,\sigma_2^x\sigma_{3}^x,\ldots,\sigma_L^x\sigma_{1}^x$). Then the block of $\mathbf{M}_{d}$ corresponding to the signature $(1,1)$ is of the block form:
\begin{equation}
    G\coloneqq\begin{bmatrix}
    1&\bc_1^{\intercal}&\bc_2^{\intercal}&\cdots&\bc_t^{\intercal}\\
    \bc_1&G_{1,1}&G_{1,2}&\cdots&G_{1,t}\\
    \bc_2&G_{2,1}&G_{2,2}&\cdots&G_{2,t}\\
    \vdots&\vdots&\vdots&\ddots&\vdots\\
    \bc_t&G_{t,1}&G_{t,2}&\cdots&G_{t,t}\\
    \end{bmatrix},
\end{equation}
where $\bc_j\coloneqq(c_j,\ldots,c_j)\in\bR^L$ and each $G_{j,k}$ is a circulant matrix. Let $U_G=\diag(1,P,\ldots,P)\in\C^{(1+Lt)\times(1+Lt)}$. We have $G=U_GD_GU_G^{\dagger}$ where 
\begin{equation}
    D_G\coloneqq\begin{bmatrix}
    1&\bd_1^{\intercal}&\bd_2^{\intercal}&\cdots&\bd_t^{\intercal}\\
    \bd_1&D_G^{1,1}&D_G^{1,2}&\cdots&D_G^{1,t}\\
    \bd_2&D_G^{2,1}&D_G^{2,2}&\cdots&D_G^{2,t}\\
    \vdots&\vdots&\vdots&\ddots&\vdots\\
    \bd_t&D_G^{t,1}&D_G^{t,2}&\cdots&D_G^{t,t}\\
    \end{bmatrix}
\end{equation}
with $\bd_j\coloneqq(c_j\sqrt{L},0,\ldots,0)\in\bR^L$ and diagonal matrices $D_G^{j,k}$.
Moreover, the other three blocks of $\mathbf{M}_{d}$ respectively corresponding to the signatures $(1,-1),(-1,1),(-1,-1)$ are of the block form:
\begin{equation}
    H\coloneqq\begin{bmatrix}
    H_{1,1}&H_{1,2}&\cdots&H_{1,s}\\
    H_{2,1}&H_{2,2}&\cdots&H_{2,s}\\
    \vdots&\vdots&\ddots&\vdots\\
    H_{s,1}&H_{s,2}&\cdots&H_{s,s}\\
    \end{bmatrix},
\end{equation}
where each $H_{j,k}$ is a circulant matrix. Let $U_H=\diag(P,\ldots,P)\in\C^{(Ls)\times(Ls)}$. We have $H=U_HD_HU_H^{\dagger}$ where 
\begin{equation}
    D_H\coloneqq\begin{bmatrix}
    D_H^{1,1}&D_H^{1,2}&\cdots&D_H^{1,s}\\
    D_H^{2,1}&D_H^{2,2}&\cdots&D_H^{2,s}\\
    \vdots&\vdots&\ddots&\vdots\\
    D_H^{s,1}&D_H^{s,2}&\cdots&D_H^{s,s}\\
    \end{bmatrix}
\end{equation}
with diagonal matrices $D_H^{j,k}$. By reordering rows and columns, both $D_G$ and $D_H$ have a block-diagonal form. Hence, the positive semidefiniteness of $G$ and $H$ can be imposed by requiring that each diagonal block of $D_G$ and $D_H$ is positive semidefinite.
}

Note that similar block-diagonalisation techniques have also been obtained in the commutative polynomial optimisation setting; see \cite{riener2013exploiting} for more details.

\subsubsection{Permutation symmetry}\label{ps}
The permutation symmetry of \eqref{ncpop} comes from that the Hamiltonian $H$ is invariant under any permutation of $\{x,y,z\}$, which yields the following moment replacement rule on the moment matrix $\MM_{d}$ (by a similar proof as in Section~\ref{ss2}):
\begin{equation} \langle\tau(\sigma_{i_1}^{a_1}\sigma_{i_2}^{a_2}\cdots\sigma_{i_r}^{a_r})\rangle=\langle\sigma_{i_1}^{a_1}\sigma_{i_2}^{a_2}\cdots\sigma_{i_r}^{a_r}\rangle,
\end{equation}
where $1\le i_1<i_2<\cdots<i_r\le N$, $a_1,\ldots,a_r\in\{x,y,z\}$, and $\tau$ denotes any permutation of $\{x,y,z\}$.

\subsubsection{Mirror symmetry}\label{ms}
For a model supported on a 2D square lattice, there may exist an additional symmetry. Let us consider for instance the 2D Heisenberg model
\begin{equation}\label{2dHeisen}
    H=\sum_{a=x,y,z}\sum_{i=1}^L\sum_{j=1}^L\sigma_{(i, j)}^a\left(\sigma_{(i+1, j)}^a+\sigma_{(i, j+1)}^a\right)
\end{equation}
supported on an $N=L\times L$ square lattice. The mirror symmetry of this 2D Heisenberg model means that the Hamiltonian $H$ is invariant under the transformation $\omega\colon(i,j)\longrightarrow(j,i)$, which yields the following moment replacement rule on the moment matrix $\MM_{d}$:
\begin{equation}
    \langle\omega(u)\rangle=\langle u\rangle.
\end{equation}

\vspace{1em}
\revision{
To summarize, for the 1D Heisenberg model, the SDP \eqref{mom} with $d=4$ after reductions involves one positive semidefinite block of size $3r+28$, $N-1$ positive semidefinite blocks of size $3r+27$, and $3N$ positive semidefinite blocks of size $2r+28$; for the 2D Heisenberg model, the SDP \eqref{mom} with $d=4$ after reductions involves one positive semidefinite block of size $129L+1$, $L-1$ positive semidefinite blocks of size $129L$, and $3L$ positive semidefinite blocks of size $111L$.
}

\section{Improving SDP bounds by imposing an extra positivity constraint}
The bound given by the SDP relaxation for a fixed relaxation order can be improved by imposing an extra positivity constraint, namely that the $k$-body reduced density matrix $\rho_{[k]}$ is positive. This constraint can be written as:
\begin{equation}\label{positivity}
    \rho_{[k]} = \frac{1}{2^k}\sum_{a_1,\ldots,a_k}\langle\sigma_1^{a_1}\sigma_2^{a_2}\cdots\sigma_k^{a_k}\rangle\sigma_1^{a_1}\sigma_2^{a_2}\cdots\sigma_k^{a_k} \succeq 0,
\end{equation}
where $a_i\in\{0,x,y,z\}$, $i=1,\ldots,k$ (with $\sigma_i^0 = \mathbf{1}$). As Eq.~ \eqref{positivity} is linear in the moments, we can add it to the constraints of Eq.~ \eqref{cmom}. In general one expects that the larger $k$, the tighter is the resulting bound. However, as $\rho_{[k]}$ is a matrix of size $2^k \times 2^k$, a large $k$ also leads to a SDP of large size. In practice we notice that $k=8$ achieves a good balance between the computational cost and the improvement of bounds.

\section{Heisenberg chain}
\label{app_B1}
In this appendix, we provide the numerical data corresponding to Fig.~\ref{fig:B1_fig1} in the main text (section \ref{sec:B1}), namely the ground-state energy in the Heisenberg chain with PBC evaluated through both DMRG and SDP approaches (Table \ref{Tab:B1-Energies}).
\begin{table}[ht!]
\centering
\[
\begin{array}{|c|c|c|c|c|}
\hline
N & E_{\rm DMRG} & E_{\rm SDP}   & \dfrac{E_{\rm DMRG}-E_{\rm SDP}}{|E_{\rm DMRG}|} & r \\
\hline
6 & -0.467129 & -0.467129 & 0.000000 & 3 \\
10 & -0.451545 & -0.451545 & 0.000000 & 5 \\
14 & -0.447396 & -0.447403 & 0.000015 & 7 \\
18 & -0.445708 & -0.445734 & 0.000059 & 9 \\
22 & -0.444858 & -0.444898 & 0.000090 & 11 \\
26 & -0.444371 & -0.444433 & 0.000141 & 13 \\
30 & -0.444065 & -0.444151 & 0.000193 & 15 \\
34 & -0.443862 & -0.443964 & 0.000231 & 17 \\
38 & -0.443719 & -0.443833 & 0.000257 & 19 \\
42 & -0.443615 & -0.443737 & 0.000275 & 21 \\
46 & -0.443537 & -0.443666 & 0.000290 & 23 \\
50 & -0.443477 & -0.443610 & 0.000300 & 25 \\
60 & -0.443376 & -0.443517 & 0.000318 & 30 \\
80 & -0.443276 & -0.443538 & 0.000591 & 20 \\
100 & -0.443229 & -0.443593 & 0.000820 & 20 \\
\hline
\end{array}
\]
\caption{Heisenberg chain energy in a system of $N$ spins. $r$ denotes the maximal distance between spins for two-body terms in the monomial list (see main text, section \ref{sec:B1}). Note that the relative accuracy (third column) is below $10^{-3}$ for all sizes.}
\label{Tab:B1-Energies}
\end{table}

\section{Heisenberg chain with second-neighbour couplings}
\label{app_B2}
In this section, we provide numerical data related to the $J_1-J_2$ Heisenberg chain (section \ref{sec:B2} in the main text). We provide SDP and DMRG data for the ground-state energy for $N=40$ spins as a function of $J_2$ (Table \ref{Tab:Heisenberg-2nd-Energies-L40}, corresponding to Fig.~\ref{fig:B2_fig1} in the main text)), as well as SDP data for $N=100$ spins (Table \ref{Tab:Heisenberg-2nd-Energies-L100}). In both tables, the last column indicates the degree $d$ of the monomials to construct the moment matrix, as explained in Appendix \ref{app_algorithmic_tricks}.

\begin{table}[ht!]
\centering
\begin{tabular}{|c|c|c|c|c|}
\hline
$J_2$ & $E_{\rm DMRG}$ & $E_{\rm SDP}$ & $\frac{E_{\rm DMRG}-E_{\rm SDP}}{|E_{\rm DMRG}|}$ & $d$\\
\hline
        0.1 & $-0.42581$ & $-0.42585$ & 0.00011 & 4\\
        0.2 & $-0.40892$ & $-0.40893$ & 0.00002& 4\\
        0.24117 & $-0.40233$ & $-0.40234$ & 0.00002 & 4 \\
        0.3 & $-0.39342$ & $-0.39346$ & 0.00012 & 4 \\
        0.4 & $-0.38055$ & $-0.38092$ & 0.00098 & 4\\
        0.5 & $-0.37500$ & $-0.37500$ &0.00000& 4\\
        0.6 & $-0.38081$ & $-0.38167$ & 0.00226 & 4\\
        0.7 & $-0.39721$ & $-0.39952$ & 0.00582 & 4\\
        0.8 & $-0.42177$ & $-0.42613$ & 0.01035 & 4\\
        0.9 & $-0.45206$ & $-0.45839$ & 0.01401 & 4\\
        1.0 & $-0.48657$ & $-0.49446$ & 0.01620 & 4\\
        1.5 & $-0.68570$ & $-0.69570$ & 0.01458 & 4\\
        2.0 & $-0.90242$ & $-0.91010$ & 0.00852 & 4\\
\hline
\end{tabular}
\caption{Ground-state energy in the Heisenberg chain with second-neighbour ($J_2$) for $N=40$ spins, as evaluated by SDP and DMRG methods.} \label{Tab:Heisenberg-2nd-Energies-L40}
\end{table}

\begin{table}[ht!]
\centering
\begin{tabular}{|c|c|c|c|}
\hline
$J_2$ & $E_{\rm SDP}$ & $r$ & $d$ \\
\hline
0.1 & $-0.42558385$ & 20 & 3 \\
0.2 & $-0.40861848$ & 20 & 3 \\
0.24117 & $-0.40205147$ & 20 & 3 \\
0.3 & $-0.39326097$ & 20 & 3 \\
0.4 & $-0.38088778$ & 20 & 3 \\
0.5 & $-0.37500000$ & 20 & 3 \\
0.6 & $-0.38215931$ & 20 & 3 \\
0.7 & $-0.40039365$ & 20 & 3 \\
0.8 & $-0.42707445$ & 20 & 3 \\
0.9 & $-0.45923392$ & 20 & 3 \\
1.0 & $-0.49525052$ & 20 & 3 \\
1.5 & $-0.69630683$ & 20 & 3 \\
2.0 & $-0.91191391$ & 20 & 3 \\
\hline
\end{tabular}
\caption{Ground-state energy lower-bound in the Heisenberg chain with second-neighbour ($J_2$) for $N=100$ spins, as evaluated by the SDP algorithm.}\label{Tab:Heisenberg-2nd-Energies-L100}
\end{table}

We also provide numerical data for the spin-spin correlations in a system with $N=40$ spins with PBC. 

In Table \ref{tab:B2-C1} we display SDP bounds on the first-neighbour correlation $C(1)=(1/4)\langle \sigma_i^x \sigma_{i+1}^x \rangle$ sandwiching the DMRG value (Fig.~\ref{fig:B2_figC1} in the main text).

\begin{table}[ht!]
\centering
\begin{tabular}{|c|c|c|c|}
\hline
$J_2$ & SDP Lower Bound & $C(1)_{\rm DMRG}$ & SDP Upper Bound \\
\hline
0.1 & $-0.14786074$ & $-0.1477430325$ & $-0.14765720$ \\ 
0.2 & $-0.14729208$ & $-0.147169525$ & $-0.14706027$ \\ 
0.241167 & $-0.14686185$ & $-0.1467200175$ & $-0.14655361$ \\ 
0.3 & $-0.14603355$ & $-0.14567769$ & $-0.14503139$ \\ 
0.4 & $-0.14304080$ & $-0.14074337$ & $-0.13719501$ \\ 
0.5 & $-0.12586317$ & $-0.125$ & $-0.12412305$ \\ 
0.6 & $-0.11102647$ & $-0.106575027$ & $-0.09321750$ \\ 
0.7 & $-0.09475276$ & $-0.0837010968$ & $-0.06511861$ \\ 
0.8 & $-0.08121743$ & $-0.0662572234$ & $-0.04425710$ \\ 
0.9 & $-0.07047691$ & $-0.0539501872$ & $-0.03055534$ \\ 
1.0 & $-0.06230067$ & $-0.0413538$ & $-0.02283913$ \\ 
1.5 & $-0.03687337$ & $-0.01515181$ & $-0.00762655$ \\ 
2.0 & $-0.02456216$ & $-0.00916219$ & $-0.00477422$ \\
\hline
\end{tabular}
\caption{Heisenberg chain with second-neighbour couplings. Spin-spin correlation at first neighbour ($N=40$).}
\label{tab:B2-C1}
\end{table}

In Table \ref{tab:B2-C2} we display similarly SDP bounds on the second-neighbour correlation $C(2)=(1/4)\langle \sigma_i^x \sigma_{i+2}^x \rangle$ sandwiching the DMRG value (Fig.~\ref{fig:B2_figC2} in the main text).

\begin{table}[ht!]
\centering
\begin{tabular}{|c|c|c|c|}
\hline
$J_2$ & SDP Lower Bound & $C(2)_{\rm DMRG}$ & SDP Upper Bound \\
\hline
0.1 & 0.05721213 & 0.058070785 & 0.05925762 \\
0.2 & 0.05377170 & 0.0543200625 & 0.05493536 \\
0.241167 & 0.05159693 & 0.0522883225 & 0.05287970 \\
0.3 & 0.04630329 & 0.0484616925 & 0.04965045 \\
0.4 & 0.02586365 & 0.034734385 & 0.04048043 \\
0.5 & $-0.00175465$ & 0 & 0.00173068 \\
0.6 & $-0.05619797$ & $-0.0381662752$ & $-0.02651573$ \\
0.7 & $-0.09611990$ & $-0.0697203904$ & $-0.05378431$ \\
0.8 & $-0.12041565$ & $-0.0929827357$ & $-0.07421526$ \\
0.9 & $-0.13347897$ & $-0.107586145$ & $-0.08912191$ \\
1.0 & $-0.13935193$ & $-0.12083676$ & $-0.09989024$ \\
1.5 & $-0.14729411$ & $-0.14224963$ & $-0.12779654$ \\
2.0 & $-0.14801455$ & $-0.14578753$ & $-0.13812159$ \\
\hline
\end{tabular}
\caption{Heisenberg chain with second-neighbour couplings. Spin-spin correlation at second neighbour ($N=40$).}
\label{tab:B2-C2}
\end{table}

We then provide the numerical data for the spin-spin correlation as a function of distance, for both $J_2=0.2$ (Table \ref{tab:B2-spinspin-J202} and main Fig.~\ref{fig:B2_spinspin-J202}) and $J_2=1.0$ (Table \ref{tab:B2-spinspin-J21} and main Fig.~\ref{fig:B2_spinspin-J21}).

\begin{table}[ht!]
\centering
\begin{tabular}{|c|c|c|c|}
\hline
$i$ & SDP Lower Bound & $C(i)_{\rm DMRG}$ & SDP Upper Bound \\
\hline
1 & $-0.14729208$ & $-0.14716953$ &  $-0.14706027$ \\
2 & 0.05377170 & 0.05432006 & 0.05493536 \\
3 & $-0.04223742$ & $-0.04150683$ & $-0.04071656$ \\
4 & 0.02697955 & 0.02788165 & 0.02875245 \\
5 & $-0.02608222$ & $-0.02506805$ & $-0.02391077$ \\
6 & 0.01828035 & 0.01960574 & 0.02097705 \\
7 & $-0.02014718$ & $-0.01837402$ & $-0.01667832$ \\
8 & 0.01339107 & 0.01542562 & 0.01760283 \\
9 & $-0.01748701$ & $-0.01480768$ & $-0.01235376$ \\
10 & 0.01014856 & 0.0129733 & 0.01597931 \\
11 & $-0.01610359$ & $-0.01266906$ & $-0.00949348$ \\
12 & 0.00792242 & 0.01142796 & 0.01513887 \\
13 & $-0.01540711$ & $-0.0113155$ & $-0.00748512$ \\
14 & 0.00627837 & 0.01043121 & 0.01476372 \\
15 & $-0.01513636$ & $-0.010455$ & $-0.00599421$ \\
16 & 0.00504368 & 0.00980607 & 0.01471382 \\
17 & $-0.01518012$ & $-0.00994131$ & $-0.00490270$ \\
18 & 0.00414641 & 0.00946065 & 0.01491181 \\
19 & $-0.01547775$ & $-0.00970028$ &  $-0.00413067$ \\
20 & 0.00358767 & 0.00935002 & 0.01525790 \\
\hline
\end{tabular}
\caption{Lower and upper SDP bounds for the spin-spin correlator at distance $i$ in the Heisenberg chain with second-neighbour couplings ($J_2=0.2$ and size $N=40$).}
\label{tab:B2-spinspin-J202}
\end{table}

\begin{table}[ht!]
    \centering
    \begin{tabular}{|c|c|c|c|}
\hline
$i$ & SDP Lower Bound & $C(i)_{\rm DMRG}$ & SDP Upper Bound \\
\hline
         1 & $-0.06230067 $ & $-0.0413538 $ & $-0.02283913$ \\
         2 & $-0.13935193 $ & $-0.12083676$ & $-0.09989024$ \\
         3 & $ 0.01884823 $ & $ 0.03311709$ & $ 0.05421466$ \\
         4 & $ 0.00971664 $ & $ 0.03623002$ & $ 0.05751094$ \\
         5 & $-0.06031966 $ & $-0.02670336$ & $-0.00682832$ \\
         6 & $-0.05338124 $ & $-0.01955047$ & $ 0.01661007$ \\
         7 & $-0.00390555 $ & $ 0.01860728$ & $ 0.05169353$ \\
         8 & $-0.04111449 $ & $ 0.00624808$ & $ 0.03992840$ \\
         9 & $-0.05681356 $ & $-0.01321554$ & $ 0.02103763$ \\
        10 & $-0.04328585 $ & $-0.00198039$ & $ 0.04430916$ \\
        11 & $-0.03371714 $ & $ 0.00905963$ & $ 0.04936674$ \\
        12 & $-0.04668599 $ & $-0.00176414$ & $ 0.03943488$ \\
        13 & $-0.05242524 $ & $-0.00596366$ & $ 0.03949383$ \\
        14 & $-0.04496886 $ & $ 0.00272444$ & $ 0.03852069$ \\
        15 & $-0.04945007 $ & $ 0.00389108$ & $ 0.04193050$ \\
        16 & $-0.05042092 $ & $-0.00398912$ & $ 0.04059603$ \\
        17 & $-0.042316503$ & $-0.00202086$ & $ 0.03886868$ \\
        18 & $-0.05358065 $ & $ 0.00395704$ & $ 0.05037295$ \\
        19 & $-0.02405042 $ & $ 0.00076097$ & $ 0.02054356$ \\
        20 & $-0.06885791 $ & $-0.00445211$ & $ 0.05499198$ \\
        \hline
    \end{tabular}
    \caption{Lower and upper SDP bounds for the spin-spin correlator at distance $i$ in the Heisenberg chain with second-neighbour couplings ($J_2=1.0$ and size $N=40$).}
    \label{tab:B2-spinspin-J21}
\end{table}

\clearpage

\section{Square lattice Heisenberg model}
\label{app_B3}
In this appendix, we provide numerical data on the square lattice Heisenberg model with PBC. We compare the SDP bounds with quantum Monte Carlo results of ref.~\cite{sandvik1997}. In Table \ref{tab:B3-Energies} we provide data for the ground-state energy as a function of the system size ($N=L\times L$ with $L=4,6,8,10$, main Fig.~\ref{fig:B3_Energies}).

\begin{table}[ht!]
    \centering
    \begin{tabular}{|c|c|c|c|}
\hline
$L$ & $E_{\rm SDP}$ & $E_{\rm MC}$& $\frac{E_{\rm MC}-E_{\rm SDP}}{|E_{\rm MC}|}$\\
\hline
    4 & $-0.70305078$ & $-0.7017777$&0.0018141\\
    6 & $-0.68317181$ & $-0.6788734$&0.0063317\\
    8 & $-0.67967080$ & $-0.6734875$&0.0091810\\
    10 & $-0.68003093$ & $-0.6715494$&0.0126298\\
    \hline
    \end{tabular}
    \caption{SDP lower bound on the energy of the square lattice Heisenberg model as compared to quantum Monte Carlo results.}
    \label{tab:B3-Energies}
\end{table}

In Table \ref{tab:B3-CL2} we provide data on the spin correlation $C(L/2,L/2)$, namely at maximal distance along the diagonal of the square lattice (main Fig.~\ref{fig:B3_CL2}). It is expected that this correlation remains nonzero in the thermodynamic limit, corresponding to antiferromagnetic long-range order in the ground state.

\begin{table}[ht!]
    \centering
    \begin{tabular}{|c|c|c|c|}
\hline
$L$ & SDP Lower Bound & $C(L/2,L/2)_{\rm MC}$ & SDP Upper Bound\\
\hline
  4 & 0.05227666 & 0.059872 & 0.06519277 \\
  6 & 0.02626831 & 0.050856 & 0.06314557 \\
  8 & 0.00199976 & 0.045867 & 0.07006021 \\
\hline
    \end{tabular}
    \caption{SDP lower and upper bound for the spin correlations at maximum distance in the square lattice Heisenberg model, sandwiching quantum Monte Carlo results.}
    \label{tab:B3-CL2}
\end{table}

\section{Square-lattice $J_1-J_2$ Heisenberg model}
In this appendix, we provide numeral data regarding the $J_1-J_2$ Heisenberg model on a square lattice (size $N=L\times L$ and PBC), as discussed in Section \ref{sec:B4} in the main text.

\subsection{Ground-state energy}
\label{app_B4}
We first present data for the ground-state energy, as compared with various variational methods employed in previous works in the literature.

\begin{figure}[!ht]
	\centering
\includegraphics[width=1\linewidth]{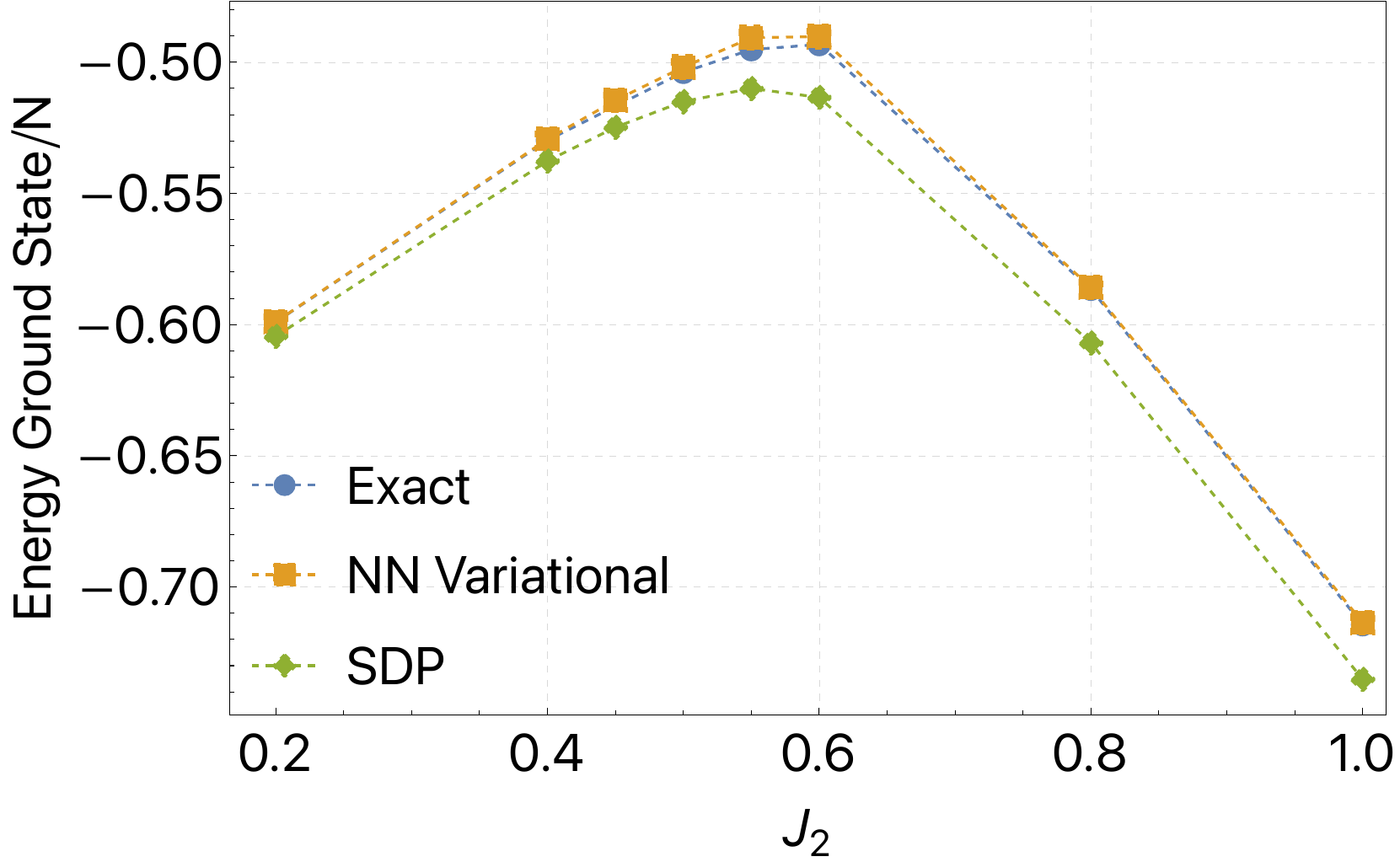}
	\caption{Energy lower bounds for the $2D$ $J_1-J_2$ Heisenberg model on a square with $L=6$ (data in Table \ref{tab:B4-L6}).}
	\label{fig:B4_1}
\end{figure}

In Table \ref{tab:B4-L6} we present data for $L=6$. We compare the SDP lower-bound with neural-network (NN) Ansatz wavefunctions  \cite{chooetal2019} and exact results \cite{Schulz96}.The same data are plotted in Fig.~\ref{fig:B4_1}.
\begin{table}[ht!]
    \centering
    \begin{tabular}{|c|c|c|c|c|}
\hline
$J_2$ & $E_{\rm SDP}$ & $E_{\rm NN}$ & $E_{\rm exact}$& $\frac{E_{\rm exact}-E_{\rm SDP}}{|E_{\rm exact}|}$\\
    \hline
    0.2 & $-0.60446854$ & $-0.59895$ & $-0.599046$ &0.00905\\
    0.4 & $-0.53763182$ & $-0.52936$ & $-0.529745$ &0.01489\\
    0.45 & $-0.52479952$ & $-0.51452$ &   &0.01998\\
    0.5 & $-0.51495867$ & $-0.50185$ &  $-0.503810$&0.02213\\
    0.55 & $-0.50999811$ & $-0.49067$ &  $-0.495178$&0.02993\\
    0.6 & $-0.51339892$ & $-0.49023$ &  $-0.493239$&0.04087\\
    0.8 & $-0.60697786$ & $-0.58590$ &  $-0.586487$&0.03495\\
    1.0 & $-0.73517835$ & $-0.71351$ &  $-0.714360$&0.02914\\
    \hline
    \end{tabular}
    \caption{Ground-state energy for the square-lattice $J_1-J_2$ Heisenberg model ($L=6$). Last column: relative difference between the best variational upper bound and the SDP lower bound.}
    \label{tab:B4-L6}
\end{table}

\begin{figure}[!ht]
	\centering
\includegraphics[width=1\linewidth]{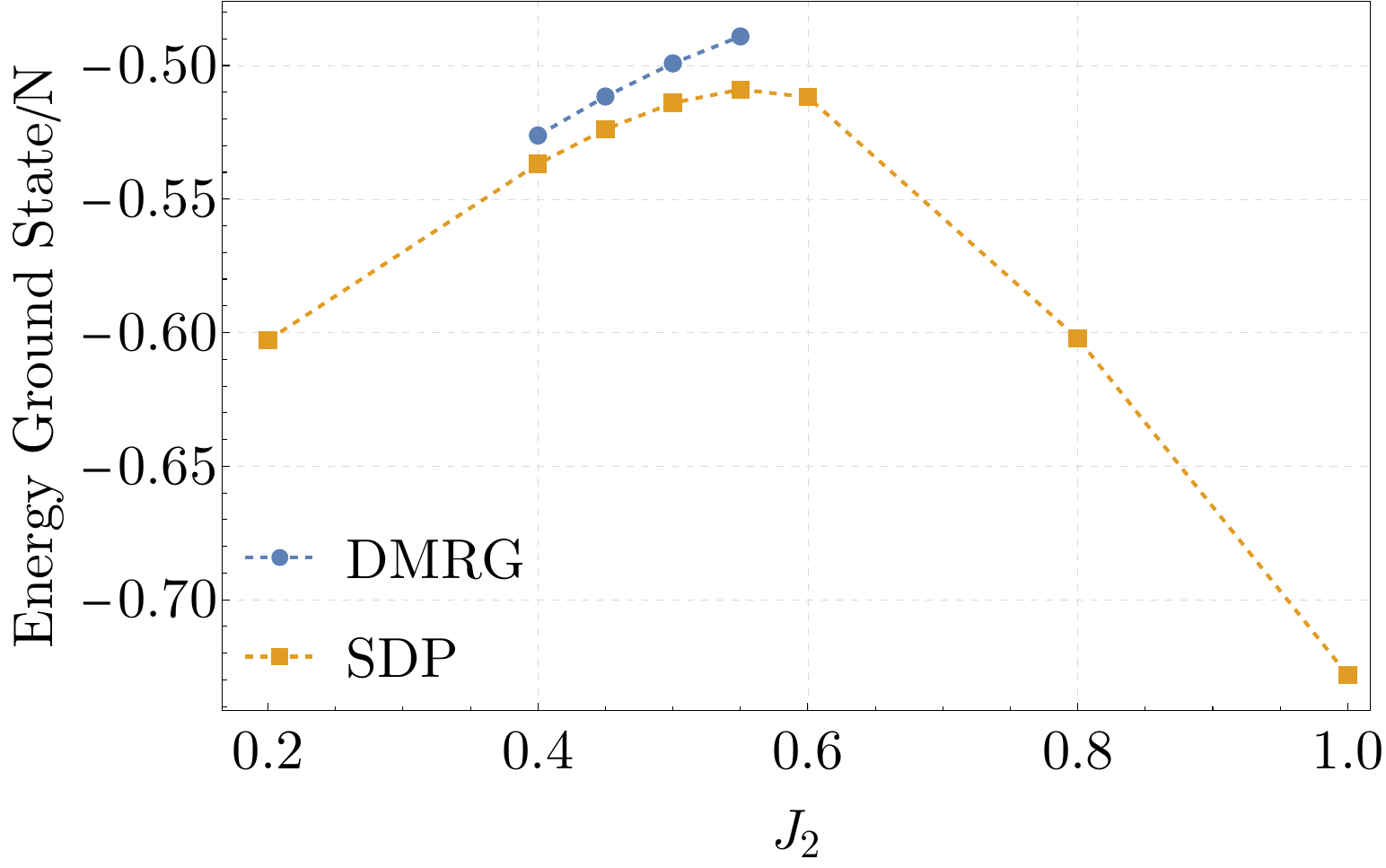}
	\caption{Energy lower bounds for the square-lattice $J_1-J_2$ Heisenberg model with $L=8$ (data in Table \ref{tab:B4-L8}).}
	\label{fig:B4_2}
\end{figure}

In Table \ref{tab:B4-L8} we present data for $L=8$. We compare the SDP lower-bound with variational Monte Carlo on Ansatz wavefunctions (VMC) and DMRG computations \cite{gongetal2014}. The same data are plotted in Fig.~\ref{fig:B4_2}.

\begin{table}[ht!]
    \centering
    \begin{tabular}{|c|c|c|c|c|}
\hline
$J_2$ & $E_{\rm SDP}$ & $E_{\rm VMC}$ & $E_{\rm DMRG}$& $\frac{E_{\rm var}-E_{\rm SDP}}{|E_{\rm var}|}$\\
\hline
0.2     & $-0.60284236$    & & &\\
0.4     & $-0.53682283$   & $-0.52556$  & $-0.5262$  & 0.02019\\
0.45    & $-0.52391112$   & $-0.51140$  & $-0.5116$  & 0.02406\\
0.5     & $-0.51398956$   & $-0.49906$  & $-0.4992$  & 0.02963\\
0.55    & $-0.50899192$   & $-0.48894$  & $-0.4891$  & 0.04067\\
0.6     & $-0.51182820$   & &    &\\
0.8     & $-0.60221175$   & &    &\\
1.0     & $-0.72821272$   & &    &\\
\hline
    \end{tabular}
    \caption{Energy lower bounds for the square-lattice $J_1-J_2$ Heisenberg model ($L=8$). Last column: relative difference between the best variational upper bound and the SDP lower bound.}
    \label{tab:B4-L8}
\end{table}

In Table \ref{tab:B4-L10} we provide numerical data for Fig.~\ref{fig:B4_3} in the main text, namely ground-state energy for $L=10$ for both SDP, NN and DMRG methods.

\begin{table}[ht!]
    \centering
    \begin{tabular}{|c|c|c|c|c|}
\hline
$J_2$ & $E_{\rm SDP}$ & $E_{\rm NN}$ & $E_{\rm DMRG}$ & $\frac{E_{\rm var}-E_{\rm SDP}}{|E_{\rm var}|}$\\
\hline
    0.2     & $-0.60308301$ & $-0.59275$ &  & 0.01743\\
    0.4     & $-0.53747136$ & $-0.52371$ & $-0.5253$ & 0.02317 \\
    0.45    & $-0.52464206$ & $-0.50905$ & $-0.5110$ & 0.02670\\
    0.5     & $-0.51462802$ & $-0.49516$ & $-0.4988$ & 0.03173\\
    0.55    & $-0.50930924$ & $-0.48277$ & $-0.4880$ & 0.04367\\
    0.6     & $-0.51136063$ & $-0.47604$ &  &  0.07420\\
    0.8     & $-0.59945283$ & $-0.57383$ &  &  0.04465\\
    1.0     & $-0.72475248$ & $-0.69636$ &  & 0.04077\\
\hline
    \end{tabular}
    \caption{Energy lower bound (SDP) and upper bounds (NN and DMRG) for the square-lattice $J_1-J_2$ Heisenberg model ($L=10$). Last column: relative difference between the best variational upper bound and the SDP lower bound.}
    \label{tab:B4-L10}
\end{table}

\subsection{First- and second-neighbour spin-spin correlations}
We finally provide SDP bounds for first- and second-neighbour spin-spin correlations as a function of the second-neighbour coupling $J_2$, for a system of size $L=10$. Data are respectively displayed in Table \ref{tab:B4-C01-L10} (corresponding to Fig.~\ref{fig:B4_bounds-C01-L10} in the main text) and Table \ref{tab:B4-C11-L10} (corresponding to Fig.~\ref{fig:B4_bounds-C11-L10} in the main text).

\begin{table}[ht!]
    \centering
    \begin{tabular}{|c|c|c|}
\hline
$J_2$ & min $C(0,1)$ & max $C(0,1)$\\
\hline
    0.2  & $-0.11333849$ & $-0.10800495$ \\
    0.4  & $-0.11324885$ & $-0.09395090$ \\
    0.45 & $-0.11296774$ & $-0.08486804$ \\
    0.5  & $-0.11249860$ & $-0.06910228$ \\
    0.55 & $-0.11175223$ & $-0.04379525$ \\
    0.6  & $-0.10979021$ & $-0.01789049$ \\
    0.8  & $-0.05065273$ & $-0.00391276$ \\
    1.0  & $-0.03156633$ & $-0.00055923$ \\
\hline
    \end{tabular}
    \caption{Bounds for the correlations $C(0,1)$ of the square-lattice $J_1-J_2$ Heisenberg model ($L=10$).}
    \label{tab:B4-C01-L10}
\end{table}

\begin{table}[ht!]
    \centering
    \begin{tabular}{|c|c|c|}
\hline
$J_2$ & min $C(0,1)$ & max $C(0,1)$\\
\hline
    0.2 & 0.04606643 & 0.07095733 \\
    0.4 & 0.01666475 & 0.06490964 \\
    0.45 & 0.00005861 & 0.06250239 \\
    0.5 & $-0.02684876$ & 0.05994387 \\
    0.55 & $-0.06666621$ & 0.05689195 \\
    0.6 & $-0.10979021$ & $-0.01789049$ \\
    0.8 & $-0.05065273$ & $-0.00391276$ \\
    1.0 & $-0.03156633$ & $-0.00055923$ \\
\hline
    \end{tabular}
    \caption{Bounds for the correlations $C(1,1)$ of the square-lattice $J_1-J_2$ Heisenberg model ($L=10$).}
    \label{tab:B4-C11-L10}
\end{table}

\revision{
\section{Implementation details}
All experiments were performed with one CPU core and 128G memory. In practice, we solve the dual problem of the SDP \eqref{mom} (i.e., the sum-of-Hermitian-squares problem) since it contains fewer linear constraints and so can be solved more efficiently. We rely on the Julia package \href{https://github.com/wangjie212/QMBCertify}{{\tt QMBCertify}} to create the optimization model and employ {\tt Mosek 10.0} as an underlying SDP solver. For the 1D model with $N=40,100$, it takes around 3.3h, 12h to generate a data point, respectively; for the 2D model model with $L=6,8,10$, it takes around 1.8h, 9.7h, 21h to generate a data point, respectively.
}

\end{appendix}
\clearpage

\bibliography{refer}

\begin{thebibliography}{73}%
\makeatletter
\providecommand \@ifxundefined [1]{%
 \@ifx{#1\undefined}
}%
\providecommand \@ifnum [1]{%
 \ifnum #1\expandafter \@firstoftwo
 \else \expandafter \@secondoftwo
 \fi
}%
\providecommand \@ifx [1]{%
 \ifx #1\expandafter \@firstoftwo
 \else \expandafter \@secondoftwo
 \fi
}%
\providecommand \natexlab [1]{#1}%
\providecommand \enquote  [1]{``#1''}%
\providecommand \bibnamefont  [1]{#1}%
\providecommand \bibfnamefont [1]{#1}%
\providecommand \citenamefont [1]{#1}%
\providecommand \href@noop [0]{\@secondoftwo}%
\providecommand \href [0]{\begingroup \@sanitize@url \@href}%
\providecommand \@href[1]{\@@startlink{#1}\@@href}%
\providecommand \@@href[1]{\endgroup#1\@@endlink}%
\providecommand \@sanitize@url [0]{\catcode `\\12\catcode `\$12\catcode
  `\&12\catcode `\#12\catcode `\^12\catcode `\_12\catcode `\%12\relax}%
\providecommand \@@startlink[1]{}%
\providecommand \@@endlink[0]{}%
\providecommand \url  [0]{\begingroup\@sanitize@url \@url }%
\providecommand \@url [1]{\endgroup\@href {#1}{\urlprefix }}%
\providecommand \urlprefix  [0]{URL }%
\providecommand \Eprint [0]{\href }%
\providecommand \doibase [0]{https://doi.org/}%
\providecommand \selectlanguage [0]{\@gobble}%
\providecommand \bibinfo  [0]{\@secondoftwo}%
\providecommand \bibfield  [0]{\@secondoftwo}%
\providecommand \translation [1]{[#1]}%
\providecommand \BibitemOpen [0]{}%
\providecommand \bibitemStop [0]{}%
\providecommand \bibitemNoStop [0]{.\EOS\space}%
\providecommand \EOS [0]{\spacefactor3000\relax}%
\providecommand \BibitemShut  [1]{\csname bibitem#1\endcsname}%
\let\auto@bib@innerbib\@empty
\bibitem [{\citenamefont {Sachdev}(2011)}]{sachdev2011quantum}%
  \BibitemOpen
  \bibfield  {author} {\bibinfo {author} {\bibfnamefont {S.}~\bibnamefont
  {Sachdev}},\ }\href
  {https://www.cambridge.org/nl/universitypress/subjects/physics/condensed-matter-physics-nanoscience-and-mesoscopic-physics/quantum-phase-transitions-2nd-edition?format=HB&isbn=9780521514682}
  {\emph {\bibinfo {title} {Quantum Phase Transitions}}}\ (\bibinfo
  {publisher} {Cambridge University Press},\ \bibinfo {year}
  {2011})\BibitemShut {NoStop}%
\bibitem [{\citenamefont {Wu}\ \emph {et~al.}(2023)\citenamefont {Wu},
  \citenamefont {Rossi}, \citenamefont {Vicentini}, \citenamefont
  {Astrakhantsev}, \citenamefont {Becca}, \citenamefont {Cao}, \citenamefont
  {Carrasquilla}, \citenamefont {Ferrari}, \citenamefont {Georges},
  \citenamefont {Hibat-Allah}, \citenamefont {Imada}, \citenamefont {Läuchli},
  \citenamefont {Mazzola}, \citenamefont {Mezzacapo}, \citenamefont {Millis},
  \citenamefont {Moreno}, \citenamefont {Neupert}, \citenamefont {Nomura},
  \citenamefont {Nys}, \citenamefont {Parcollet}, \citenamefont {Pohle},
  \citenamefont {Romero}, \citenamefont {Schmid}, \citenamefont {Silvester},
  \citenamefont {Sorella}, \citenamefont {Tocchio}, \citenamefont {Wang},
  \citenamefont {White}, \citenamefont {Wietek}, \citenamefont {Yang},
  \citenamefont {Yang}, \citenamefont {Zhang},\ and\ \citenamefont
  {Carleo}}]{wu2023variational}%
  \BibitemOpen
  \bibfield  {author} {\bibinfo {author} {\bibfnamefont {D.}~\bibnamefont
  {Wu}}, \bibinfo {author} {\bibfnamefont {R.}~\bibnamefont {Rossi}}, \bibinfo
  {author} {\bibfnamefont {F.}~\bibnamefont {Vicentini}}, \bibinfo {author}
  {\bibfnamefont {N.}~\bibnamefont {Astrakhantsev}}, \bibinfo {author}
  {\bibfnamefont {F.}~\bibnamefont {Becca}}, \bibinfo {author} {\bibfnamefont
  {X.}~\bibnamefont {Cao}}, \bibinfo {author} {\bibfnamefont {J.}~\bibnamefont
  {Carrasquilla}}, \bibinfo {author} {\bibfnamefont {F.}~\bibnamefont
  {Ferrari}}, \bibinfo {author} {\bibfnamefont {A.}~\bibnamefont {Georges}},
  \bibinfo {author} {\bibfnamefont {M.}~\bibnamefont {Hibat-Allah}}, \bibinfo
  {author} {\bibfnamefont {M.}~\bibnamefont {Imada}}, \bibinfo {author}
  {\bibfnamefont {A.~M.}\ \bibnamefont {Läuchli}}, \bibinfo {author}
  {\bibfnamefont {G.}~\bibnamefont {Mazzola}}, \bibinfo {author} {\bibfnamefont
  {A.}~\bibnamefont {Mezzacapo}}, \bibinfo {author} {\bibfnamefont
  {A.}~\bibnamefont {Millis}}, \bibinfo {author} {\bibfnamefont {J.~R.}\
  \bibnamefont {Moreno}}, \bibinfo {author} {\bibfnamefont {T.}~\bibnamefont
  {Neupert}}, \bibinfo {author} {\bibfnamefont {Y.}~\bibnamefont {Nomura}},
  \bibinfo {author} {\bibfnamefont {J.}~\bibnamefont {Nys}}, \bibinfo {author}
  {\bibfnamefont {O.}~\bibnamefont {Parcollet}}, \bibinfo {author}
  {\bibfnamefont {R.}~\bibnamefont {Pohle}}, \bibinfo {author} {\bibfnamefont
  {I.}~\bibnamefont {Romero}}, \bibinfo {author} {\bibfnamefont
  {M.}~\bibnamefont {Schmid}}, \bibinfo {author} {\bibfnamefont {J.~M.}\
  \bibnamefont {Silvester}}, \bibinfo {author} {\bibfnamefont {S.}~\bibnamefont
  {Sorella}}, \bibinfo {author} {\bibfnamefont {L.~F.}\ \bibnamefont
  {Tocchio}}, \bibinfo {author} {\bibfnamefont {L.}~\bibnamefont {Wang}},
  \bibinfo {author} {\bibfnamefont {S.~R.}\ \bibnamefont {White}}, \bibinfo
  {author} {\bibfnamefont {A.}~\bibnamefont {Wietek}}, \bibinfo {author}
  {\bibfnamefont {Q.}~\bibnamefont {Yang}}, \bibinfo {author} {\bibfnamefont
  {Y.}~\bibnamefont {Yang}}, \bibinfo {author} {\bibfnamefont {S.}~\bibnamefont
  {Zhang}},\ and\ \bibinfo {author} {\bibfnamefont {G.}~\bibnamefont
  {Carleo}},\ }\href {https://doi.org/10.48550/arXiv.2302.04919} {\bibinfo
  {title} {Variational benchmarks for quantum many-body problems}} (\bibinfo
  {year} {2023}),\ \Eprint {https://arxiv.org/abs/2302.04919} {arXiv:2302.04919
  [quant-ph]} \BibitemShut {NoStop}%
\bibitem [{\citenamefont {Zheng}\ \emph {et~al.}(2017)\citenamefont {Zheng},
  \citenamefont {Chung}, \citenamefont {Corboz}, \citenamefont {Ehlers},
  \citenamefont {Qin}, \citenamefont {Noack}, \citenamefont {Shi},
  \citenamefont {White}, \citenamefont {Zhang},\ and\ \citenamefont
  {Chan}}]{zhengetal2017}%
  \BibitemOpen
  \bibfield  {author} {\bibinfo {author} {\bibfnamefont {B.-X.}\ \bibnamefont
  {Zheng}}, \bibinfo {author} {\bibfnamefont {C.-M.}\ \bibnamefont {Chung}},
  \bibinfo {author} {\bibfnamefont {P.}~\bibnamefont {Corboz}}, \bibinfo
  {author} {\bibfnamefont {G.}~\bibnamefont {Ehlers}}, \bibinfo {author}
  {\bibfnamefont {M.-P.}\ \bibnamefont {Qin}}, \bibinfo {author} {\bibfnamefont
  {R.~M.}\ \bibnamefont {Noack}}, \bibinfo {author} {\bibfnamefont
  {H.}~\bibnamefont {Shi}}, \bibinfo {author} {\bibfnamefont {S.~R.}\
  \bibnamefont {White}}, \bibinfo {author} {\bibfnamefont {S.}~\bibnamefont
  {Zhang}},\ and\ \bibinfo {author} {\bibfnamefont {G.~K.-L.}\ \bibnamefont
  {Chan}},\ }\bibfield  {title} {\bibinfo {title} {Stripe order in the
  underdoped region of the two-dimensional hubbard model},\ }\href
  {https://doi.org/10.1126/science.aam7127} {\bibfield  {journal} {\bibinfo
  {journal} {Science}\ }\textbf {\bibinfo {volume} {358}},\ \bibinfo {pages}
  {1155} (\bibinfo {year} {2017})},\ \Eprint
  {https://arxiv.org/abs/https://www.science.org/doi/pdf/10.1126/science.aam7127}
  {https://www.science.org/doi/pdf/10.1126/science.aam7127} \BibitemShut
  {NoStop}%
\bibitem [{\citenamefont {Nakata}\ \emph {et~al.}(2001)\citenamefont {Nakata},
  \citenamefont {Nakatsuji}, \citenamefont {Ehara}, \citenamefont {Fukuda},
  \citenamefont {Nakata},\ and\ \citenamefont {Fujisawa}}]{Nakata01}%
  \BibitemOpen
  \bibfield  {author} {\bibinfo {author} {\bibfnamefont {M.}~\bibnamefont
  {Nakata}}, \bibinfo {author} {\bibfnamefont {H.}~\bibnamefont {Nakatsuji}},
  \bibinfo {author} {\bibfnamefont {M.}~\bibnamefont {Ehara}}, \bibinfo
  {author} {\bibfnamefont {M.}~\bibnamefont {Fukuda}}, \bibinfo {author}
  {\bibfnamefont {K.}~\bibnamefont {Nakata}},\ and\ \bibinfo {author}
  {\bibfnamefont {K.}~\bibnamefont {Fujisawa}},\ }\bibfield  {title} {\bibinfo
  {title} {{Variational calculations of fermion second-order reduced density
  matrices by semidefinite programming algorithm}},\ }\href
  {https://doi.org/10.1063/1.1360199} {\bibfield  {journal} {\bibinfo
  {journal} {The Journal of Chemical Physics}\ }\textbf {\bibinfo {volume}
  {114}},\ \bibinfo {pages} {8282} (\bibinfo {year} {2001})},\ \Eprint
  {https://arxiv.org/abs/https://pubs.aip.org/aip/jcp/article-pdf/114/19/8282/10831897/8282\_1\_online.pdf}
  {https://pubs.aip.org/aip/jcp/article-pdf/114/19/8282/10831897/8282\_1\_online.pdf}
  \BibitemShut {NoStop}%
\bibitem [{\citenamefont {Mazziotti}\ and\ \citenamefont
  {Erdahl}(2001)}]{Mazziotti01}%
  \BibitemOpen
  \bibfield  {author} {\bibinfo {author} {\bibfnamefont {D.~A.}\ \bibnamefont
  {Mazziotti}}\ and\ \bibinfo {author} {\bibfnamefont {R.~M.}\ \bibnamefont
  {Erdahl}},\ }\bibfield  {title} {\bibinfo {title} {Uncertainty relations and
  reduced density matrices: Mapping many-body quantum mechanics onto four
  particles},\ }\href {https://doi.org/10.1103/PhysRevA.63.042113} {\bibfield
  {journal} {\bibinfo  {journal} {Phys. Rev. A}\ }\textbf {\bibinfo {volume}
  {63}},\ \bibinfo {pages} {042113} (\bibinfo {year} {2001})}\BibitemShut
  {NoStop}%
\bibitem [{\citenamefont {Barthel}\ and\ \citenamefont
  {H\"ubener}(2012)}]{BarthelH2012}%
  \BibitemOpen
  \bibfield  {author} {\bibinfo {author} {\bibfnamefont {T.}~\bibnamefont
  {Barthel}}\ and\ \bibinfo {author} {\bibfnamefont {R.}~\bibnamefont
  {H\"ubener}},\ }\bibfield  {title} {\bibinfo {title} {Solving
  condensed-matter ground-state problems by semidefinite relaxations},\ }\href
  {https://doi.org/10.1103/PhysRevLett.108.200404} {\bibfield  {journal}
  {\bibinfo  {journal} {Phys. Rev. Lett.}\ }\textbf {\bibinfo {volume} {108}},\
  \bibinfo {pages} {200404} (\bibinfo {year} {2012})}\BibitemShut {NoStop}%
\bibitem [{\citenamefont {Baumgratz}\ and\ \citenamefont
  {Plenio}(2012)}]{Baumgratz_2012}%
  \BibitemOpen
  \bibfield  {author} {\bibinfo {author} {\bibfnamefont {T.}~\bibnamefont
  {Baumgratz}}\ and\ \bibinfo {author} {\bibfnamefont {M.~B.}\ \bibnamefont
  {Plenio}},\ }\bibfield  {title} {\bibinfo {title} {Lower bounds for ground
  states of condensed matter systems},\ }\href
  {https://doi.org/10.1088/1367-2630/14/2/023027} {\bibfield  {journal}
  {\bibinfo  {journal} {New Journal of Physics}\ }\textbf {\bibinfo {volume}
  {14}},\ \bibinfo {pages} {023027} (\bibinfo {year} {2012})}\BibitemShut
  {NoStop}%
\bibitem [{\citenamefont {Navascu{\'e}s}\ \emph {et~al.}(2008)\citenamefont
  {Navascu{\'e}s}, \citenamefont {Pironio},\ and\ \citenamefont
  {Ac{\'i}n}}]{navascues2008convergent}%
  \BibitemOpen
  \bibfield  {author} {\bibinfo {author} {\bibfnamefont {M.}~\bibnamefont
  {Navascu{\'e}s}}, \bibinfo {author} {\bibfnamefont {S.}~\bibnamefont
  {Pironio}},\ and\ \bibinfo {author} {\bibfnamefont {A.}~\bibnamefont
  {Ac{\'i}n}},\ }\bibfield  {title} {\bibinfo {title} {A convergent hierarchy
  of semidefinite programs characterizing the set of quantum correlations},\
  }\href {https://dx.doi.org/10.1088/1367-2630/10/7/073013} {\bibfield
  {journal} {\bibinfo  {journal} {New Journal of Physics}\ }\textbf {\bibinfo
  {volume} {10}},\ \bibinfo {pages} {073013} (\bibinfo {year}
  {2008})}\BibitemShut {NoStop}%
\bibitem [{\citenamefont {Pironio}\ \emph {et~al.}(2010)\citenamefont
  {Pironio}, \citenamefont {Navascu{\'e}s},\ and\ \citenamefont
  {Ac{\'{i}}n}}]{pironio2010convergent}%
  \BibitemOpen
  \bibfield  {author} {\bibinfo {author} {\bibfnamefont {S.}~\bibnamefont
  {Pironio}}, \bibinfo {author} {\bibfnamefont {M.}~\bibnamefont
  {Navascu{\'e}s}},\ and\ \bibinfo {author} {\bibfnamefont {A.}~\bibnamefont
  {Ac{\'{i}}n}},\ }\bibfield  {title} {\bibinfo {title} {Convergent relaxations
  of polynomial optimization problems with noncommuting variables},\ }\href
  {https://doi.org/10.1137/090760155} {\bibfield  {journal} {\bibinfo
  {journal} {SIAM Journal on Optimization}\ }\textbf {\bibinfo {volume} {20}},\
  \bibinfo {pages} {2157} (\bibinfo {year} {2010})}\BibitemShut {NoStop}%
\bibitem [{\citenamefont {Helton}\ and\ \citenamefont
  {McCullough}(2004)}]{helton2004positivstellensatz}%
  \BibitemOpen
  \bibfield  {author} {\bibinfo {author} {\bibfnamefont {J.}~\bibnamefont
  {Helton}}\ and\ \bibinfo {author} {\bibfnamefont {S.}~\bibnamefont
  {McCullough}},\ }\bibfield  {title} {\bibinfo {title} {A positivstellensatz
  for non-commutative polynomials},\ }\href
  {https://doi.org/10.1090/S0002-9947-04-03433-6} {\bibfield  {journal}
  {\bibinfo  {journal} {Transactions of the American Mathematical Society}\
  }\textbf {\bibinfo {volume} {356}},\ \bibinfo {pages} {3721} (\bibinfo {year}
  {2004})}\BibitemShut {NoStop}%
\bibitem [{\citenamefont {Doherty}\ \emph {et~al.}(2008)\citenamefont
  {Doherty}, \citenamefont {Liang}, \citenamefont {Toner},\ and\ \citenamefont
  {Wehner}}]{doherty08}%
  \BibitemOpen
  \bibfield  {author} {\bibinfo {author} {\bibfnamefont {A.~C.}\ \bibnamefont
  {Doherty}}, \bibinfo {author} {\bibfnamefont {Y.-C.}\ \bibnamefont {Liang}},
  \bibinfo {author} {\bibfnamefont {B.}~\bibnamefont {Toner}},\ and\ \bibinfo
  {author} {\bibfnamefont {S.}~\bibnamefont {Wehner}},\ }\bibfield  {title}
  {\bibinfo {title} {The quantum moment problem and bounds on entangled
  multi-prover games},\ }in\ \href {https://doi.org/10.1109/CCC.2008.26} {\emph
  {\bibinfo {booktitle} {Proceedings of the 2008 IEEE 23rd Annual Conference on
  Computational Complexity}}},\ \bibinfo {series and number} {CCC '08}\
  (\bibinfo  {publisher} {IEEE Computer Society},\ \bibinfo {address} {USA},\
  \bibinfo {year} {2008})\ p.\ \bibinfo {pages} {199–210}\BibitemShut
  {NoStop}%
\bibitem [{\citenamefont {Han}(2020)}]{han2020quantum}%
  \BibitemOpen
  \bibfield  {author} {\bibinfo {author} {\bibfnamefont {X.}~\bibnamefont
  {Han}},\ }\href@noop {} {\bibinfo {title} {Quantum many-body bootstrap}}
  (\bibinfo {year} {2020}),\ \Eprint {https://arxiv.org/abs/2006.06002}
  {arXiv:2006.06002 [cond-mat.str-el]} \BibitemShut {NoStop}%
\bibitem [{\citenamefont {White}(1992)}]{WhiteDMRG}%
  \BibitemOpen
  \bibfield  {author} {\bibinfo {author} {\bibfnamefont {S.~R.}\ \bibnamefont
  {White}},\ }\bibfield  {title} {\bibinfo {title} {Density matrix formulation
  for quantum renormalization groups},\ }\href
  {https://doi.org/10.1103/PhysRevLett.69.2863} {\bibfield  {journal} {\bibinfo
   {journal} {Phys. Rev. Lett.}\ }\textbf {\bibinfo {volume} {69}},\ \bibinfo
  {pages} {2863} (\bibinfo {year} {1992})}\BibitemShut {NoStop}%
\bibitem [{\citenamefont {Schollw\"ock}(2005)}]{schollwock_DMRG}%
  \BibitemOpen
  \bibfield  {author} {\bibinfo {author} {\bibfnamefont {U.}~\bibnamefont
  {Schollw\"ock}},\ }\bibfield  {title} {\bibinfo {title} {The density-matrix
  renormalization group},\ }\href {https://doi.org/10.1103/RevModPhys.77.259}
  {\bibfield  {journal} {\bibinfo  {journal} {Rev. Mod. Phys.}\ }\textbf
  {\bibinfo {volume} {77}},\ \bibinfo {pages} {259} (\bibinfo {year}
  {2005})}\BibitemShut {NoStop}%
\bibitem [{\citenamefont {\"Ostlund}\ and\ \citenamefont
  {Rommer}(1995)}]{OstlundRommer}%
  \BibitemOpen
  \bibfield  {author} {\bibinfo {author} {\bibfnamefont {S.}~\bibnamefont
  {\"Ostlund}}\ and\ \bibinfo {author} {\bibfnamefont {S.}~\bibnamefont
  {Rommer}},\ }\bibfield  {title} {\bibinfo {title} {Thermodynamic limit of
  density matrix renormalization},\ }\href
  {https://doi.org/10.1103/PhysRevLett.75.3537} {\bibfield  {journal} {\bibinfo
   {journal} {Phys. Rev. Lett.}\ }\textbf {\bibinfo {volume} {75}},\ \bibinfo
  {pages} {3537} (\bibinfo {year} {1995})}\BibitemShut {NoStop}%
\bibitem [{\citenamefont {Schollwöck}(2011)}]{SCHOLLWOCK201196}%
  \BibitemOpen
  \bibfield  {author} {\bibinfo {author} {\bibfnamefont {U.}~\bibnamefont
  {Schollwöck}},\ }\bibfield  {title} {\bibinfo {title} {The density-matrix
  renormalization group in the age of matrix product states},\ }\href
  {https://doi.org/https://doi.org/10.1016/j.aop.2010.09.012} {\bibfield
  {journal} {\bibinfo  {journal} {Annals of Physics}\ }\textbf {\bibinfo
  {volume} {326}},\ \bibinfo {pages} {96} (\bibinfo {year} {2011})},\ \bibinfo
  {note} {january 2011 Special Issue}\BibitemShut {NoStop}%
\bibitem [{\citenamefont {Verstraete}\ \emph {et~al.}(2008)\citenamefont
  {Verstraete}, \citenamefont {Murg},\ and\ \citenamefont
  {Cirac}}]{doi:10.1080/14789940801912366}%
  \BibitemOpen
  \bibfield  {author} {\bibinfo {author} {\bibfnamefont {F.}~\bibnamefont
  {Verstraete}}, \bibinfo {author} {\bibfnamefont {V.}~\bibnamefont {Murg}},\
  and\ \bibinfo {author} {\bibfnamefont {J.}~\bibnamefont {Cirac}},\ }\bibfield
   {title} {\bibinfo {title} {Matrix product states, projected entangled pair
  states, and variational renormalization group methods for quantum spin
  systems},\ }\href {https://doi.org/10.1080/14789940801912366} {\bibfield
  {journal} {\bibinfo  {journal} {Advances in Physics}\ }\textbf {\bibinfo
  {volume} {57}},\ \bibinfo {pages} {143} (\bibinfo {year} {2008})},\ \Eprint
  {https://arxiv.org/abs/https://doi.org/10.1080/14789940801912366}
  {https://doi.org/10.1080/14789940801912366} \BibitemShut {NoStop}%
\bibitem [{\citenamefont {Cirac}\ \emph {et~al.}(2021)\citenamefont {Cirac},
  \citenamefont {P\'erez-Garc\'{\i}a}, \citenamefont {Schuch},\ and\
  \citenamefont {Verstraete}}]{Ciracetal2021}%
  \BibitemOpen
  \bibfield  {author} {\bibinfo {author} {\bibfnamefont {J.~I.}\ \bibnamefont
  {Cirac}}, \bibinfo {author} {\bibfnamefont {D.}~\bibnamefont
  {P\'erez-Garc\'{\i}a}}, \bibinfo {author} {\bibfnamefont {N.}~\bibnamefont
  {Schuch}},\ and\ \bibinfo {author} {\bibfnamefont {F.}~\bibnamefont
  {Verstraete}},\ }\bibfield  {title} {\bibinfo {title} {Matrix product states
  and projected entangled pair states: Concepts, symmetries, theorems},\ }\href
  {https://doi.org/10.1103/RevModPhys.93.045003} {\bibfield  {journal}
  {\bibinfo  {journal} {Rev. Mod. Phys.}\ }\textbf {\bibinfo {volume} {93}},\
  \bibinfo {pages} {045003} (\bibinfo {year} {2021})}\BibitemShut {NoStop}%
\bibitem [{\citenamefont {Orús}(2014)}]{ORUS2014117}%
  \BibitemOpen
  \bibfield  {author} {\bibinfo {author} {\bibfnamefont {R.}~\bibnamefont
  {Orús}},\ }\bibfield  {title} {\bibinfo {title} {A practical introduction to
  tensor networks: Matrix product states and projected entangled pair states},\
  }\href {https://doi.org/https://doi.org/10.1016/j.aop.2014.06.013} {\bibfield
   {journal} {\bibinfo  {journal} {Annals of Physics}\ }\textbf {\bibinfo
  {volume} {349}},\ \bibinfo {pages} {117} (\bibinfo {year}
  {2014})}\BibitemShut {NoStop}%
\bibitem [{\citenamefont {Vidal}(2007{\natexlab{a}})}]{vidal1}%
  \BibitemOpen
  \bibfield  {author} {\bibinfo {author} {\bibfnamefont {G.}~\bibnamefont
  {Vidal}},\ }\bibfield  {title} {\bibinfo {title} {Classical simulation of
  infinite-size quantum lattice systems in one spatial dimension},\ }\href
  {https://doi.org/10.1103/PhysRevLett.98.070201} {\bibfield  {journal}
  {\bibinfo  {journal} {Phys. Rev. Lett.}\ }\textbf {\bibinfo {volume} {98}},\
  \bibinfo {pages} {070201} (\bibinfo {year} {2007}{\natexlab{a}})}\BibitemShut
  {NoStop}%
\bibitem [{\citenamefont {Vidal}(2007{\natexlab{b}})}]{vidal2}%
  \BibitemOpen
  \bibfield  {author} {\bibinfo {author} {\bibfnamefont {G.}~\bibnamefont
  {Vidal}},\ }\bibfield  {title} {\bibinfo {title} {Entanglement
  renormalization},\ }\href {https://doi.org/10.1103/PhysRevLett.99.220405}
  {\bibfield  {journal} {\bibinfo  {journal} {Phys. Rev. Lett.}\ }\textbf
  {\bibinfo {volume} {99}},\ \bibinfo {pages} {220405} (\bibinfo {year}
  {2007}{\natexlab{b}})}\BibitemShut {NoStop}%
\bibitem [{\citenamefont {Evenbly}\ and\ \citenamefont
  {Vidal}(2014)}]{Evenbly_2014}%
  \BibitemOpen
  \bibfield  {author} {\bibinfo {author} {\bibfnamefont {G.}~\bibnamefont
  {Evenbly}}\ and\ \bibinfo {author} {\bibfnamefont {G.}~\bibnamefont
  {Vidal}},\ }\bibfield  {title} {\bibinfo {title} {Algorithms for entanglement
  renormalization: Boundaries, impurities and interfaces},\ }\href
  {https://doi.org/10.1007/s10955-014-0983-1} {\bibfield  {journal} {\bibinfo
  {journal} {Journal of Statistical Physics}\ }\textbf {\bibinfo {volume}
  {157}},\ \bibinfo {pages} {931} (\bibinfo {year} {2014})}\BibitemShut
  {NoStop}%
\bibitem [{\citenamefont {Liang}\ \emph {et~al.}(1988)\citenamefont {Liang},
  \citenamefont {Doucot},\ and\ \citenamefont {Anderson}}]{RVB}%
  \BibitemOpen
  \bibfield  {author} {\bibinfo {author} {\bibfnamefont {S.}~\bibnamefont
  {Liang}}, \bibinfo {author} {\bibfnamefont {B.}~\bibnamefont {Doucot}},\ and\
  \bibinfo {author} {\bibfnamefont {P.~W.}\ \bibnamefont {Anderson}},\
  }\bibfield  {title} {\bibinfo {title} {Some new variational
  resonating-valence-bond-type wave functions for the spin-\textonehalf{}
  antiferromagnetic heisenberg model on a square lattice},\ }\href
  {https://doi.org/10.1103/PhysRevLett.61.365} {\bibfield  {journal} {\bibinfo
  {journal} {Phys. Rev. Lett.}\ }\textbf {\bibinfo {volume} {61}},\ \bibinfo
  {pages} {365} (\bibinfo {year} {1988})}\BibitemShut {NoStop}%
\bibitem [{\citenamefont {Carleo}\ and\ \citenamefont
  {Troyer}(2017)}]{carleo_treuer}%
  \BibitemOpen
  \bibfield  {author} {\bibinfo {author} {\bibfnamefont {G.}~\bibnamefont
  {Carleo}}\ and\ \bibinfo {author} {\bibfnamefont {M.}~\bibnamefont
  {Troyer}},\ }\bibfield  {title} {\bibinfo {title} {Solving the quantum
  many-body problem with artificial neural networks},\ }\href
  {https://doi.org/10.1126/science.aag2302} {\bibfield  {journal} {\bibinfo
  {journal} {Science}\ }\textbf {\bibinfo {volume} {355}},\ \bibinfo {pages}
  {602} (\bibinfo {year} {2017})},\ \Eprint
  {https://arxiv.org/abs/https://www.science.org/doi/pdf/10.1126/science.aag2302}
  {https://www.science.org/doi/pdf/10.1126/science.aag2302} \BibitemShut
  {NoStop}%
\bibitem [{\citenamefont {Mezzacapo}\ \emph {et~al.}(2009)\citenamefont
  {Mezzacapo}, \citenamefont {Schuch}, \citenamefont {Boninsegni},\ and\
  \citenamefont {Cirac}}]{Mezzacapo_2009}%
  \BibitemOpen
  \bibfield  {author} {\bibinfo {author} {\bibfnamefont {F.}~\bibnamefont
  {Mezzacapo}}, \bibinfo {author} {\bibfnamefont {N.}~\bibnamefont {Schuch}},
  \bibinfo {author} {\bibfnamefont {M.}~\bibnamefont {Boninsegni}},\ and\
  \bibinfo {author} {\bibfnamefont {J.~I.}\ \bibnamefont {Cirac}},\ }\bibfield
  {title} {\bibinfo {title} {Ground-state properties of quantum many-body
  systems: entangled-plaquette states and variational monte carlo},\ }\href
  {https://doi.org/10.1088/1367-2630/11/8/083026} {\bibfield  {journal}
  {\bibinfo  {journal} {New Journal of Physics}\ }\textbf {\bibinfo {volume}
  {11}},\ \bibinfo {pages} {083026} (\bibinfo {year} {2009})}\BibitemShut
  {NoStop}%
\bibitem [{\citenamefont {Changlani}\ \emph {et~al.}(2009)\citenamefont
  {Changlani}, \citenamefont {Kinder}, \citenamefont {Umrigar},\ and\
  \citenamefont {Chan}}]{PhysRevB.80.245116}%
  \BibitemOpen
  \bibfield  {author} {\bibinfo {author} {\bibfnamefont {H.~J.}\ \bibnamefont
  {Changlani}}, \bibinfo {author} {\bibfnamefont {J.~M.}\ \bibnamefont
  {Kinder}}, \bibinfo {author} {\bibfnamefont {C.~J.}\ \bibnamefont
  {Umrigar}},\ and\ \bibinfo {author} {\bibfnamefont {G.~K.-L.}\ \bibnamefont
  {Chan}},\ }\bibfield  {title} {\bibinfo {title} {Approximating strongly
  correlated wave functions with correlator product states},\ }\href
  {https://doi.org/10.1103/PhysRevB.80.245116} {\bibfield  {journal} {\bibinfo
  {journal} {Phys. Rev. B}\ }\textbf {\bibinfo {volume} {80}},\ \bibinfo
  {pages} {245116} (\bibinfo {year} {2009})}\BibitemShut {NoStop}%
\bibitem [{\citenamefont {Thibaut}\ \emph {et~al.}(2019)\citenamefont
  {Thibaut}, \citenamefont {Roscilde},\ and\ \citenamefont
  {Mezzacapo}}]{PhysRevB.100.155148}%
  \BibitemOpen
  \bibfield  {author} {\bibinfo {author} {\bibfnamefont {J.}~\bibnamefont
  {Thibaut}}, \bibinfo {author} {\bibfnamefont {T.}~\bibnamefont {Roscilde}},\
  and\ \bibinfo {author} {\bibfnamefont {F.}~\bibnamefont {Mezzacapo}},\
  }\bibfield  {title} {\bibinfo {title} {Long-range entangled-plaquette states
  for critical and frustrated quantum systems on a lattice},\ }\href
  {https://doi.org/10.1103/PhysRevB.100.155148} {\bibfield  {journal} {\bibinfo
   {journal} {Phys. Rev. B}\ }\textbf {\bibinfo {volume} {100}},\ \bibinfo
  {pages} {155148} (\bibinfo {year} {2019})}\BibitemShut {NoStop}%
\bibitem [{\citenamefont {Levin}\ and\ \citenamefont
  {Wen}(2006)}]{LevinWen2006}%
  \BibitemOpen
  \bibfield  {author} {\bibinfo {author} {\bibfnamefont {M.}~\bibnamefont
  {Levin}}\ and\ \bibinfo {author} {\bibfnamefont {X.-G.}\ \bibnamefont
  {Wen}},\ }\bibfield  {title} {\bibinfo {title} {Detecting topological order
  in a ground state wave function},\ }\href
  {https://doi.org/10.1103/PhysRevLett.96.110405} {\bibfield  {journal}
  {\bibinfo  {journal} {Phys. Rev. Lett.}\ }\textbf {\bibinfo {volume} {96}},\
  \bibinfo {pages} {110405} (\bibinfo {year} {2006})}\BibitemShut {NoStop}%
\bibitem [{\citenamefont {Wen}(2017)}]{wen2017}%
  \BibitemOpen
  \bibfield  {author} {\bibinfo {author} {\bibfnamefont {X.-G.}\ \bibnamefont
  {Wen}},\ }\bibfield  {title} {\bibinfo {title} {Colloquium: Zoo of
  quantum-topological phases of matter},\ }\href
  {https://doi.org/10.1103/RevModPhys.89.041004} {\bibfield  {journal}
  {\bibinfo  {journal} {Rev. Mod. Phys.}\ }\textbf {\bibinfo {volume} {89}},\
  \bibinfo {pages} {041004} (\bibinfo {year} {2017})}\BibitemShut {NoStop}%
\bibitem [{\citenamefont {Landau}\ \emph {et~al.}(2015)\citenamefont {Landau},
  \citenamefont {Vazirani},\ and\ \citenamefont {Vidick}}]{Efficient1D}%
  \BibitemOpen
  \bibfield  {author} {\bibinfo {author} {\bibfnamefont {Z.}~\bibnamefont
  {Landau}}, \bibinfo {author} {\bibfnamefont {U.}~\bibnamefont {Vazirani}},\
  and\ \bibinfo {author} {\bibfnamefont {T.}~\bibnamefont {Vidick}},\
  }\bibfield  {title} {\bibinfo {title} {A polynomial time algorithm for the
  ground state of one-dimensional gapped local hamiltonians},\ }\href
  {https://doi.org/10.1038/nphys3345} {\bibfield  {journal} {\bibinfo
  {journal} {Nature Physics}\ }\textbf {\bibinfo {volume} {11}},\ \bibinfo
  {pages} {566} (\bibinfo {year} {2015})}\BibitemShut {NoStop}%
\bibitem [{\citenamefont {Schuch}\ \emph {et~al.}(2007)\citenamefont {Schuch},
  \citenamefont {Wolf}, \citenamefont {Verstraete},\ and\ \citenamefont
  {Cirac}}]{ComplexPEPS}%
  \BibitemOpen
  \bibfield  {author} {\bibinfo {author} {\bibfnamefont {N.}~\bibnamefont
  {Schuch}}, \bibinfo {author} {\bibfnamefont {M.~M.}\ \bibnamefont {Wolf}},
  \bibinfo {author} {\bibfnamefont {F.}~\bibnamefont {Verstraete}},\ and\
  \bibinfo {author} {\bibfnamefont {J.~I.}\ \bibnamefont {Cirac}},\ }\bibfield
  {title} {\bibinfo {title} {Computational complexity of projected entangled
  pair states},\ }\href {https://doi.org/10.1103/PhysRevLett.98.140506}
  {\bibfield  {journal} {\bibinfo  {journal} {Phys. Rev. Lett.}\ }\textbf
  {\bibinfo {volume} {98}},\ \bibinfo {pages} {140506} (\bibinfo {year}
  {2007})}\BibitemShut {NoStop}%
\bibitem [{Note1()}]{Note1}%
  \BibitemOpen
  \bibinfo {note} {Notice that ${\protect \mathcal {M}}_Q$ forms a convex set:
  this property follows directly from the convexity of the set of quantum
  density matrices.}\BibitemShut {Stop}%
\bibitem [{\citenamefont {Liu}\ \emph {et~al.}(2007)\citenamefont {Liu},
  \citenamefont {Christandl},\ and\ \citenamefont {Verstraete}}]{QMA_Repr}%
  \BibitemOpen
  \bibfield  {author} {\bibinfo {author} {\bibfnamefont {Y.-K.}\ \bibnamefont
  {Liu}}, \bibinfo {author} {\bibfnamefont {M.}~\bibnamefont {Christandl}},\
  and\ \bibinfo {author} {\bibfnamefont {F.}~\bibnamefont {Verstraete}},\
  }\bibfield  {title} {\bibinfo {title} {Quantum computational complexity of
  the $n$-representability problem: Qma complete},\ }\href
  {https://doi.org/10.1103/PhysRevLett.98.110503} {\bibfield  {journal}
  {\bibinfo  {journal} {Phys. Rev. Lett.}\ }\textbf {\bibinfo {volume} {98}},\
  \bibinfo {pages} {110503} (\bibinfo {year} {2007})}\BibitemShut {NoStop}%
\bibitem [{\citenamefont {Anderson}(1951)}]{AndersonBound}%
  \BibitemOpen
  \bibfield  {author} {\bibinfo {author} {\bibfnamefont {P.~W.}\ \bibnamefont
  {Anderson}},\ }\bibfield  {title} {\bibinfo {title} {Limits on the energy of
  the antiferromagnetic ground state},\ }\href
  {https://doi.org/10.1103/PhysRev.83.1260} {\bibfield  {journal} {\bibinfo
  {journal} {Phys. Rev.}\ }\textbf {\bibinfo {volume} {83}},\ \bibinfo {pages}
  {1260} (\bibinfo {year} {1951})}\BibitemShut {NoStop}%
\bibitem [{\citenamefont {Slofstra}(2019)}]{slofstra_2019}%
  \BibitemOpen
  \bibfield  {author} {\bibinfo {author} {\bibfnamefont {W.}~\bibnamefont
  {Slofstra}},\ }\bibfield  {title} {\bibinfo {title} {The set of quantum
  correlations is not closed},\ }\href {https://doi.org/10.1017/fmp.2018.3}
  {\bibfield  {journal} {\bibinfo  {journal} {Forum of Mathematics, Pi}\
  }\textbf {\bibinfo {volume} {7}},\ \bibinfo {pages} {e1} (\bibinfo {year}
  {2019})}\BibitemShut {NoStop}%
\bibitem [{\citenamefont {Han}\ \emph {et~al.}(2020)\citenamefont {Han},
  \citenamefont {Hartnoll},\ and\ \citenamefont {Kruthoff}}]{conformal}%
  \BibitemOpen
  \bibfield  {author} {\bibinfo {author} {\bibfnamefont {X.}~\bibnamefont
  {Han}}, \bibinfo {author} {\bibfnamefont {S.~A.}\ \bibnamefont {Hartnoll}},\
  and\ \bibinfo {author} {\bibfnamefont {J.}~\bibnamefont {Kruthoff}},\
  }\bibfield  {title} {\bibinfo {title} {Bootstrapping matrix quantum
  mechanics},\ }\href {https://doi.org/10.1103/PhysRevLett.125.041601}
  {\bibfield  {journal} {\bibinfo  {journal} {Phys. Rev. Lett.}\ }\textbf
  {\bibinfo {volume} {125}},\ \bibinfo {pages} {041601} (\bibinfo {year}
  {2020})}\BibitemShut {NoStop}%
\bibitem [{\citenamefont {Lasserre}(2001)}]{lasserre2001global}%
  \BibitemOpen
  \bibfield  {author} {\bibinfo {author} {\bibfnamefont {J.~B.}\ \bibnamefont
  {Lasserre}},\ }\bibfield  {title} {\bibinfo {title} {Global optimization with
  polynomials and the problem of moments},\ }\href@noop {} {\bibfield
  {journal} {\bibinfo  {journal} {SIAM Journal on optimization}\ }\textbf
  {\bibinfo {volume} {11}},\ \bibinfo {pages} {796} (\bibinfo {year}
  {2001})}\BibitemShut {NoStop}%
\bibitem [{Note2()}]{Note2}%
  \BibitemOpen
  \bibinfo {note} {Our codes for reproducing the results are available at
  \protect \url {https://github.com/wangjie212/QMBCertify}. See \protect \url
  {https://github.com/blegat/CondensedMatterSOS.jl} for other related
  codes.}\BibitemShut {Stop}%
\bibitem [{\citenamefont {Wang}(2023)}]{wang2023efficient}%
  \BibitemOpen
  \bibfield  {author} {\bibinfo {author} {\bibfnamefont {J.}~\bibnamefont
  {Wang}},\ }\href@noop {} {\bibinfo {title} {A more efficient reformulation of
  complex {SDP} as real {SDP}}} (\bibinfo {year} {2023}),\ \bibinfo {note}
  {preprint \url{https:arXiv.org/abs/2307.11599}}\BibitemShut {NoStop}%
\bibitem [{\citenamefont {Giamarchi}(2003)}]{giamarchi2003}%
  \BibitemOpen
  \bibfield  {author} {\bibinfo {author} {\bibfnamefont {T.}~\bibnamefont
  {Giamarchi}},\ }\href
  {https://doi.org/10.1093/acprof:oso/9780198525004.001.0001} {\emph {\bibinfo
  {title} {{Quantum Physics in One Dimension}}}}\ (\bibinfo  {publisher}
  {Oxford University Press},\ \bibinfo {year} {2003})\BibitemShut {NoStop}%
\bibitem [{\citenamefont {Haim}\ \emph {et~al.}(2020)\citenamefont {Haim},
  \citenamefont {Kueng},\ and\ \citenamefont
  {Refael}}]{haim2020variationalcorrelations}%
  \BibitemOpen
  \bibfield  {author} {\bibinfo {author} {\bibfnamefont {A.}~\bibnamefont
  {Haim}}, \bibinfo {author} {\bibfnamefont {R.}~\bibnamefont {Kueng}},\ and\
  \bibinfo {author} {\bibfnamefont {G.}~\bibnamefont {Refael}},\ }\href@noop {}
  {\bibinfo {title} {Variational-correlations approach to quantum many-body
  problems}} (\bibinfo {year} {2020}),\ \Eprint
  {https://arxiv.org/abs/2001.06510} {arXiv:2001.06510 [cond-mat.str-el]}
  \BibitemShut {NoStop}%
\bibitem [{\citenamefont {White}\ and\ \citenamefont
  {Affleck}(1996)}]{white_affleck_1996}%
  \BibitemOpen
  \bibfield  {author} {\bibinfo {author} {\bibfnamefont {S.~R.}\ \bibnamefont
  {White}}\ and\ \bibinfo {author} {\bibfnamefont {I.}~\bibnamefont
  {Affleck}},\ }\bibfield  {title} {\bibinfo {title} {Dimerization and
  incommensurate spiral spin correlations in the zigzag spin chain: Analogies
  to the kondo lattice},\ }\href {https://doi.org/10.1103/PhysRevB.54.9862}
  {\bibfield  {journal} {\bibinfo  {journal} {Phys. Rev. B}\ }\textbf {\bibinfo
  {volume} {54}},\ \bibinfo {pages} {9862} (\bibinfo {year}
  {1996})}\BibitemShut {NoStop}%
\bibitem [{\citenamefont {Majumdar}\ and\ \citenamefont
  {Ghosh}(1969)}]{majumdar1969next}%
  \BibitemOpen
  \bibfield  {author} {\bibinfo {author} {\bibfnamefont {C.~K.}\ \bibnamefont
  {Majumdar}}\ and\ \bibinfo {author} {\bibfnamefont {D.~K.}\ \bibnamefont
  {Ghosh}},\ }\bibfield  {title} {\bibinfo {title} {On next-nearest-neighbor
  interaction in linear chain. i},\ }\href@noop {} {\bibfield  {journal}
  {\bibinfo  {journal} {Journal of Mathematical Physics}\ }\textbf {\bibinfo
  {volume} {10}},\ \bibinfo {pages} {1388} (\bibinfo {year}
  {1969})}\BibitemShut {NoStop}%
\bibitem [{\citenamefont {Sandvik}(1997)}]{sandvik1997}%
  \BibitemOpen
  \bibfield  {author} {\bibinfo {author} {\bibfnamefont {A.~W.}\ \bibnamefont
  {Sandvik}},\ }\bibfield  {title} {\bibinfo {title} {Finite-size scaling of
  the ground-state parameters of the two-dimensional heisenberg model},\ }\href
  {https://doi.org/10.1103/PhysRevB.56.11678} {\bibfield  {journal} {\bibinfo
  {journal} {Phys. Rev. B}\ }\textbf {\bibinfo {volume} {56}},\ \bibinfo
  {pages} {11678} (\bibinfo {year} {1997})}\BibitemShut {NoStop}%
\bibitem [{\citenamefont {Zhitomirsky}\ and\ \citenamefont
  {Ueda}(1996)}]{zhitomirskyU1996}%
  \BibitemOpen
  \bibfield  {author} {\bibinfo {author} {\bibfnamefont {M.~E.}\ \bibnamefont
  {Zhitomirsky}}\ and\ \bibinfo {author} {\bibfnamefont {K.}~\bibnamefont
  {Ueda}},\ }\bibfield  {title} {\bibinfo {title} {Valence-bond crystal phase
  of a frustrated spin-1/2 square-lattice antiferromagnet},\ }\href
  {https://doi.org/10.1103/PhysRevB.54.9007} {\bibfield  {journal} {\bibinfo
  {journal} {Phys. Rev. B}\ }\textbf {\bibinfo {volume} {54}},\ \bibinfo
  {pages} {9007} (\bibinfo {year} {1996})}\BibitemShut {NoStop}%
\bibitem [{\citenamefont {Capriotti}\ and\ \citenamefont
  {Sorella}(2000)}]{capriottiS2000}%
  \BibitemOpen
  \bibfield  {author} {\bibinfo {author} {\bibfnamefont {L.}~\bibnamefont
  {Capriotti}}\ and\ \bibinfo {author} {\bibfnamefont {S.}~\bibnamefont
  {Sorella}},\ }\bibfield  {title} {\bibinfo {title} {Spontaneous plaquette
  dimerization in the ${\mathit{j}}_{1}--{\mathit{j}}_{2}$ heisenberg model},\
  }\href {https://doi.org/10.1103/PhysRevLett.84.3173} {\bibfield  {journal}
  {\bibinfo  {journal} {Phys. Rev. Lett.}\ }\textbf {\bibinfo {volume} {84}},\
  \bibinfo {pages} {3173} (\bibinfo {year} {2000})}\BibitemShut {NoStop}%
\bibitem [{\citenamefont {Jiang}\ \emph {et~al.}(2012)\citenamefont {Jiang},
  \citenamefont {Yao},\ and\ \citenamefont {Balents}}]{jiangetal2012}%
  \BibitemOpen
  \bibfield  {author} {\bibinfo {author} {\bibfnamefont {H.-C.}\ \bibnamefont
  {Jiang}}, \bibinfo {author} {\bibfnamefont {H.}~\bibnamefont {Yao}},\ and\
  \bibinfo {author} {\bibfnamefont {L.}~\bibnamefont {Balents}},\ }\bibfield
  {title} {\bibinfo {title} {Spin liquid ground state of the spin-$\frac{1}{2}$
  square ${J}_{1}$-${J}_{2}$ heisenberg model},\ }\href
  {https://doi.org/10.1103/PhysRevB.86.024424} {\bibfield  {journal} {\bibinfo
  {journal} {Phys. Rev. B}\ }\textbf {\bibinfo {volume} {86}},\ \bibinfo
  {pages} {024424} (\bibinfo {year} {2012})}\BibitemShut {NoStop}%
\bibitem [{\citenamefont {Wang}\ \emph {et~al.}(2013)\citenamefont {Wang},
  \citenamefont {Poilblanc}, \citenamefont {Gu}, \citenamefont {Wen},\ and\
  \citenamefont {Verstraete}}]{wangetal2013}%
  \BibitemOpen
  \bibfield  {author} {\bibinfo {author} {\bibfnamefont {L.}~\bibnamefont
  {Wang}}, \bibinfo {author} {\bibfnamefont {D.}~\bibnamefont {Poilblanc}},
  \bibinfo {author} {\bibfnamefont {Z.-C.}\ \bibnamefont {Gu}}, \bibinfo
  {author} {\bibfnamefont {X.-G.}\ \bibnamefont {Wen}},\ and\ \bibinfo {author}
  {\bibfnamefont {F.}~\bibnamefont {Verstraete}},\ }\bibfield  {title}
  {\bibinfo {title} {Constructing a gapless spin-liquid state for the
  spin-$1/2$ ${J}_{1}\ensuremath{-}{J}_{2}$ heisenberg model on a square
  lattice},\ }\href {https://doi.org/10.1103/PhysRevLett.111.037202} {\bibfield
   {journal} {\bibinfo  {journal} {Phys. Rev. Lett.}\ }\textbf {\bibinfo
  {volume} {111}},\ \bibinfo {pages} {037202} (\bibinfo {year}
  {2013})}\BibitemShut {NoStop}%
\bibitem [{\citenamefont {Gong}\ \emph {et~al.}(2014)\citenamefont {Gong},
  \citenamefont {Zhu}, \citenamefont {Sheng}, \citenamefont {Motrunich},\ and\
  \citenamefont {Fisher}}]{gongetal2014}%
  \BibitemOpen
  \bibfield  {author} {\bibinfo {author} {\bibfnamefont {S.-S.}\ \bibnamefont
  {Gong}}, \bibinfo {author} {\bibfnamefont {W.}~\bibnamefont {Zhu}}, \bibinfo
  {author} {\bibfnamefont {D.~N.}\ \bibnamefont {Sheng}}, \bibinfo {author}
  {\bibfnamefont {O.~I.}\ \bibnamefont {Motrunich}},\ and\ \bibinfo {author}
  {\bibfnamefont {M.~P.~A.}\ \bibnamefont {Fisher}},\ }\bibfield  {title}
  {\bibinfo {title} {Plaquette ordered phase and quantum phase diagram in the
  spin-$\frac{1}{2}$ ${J}_{1}\text{\ensuremath{-}}{J}_{2}$ square heisenberg
  model},\ }\href {https://doi.org/10.1103/PhysRevLett.113.027201} {\bibfield
  {journal} {\bibinfo  {journal} {Phys. Rev. Lett.}\ }\textbf {\bibinfo
  {volume} {113}},\ \bibinfo {pages} {027201} (\bibinfo {year}
  {2014})}\BibitemShut {NoStop}%
\bibitem [{\citenamefont {Choo}\ \emph {et~al.}(2019)\citenamefont {Choo},
  \citenamefont {Neupert},\ and\ \citenamefont {Carleo}}]{chooetal2019}%
  \BibitemOpen
  \bibfield  {author} {\bibinfo {author} {\bibfnamefont {K.}~\bibnamefont
  {Choo}}, \bibinfo {author} {\bibfnamefont {T.}~\bibnamefont {Neupert}},\ and\
  \bibinfo {author} {\bibfnamefont {G.}~\bibnamefont {Carleo}},\ }\bibfield
  {title} {\bibinfo {title} {Two-dimensional frustrated
  ${J}_{1}\text{\ensuremath{-}}{J}_{2}$ model studied with neural network
  quantum states},\ }\href {https://doi.org/10.1103/PhysRevB.100.125124}
  {\bibfield  {journal} {\bibinfo  {journal} {Phys. Rev. B}\ }\textbf {\bibinfo
  {volume} {100}},\ \bibinfo {pages} {125124} (\bibinfo {year}
  {2019})}\BibitemShut {NoStop}%
\bibitem [{\citenamefont {Wang}\ and\ \citenamefont
  {Sandvik}(2018)}]{PhysRevLett.121.107202}%
  \BibitemOpen
  \bibfield  {author} {\bibinfo {author} {\bibfnamefont {L.}~\bibnamefont
  {Wang}}\ and\ \bibinfo {author} {\bibfnamefont {A.~W.}\ \bibnamefont
  {Sandvik}},\ }\bibfield  {title} {\bibinfo {title} {Critical level crossings
  and gapless spin liquid in the square-lattice spin-$1/2$
  ${J}_{1}\ensuremath{-}{J}_{2}$ heisenberg antiferromagnet},\ }\href
  {https://doi.org/10.1103/PhysRevLett.121.107202} {\bibfield  {journal}
  {\bibinfo  {journal} {Phys. Rev. Lett.}\ }\textbf {\bibinfo {volume} {121}},\
  \bibinfo {pages} {107202} (\bibinfo {year} {2018})}\BibitemShut {NoStop}%
\bibitem [{\citenamefont {Nomura}\ and\ \citenamefont
  {Imada}(2021)}]{PhysRevX.11.031034}%
  \BibitemOpen
  \bibfield  {author} {\bibinfo {author} {\bibfnamefont {Y.}~\bibnamefont
  {Nomura}}\ and\ \bibinfo {author} {\bibfnamefont {M.}~\bibnamefont {Imada}},\
  }\bibfield  {title} {\bibinfo {title} {Dirac-type nodal spin liquid revealed
  by refined quantum many-body solver using neural-network wave function,
  correlation ratio, and level spectroscopy},\ }\href
  {https://doi.org/10.1103/PhysRevX.11.031034} {\bibfield  {journal} {\bibinfo
  {journal} {Phys. Rev. X}\ }\textbf {\bibinfo {volume} {11}},\ \bibinfo
  {pages} {031034} (\bibinfo {year} {2021})}\BibitemShut {NoStop}%
\bibitem [{\citenamefont {Ferrari}\ and\ \citenamefont
  {Becca}(2020)}]{PhysRevB.102.014417}%
  \BibitemOpen
  \bibfield  {author} {\bibinfo {author} {\bibfnamefont {F.}~\bibnamefont
  {Ferrari}}\ and\ \bibinfo {author} {\bibfnamefont {F.}~\bibnamefont
  {Becca}},\ }\bibfield  {title} {\bibinfo {title} {Gapless spin liquid and
  valence-bond solid in the ${J}_{1}$-${J}_{2}$ heisenberg model on the square
  lattice: Insights from singlet and triplet excitations},\ }\href
  {https://doi.org/10.1103/PhysRevB.102.014417} {\bibfield  {journal} {\bibinfo
   {journal} {Phys. Rev. B}\ }\textbf {\bibinfo {volume} {102}},\ \bibinfo
  {pages} {014417} (\bibinfo {year} {2020})}\BibitemShut {NoStop}%
\bibitem [{\citenamefont {Liu}\ \emph {et~al.}(2022)\citenamefont {Liu},
  \citenamefont {Gong}, \citenamefont {Li}, \citenamefont {Poilblanc},
  \citenamefont {Chen},\ and\ \citenamefont {Gu}}]{LIU20221034}%
  \BibitemOpen
  \bibfield  {author} {\bibinfo {author} {\bibfnamefont {W.-Y.}\ \bibnamefont
  {Liu}}, \bibinfo {author} {\bibfnamefont {S.-S.}\ \bibnamefont {Gong}},
  \bibinfo {author} {\bibfnamefont {Y.-B.}\ \bibnamefont {Li}}, \bibinfo
  {author} {\bibfnamefont {D.}~\bibnamefont {Poilblanc}}, \bibinfo {author}
  {\bibfnamefont {W.-Q.}\ \bibnamefont {Chen}},\ and\ \bibinfo {author}
  {\bibfnamefont {Z.-C.}\ \bibnamefont {Gu}},\ }\bibfield  {title} {\bibinfo
  {title} {Gapless quantum spin liquid and global phase diagram of the spin-1/2
  j1-j2 square antiferromagnetic heisenberg model},\ }\href
  {https://doi.org/https://doi.org/10.1016/j.scib.2022.03.010} {\bibfield
  {journal} {\bibinfo  {journal} {Science Bulletin}\ }\textbf {\bibinfo
  {volume} {67}},\ \bibinfo {pages} {1034} (\bibinfo {year}
  {2022})}\BibitemShut {NoStop}%
\bibitem [{\citenamefont {Cho}\ \emph {et~al.}(2022)\citenamefont {Cho},
  \citenamefont {Gabai}, \citenamefont {Lin}, \citenamefont {Rodriguez},
  \citenamefont {Sandor},\ and\ \citenamefont {Yin}}]{cho2022bootstrapping}%
  \BibitemOpen
  \bibfield  {author} {\bibinfo {author} {\bibfnamefont {M.}~\bibnamefont
  {Cho}}, \bibinfo {author} {\bibfnamefont {B.}~\bibnamefont {Gabai}}, \bibinfo
  {author} {\bibfnamefont {Y.-H.}\ \bibnamefont {Lin}}, \bibinfo {author}
  {\bibfnamefont {V.~A.}\ \bibnamefont {Rodriguez}}, \bibinfo {author}
  {\bibfnamefont {J.}~\bibnamefont {Sandor}},\ and\ \bibinfo {author}
  {\bibfnamefont {X.}~\bibnamefont {Yin}},\ }\href
  {https://doi.org/10.48550/arXiv.2206.12538} {\bibinfo {title} {Bootstrapping
  the ising model on the lattice}} (\bibinfo {year} {2022}),\ \Eprint
  {https://arxiv.org/abs/2206.12538} {arXiv:2206.12538 [hep-th]} \BibitemShut
  {NoStop}%
\bibitem [{\citenamefont {Requena}\ \emph {et~al.}(2023)\citenamefont
  {Requena}, \citenamefont {Mu\~noz Gil}, \citenamefont {Lewenstein},
  \citenamefont {Dunjko},\ and\ \citenamefont {Tura}}]{MLbounds}%
  \BibitemOpen
  \bibfield  {author} {\bibinfo {author} {\bibfnamefont {B.}~\bibnamefont
  {Requena}}, \bibinfo {author} {\bibfnamefont {G.}~\bibnamefont {Mu\~noz
  Gil}}, \bibinfo {author} {\bibfnamefont {M.}~\bibnamefont {Lewenstein}},
  \bibinfo {author} {\bibfnamefont {V.}~\bibnamefont {Dunjko}},\ and\ \bibinfo
  {author} {\bibfnamefont {J.}~\bibnamefont {Tura}},\ }\bibfield  {title}
  {\bibinfo {title} {Certificates of quantum many-body properties assisted by
  machine learning},\ }\href {https://doi.org/10.1103/PhysRevResearch.5.013097}
  {\bibfield  {journal} {\bibinfo  {journal} {Phys. Rev. Res.}\ }\textbf
  {\bibinfo {volume} {5}},\ \bibinfo {pages} {013097} (\bibinfo {year}
  {2023})}\BibitemShut {NoStop}%
\bibitem [{\citenamefont {Fawzi}\ \emph
  {et~al.}(2023{\natexlab{a}})\citenamefont {Fawzi}, \citenamefont {Fawzi},\
  and\ \citenamefont {Scalet}}]{fawzi2023certified}%
  \BibitemOpen
  \bibfield  {author} {\bibinfo {author} {\bibfnamefont {H.}~\bibnamefont
  {Fawzi}}, \bibinfo {author} {\bibfnamefont {O.}~\bibnamefont {Fawzi}},\ and\
  \bibinfo {author} {\bibfnamefont {S.~O.}\ \bibnamefont {Scalet}},\ }\href
  {https://doi.org/10.48550/arXiv.2311.18706} {\bibinfo {title} {Certified
  algorithms for equilibrium states of local quantum hamiltonians}} (\bibinfo
  {year} {2023}{\natexlab{a}}),\ \Eprint {https://arxiv.org/abs/2311.18706}
  {arXiv:2311.18706 [quant-ph]} \BibitemShut {NoStop}%
\bibitem [{\citenamefont {Araújo}\ \emph {et~al.}(2023)\citenamefont
  {Araújo}, \citenamefont {Klep}, \citenamefont {Vértesi}, \citenamefont
  {Garner},\ and\ \citenamefont {Navascues}}]{araujo2023karushkuhntucker}%
  \BibitemOpen
  \bibfield  {author} {\bibinfo {author} {\bibfnamefont {M.}~\bibnamefont
  {Araújo}}, \bibinfo {author} {\bibfnamefont {I.}~\bibnamefont {Klep}},
  \bibinfo {author} {\bibfnamefont {T.}~\bibnamefont {Vértesi}}, \bibinfo
  {author} {\bibfnamefont {A.~J.~P.}\ \bibnamefont {Garner}},\ and\ \bibinfo
  {author} {\bibfnamefont {M.}~\bibnamefont {Navascues}},\ }\href
  {https://doi.org/10.48550/arXiv.2311.18707} {\bibinfo {title}
  {Karush-kuhn-tucker conditions for non-commutative optimization problems}}
  (\bibinfo {year} {2023}),\ \Eprint {https://arxiv.org/abs/2311.18707}
  {arXiv:2311.18707 [quant-ph]} \BibitemShut {NoStop}%
\bibitem [{\citenamefont {Fawzi}\ \emph
  {et~al.}(2023{\natexlab{b}})\citenamefont {Fawzi}, \citenamefont {Fawzi},\
  and\ \citenamefont {Scalet}}]{fawzi2023entropy}%
  \BibitemOpen
  \bibfield  {author} {\bibinfo {author} {\bibfnamefont {H.}~\bibnamefont
  {Fawzi}}, \bibinfo {author} {\bibfnamefont {O.}~\bibnamefont {Fawzi}},\ and\
  \bibinfo {author} {\bibfnamefont {S.~O.}\ \bibnamefont {Scalet}},\
  }\href@noop {} {\bibinfo {title} {Entropy constraints for ground energy
  optimization}} (\bibinfo {year} {2023}{\natexlab{b}}),\ \bibinfo {note}
  {preprint \url{https://arxiv.org/abs/2305.06855}}\BibitemShut {NoStop}%
\bibitem [{\citenamefont {Kull}\ \emph {et~al.}(2022)\citenamefont {Kull},
  \citenamefont {Schuch}, \citenamefont {Dive},\ and\ \citenamefont
  {Navascués}}]{kull2022lower}%
  \BibitemOpen
  \bibfield  {author} {\bibinfo {author} {\bibfnamefont {I.}~\bibnamefont
  {Kull}}, \bibinfo {author} {\bibfnamefont {N.}~\bibnamefont {Schuch}},
  \bibinfo {author} {\bibfnamefont {B.}~\bibnamefont {Dive}},\ and\ \bibinfo
  {author} {\bibfnamefont {M.}~\bibnamefont {Navascués}},\ }\href@noop {}
  {\bibinfo {title} {Lower bounding ground-state energies of local hamiltonians
  through the renormalization group}} (\bibinfo {year} {2022}),\ \Eprint
  {https://arxiv.org/abs/2212.03014} {arXiv:2212.03014 [quant-ph]} \BibitemShut
  {NoStop}%
\bibitem [{\citenamefont {Magron}\ \emph {et~al.}(2015)\citenamefont {Magron},
  \citenamefont {Allamigeon}, \citenamefont {Gaubert},\ and\ \citenamefont
  {Werner}}]{magron2015}%
  \BibitemOpen
  \bibfield  {author} {\bibinfo {author} {\bibfnamefont {V.}~\bibnamefont
  {Magron}}, \bibinfo {author} {\bibfnamefont {X.}~\bibnamefont {Allamigeon}},
  \bibinfo {author} {\bibfnamefont {S.}~\bibnamefont {Gaubert}},\ and\ \bibinfo
  {author} {\bibfnamefont {B.}~\bibnamefont {Werner}},\ }\bibfield  {title}
  {\bibinfo {title} {Formal proofs for nonlinear optimization},\ }\href
  {https://doi.org/10.6092/issn.1972-5787/4319} {\bibfield  {journal} {\bibinfo
   {journal} {Journal of Formalized Reasoning}\ }\textbf {\bibinfo {volume}
  {8}},\ \bibinfo {pages} {1–24} (\bibinfo {year} {2015})}\BibitemShut
  {NoStop}%
\bibitem [{\citenamefont {Cafuta}\ \emph {et~al.}(2015)\citenamefont {Cafuta},
  \citenamefont {Klep},\ and\ \citenamefont {Povh}}]{cafuta2015rational}%
  \BibitemOpen
  \bibfield  {author} {\bibinfo {author} {\bibfnamefont {K.}~\bibnamefont
  {Cafuta}}, \bibinfo {author} {\bibfnamefont {I.}~\bibnamefont {Klep}},\ and\
  \bibinfo {author} {\bibfnamefont {J.}~\bibnamefont {Povh}},\ }\bibfield
  {title} {\bibinfo {title} {Rational sums of hermitian squares of free
  noncommutative polynomials.},\ }\href@noop {} {\bibfield  {journal} {\bibinfo
   {journal} {Ars Math. Contemp.}\ }\textbf {\bibinfo {volume} {9}},\ \bibinfo
  {pages} {243} (\bibinfo {year} {2015})}\BibitemShut {NoStop}%
\bibitem [{\citenamefont {Wang}\ and\ \citenamefont
  {Hu}(2023)}]{wang2023solving}%
  \BibitemOpen
  \bibfield  {author} {\bibinfo {author} {\bibfnamefont {J.}~\bibnamefont
  {Wang}}\ and\ \bibinfo {author} {\bibfnamefont {L.}~\bibnamefont {Hu}},\
  }\bibfield  {title} {\bibinfo {title} {Solving low-rank semidefinite programs
  via manifold optimization},\ }\href@noop {} {\bibfield  {journal} {\bibinfo
  {journal} {arXiv preprint arXiv:2303.01722}\ } (\bibinfo {year}
  {2023})}\BibitemShut {NoStop}%
\bibitem [{\citenamefont {P{\'a}l}\ and\ \citenamefont
  {V{\'e}rtesi}(2009)}]{pal2009quantum}%
  \BibitemOpen
  \bibfield  {author} {\bibinfo {author} {\bibfnamefont {K.~F.}\ \bibnamefont
  {P{\'a}l}}\ and\ \bibinfo {author} {\bibfnamefont {T.}~\bibnamefont
  {V{\'e}rtesi}},\ }\bibfield  {title} {\bibinfo {title} {Quantum bounds on
  bell inequalities},\ }\href@noop {} {\bibfield  {journal} {\bibinfo
  {journal} {Physical Review A}\ }\textbf {\bibinfo {volume} {79}},\ \bibinfo
  {pages} {022120} (\bibinfo {year} {2009})}\BibitemShut {NoStop}%
\bibitem [{\citenamefont {Lasserre}(2006)}]{lasserre2006convergent}%
  \BibitemOpen
  \bibfield  {author} {\bibinfo {author} {\bibfnamefont {J.~B.}\ \bibnamefont
  {Lasserre}},\ }\bibfield  {title} {\bibinfo {title} {Convergent
  sdp-relaxations in polynomial optimization with sparsity},\ }\href@noop {}
  {\bibfield  {journal} {\bibinfo  {journal} {SIAM Journal on optimization}\
  }\textbf {\bibinfo {volume} {17}},\ \bibinfo {pages} {822} (\bibinfo {year}
  {2006})}\BibitemShut {NoStop}%
\bibitem [{\citenamefont {Klep}\ \emph {et~al.}(2021)\citenamefont {Klep},
  \citenamefont {Magron},\ and\ \citenamefont {Povh}}]{klep2021sparse}%
  \BibitemOpen
  \bibfield  {author} {\bibinfo {author} {\bibfnamefont {I.}~\bibnamefont
  {Klep}}, \bibinfo {author} {\bibfnamefont {V.}~\bibnamefont {Magron}},\ and\
  \bibinfo {author} {\bibfnamefont {J.}~\bibnamefont {Povh}},\ }\bibfield
  {title} {\bibinfo {title} {Sparse noncommutative polynomial optimization},\
  }\href@noop {} {\bibfield  {journal} {\bibinfo  {journal} {Mathematical
  Programming}\ ,\ \bibinfo {pages} {1}} (\bibinfo {year} {2021})}\BibitemShut
  {NoStop}%
\bibitem [{\citenamefont {Wang}\ \emph {et~al.}(2021)\citenamefont {Wang},
  \citenamefont {Magron},\ and\ \citenamefont {Lasserre}}]{wang2021tssos}%
  \BibitemOpen
  \bibfield  {author} {\bibinfo {author} {\bibfnamefont {J.}~\bibnamefont
  {Wang}}, \bibinfo {author} {\bibfnamefont {V.}~\bibnamefont {Magron}},\ and\
  \bibinfo {author} {\bibfnamefont {J.-B.}\ \bibnamefont {Lasserre}},\
  }\bibfield  {title} {\bibinfo {title} {{TSSOS}: A moment-{SOS} hierarchy that
  exploits term sparsity},\ }\href@noop {} {\bibfield  {journal} {\bibinfo
  {journal} {SIAM Journal on optimization}\ }\textbf {\bibinfo {volume} {31}},\
  \bibinfo {pages} {30} (\bibinfo {year} {2021})}\BibitemShut {NoStop}%
\bibitem [{\citenamefont {Wang}\ and\ \citenamefont
  {Magron}(2021)}]{wang2021exploiting}%
  \BibitemOpen
  \bibfield  {author} {\bibinfo {author} {\bibfnamefont {J.}~\bibnamefont
  {Wang}}\ and\ \bibinfo {author} {\bibfnamefont {V.}~\bibnamefont {Magron}},\
  }\bibfield  {title} {\bibinfo {title} {Exploiting term sparsity in
  noncommutative polynomial optimization},\ }\href@noop {} {\bibfield
  {journal} {\bibinfo  {journal} {Computational Optimization and Applications}\
  }\textbf {\bibinfo {volume} {80}},\ \bibinfo {pages} {483} (\bibinfo {year}
  {2021})}\BibitemShut {NoStop}%
\bibitem [{\citenamefont {Magron}\ and\ \citenamefont
  {Wang}(2023)}]{magron2023sparse}%
  \BibitemOpen
  \bibfield  {author} {\bibinfo {author} {\bibfnamefont {V.}~\bibnamefont
  {Magron}}\ and\ \bibinfo {author} {\bibfnamefont {J.}~\bibnamefont {Wang}},\
  }\href@noop {} {\emph {\bibinfo {title} {Sparse polynomial optimization:
  theory and practice}}}\ (\bibinfo  {publisher} {World Scientific},\ \bibinfo
  {year} {2023})\BibitemShut {NoStop}%
\bibitem [{\citenamefont {Lofberg}(2009)}]{Lofberg2009}%
  \BibitemOpen
  \bibfield  {author} {\bibinfo {author} {\bibfnamefont {J.}~\bibnamefont
  {Lofberg}},\ }\bibfield  {title} {\bibinfo {title} {Pre-and post-processing
  sum-of-squares programs in practice},\ }\href@noop {} {\bibfield  {journal}
  {\bibinfo  {journal} {IEEE Trans Autom Control}\ }\textbf {\bibinfo {volume}
  {54}},\ \bibinfo {pages} {1007} (\bibinfo {year} {2009})}\BibitemShut
  {NoStop}%
\bibitem [{\citenamefont {Hrga}\ \emph {et~al.}(2023)\citenamefont {Hrga},
  \citenamefont {Klep},\ and\ \citenamefont {Povh}}]{klep2023}%
  \BibitemOpen
  \bibfield  {author} {\bibinfo {author} {\bibfnamefont {T.}~\bibnamefont
  {Hrga}}, \bibinfo {author} {\bibfnamefont {I.}~\bibnamefont {Klep}},\ and\
  \bibinfo {author} {\bibfnamefont {J.}~\bibnamefont {Povh}},\ }\href@noop {}
  {\bibinfo {title} {Certifying optimality of bell inequality violations:
  Noncommutative polynomial optimization through semidefinite programming and
  local optimization}} (\bibinfo {year} {2023}),\ \bibinfo {note} {preprint
  \url{https://users.fmf.uni-lj.si/klep/Hrga\_Klep\_Povh\_v2.pdf}}\BibitemShut
  {NoStop}%
\bibitem [{\citenamefont {Riener}\ \emph {et~al.}(2013)\citenamefont {Riener},
  \citenamefont {Theobald}, \citenamefont {Andr{\'e}n},\ and\ \citenamefont
  {Lasserre}}]{riener2013exploiting}%
  \BibitemOpen
  \bibfield  {author} {\bibinfo {author} {\bibfnamefont {C.}~\bibnamefont
  {Riener}}, \bibinfo {author} {\bibfnamefont {T.}~\bibnamefont {Theobald}},
  \bibinfo {author} {\bibfnamefont {L.~J.}\ \bibnamefont {Andr{\'e}n}},\ and\
  \bibinfo {author} {\bibfnamefont {J.~B.}\ \bibnamefont {Lasserre}},\
  }\bibfield  {title} {\bibinfo {title} {Exploiting symmetries in
  sdp-relaxations for polynomial optimization},\ }\href@noop {} {\bibfield
  {journal} {\bibinfo  {journal} {Mathematics of Operations Research}\ }\textbf
  {\bibinfo {volume} {38}},\ \bibinfo {pages} {122} (\bibinfo {year}
  {2013})}\BibitemShut {NoStop}%
\bibitem [{\citenamefont {{H.J. Schulz}}\ \emph {et~al.}(1996)\citenamefont
  {{H.J. Schulz}}, \citenamefont {{T.A.L. Ziman}},\ and\ \citenamefont {{D.
  Poilblanc}}}]{Schulz96}%
  \BibitemOpen
  \bibfield  {author} {\bibinfo {author} {\bibnamefont {{H.J. Schulz}}},
  \bibinfo {author} {\bibnamefont {{T.A.L. Ziman}}},\ and\ \bibinfo {author}
  {\bibnamefont {{D. Poilblanc}}},\ }\bibfield  {title} {\bibinfo {title}
  {Magnetic order and disorder in the frustrated quantum heisenberg
  antiferromagnet in two dimensions},\ }\href
  {https://doi.org/10.1051/jp1:1996236} {\bibfield  {journal} {\bibinfo
  {journal} {J. Phys. I France}\ }\textbf {\bibinfo {volume} {6}},\ \bibinfo
  {pages} {675} (\bibinfo {year} {1996})}\BibitemShut {NoStop}%
\end{thebibliography}%
\end{document}